\documentclass[structabstract]{aa} 
\usepackage[dvips]{graphicx}
\usepackage{amsmath}
\usepackage{subfigure}
\usepackage{booktabs}
\usepackage{threeparttable}
\usepackage{wasysym}
\usepackage{natbib}
\usepackage{url}

\usepackage{txfonts}
\graphicspath{{figures01/}}

\begin{document}
%
\title{The 69$\mu$m forsterite band in spectra of protoplanetary disks - Results from the Herschel DIGIT programme}

\author{ B.~Sturm\inst{1} 
\and J.~Bouwman\inst{1} 
\and Th.~Henning\inst{1} 
\and N.J.~Evans~II\inst{2} 
\and L.B.F.M.~Waters\inst{3,8,11} 
\and E.F.~van~Dishoeck\inst{4,6}
\and J.D.~Green \inst{2} 
\and J.~Olofsson\inst{1}
\and G.~Meeus \inst{5} 
\and K.~Maaskant \inst{3}
\and C.~Dominik\inst{3}
\and J.~C.~Augereau\inst{7}
\and G.D.~Mulders\inst{3,11} 
\and B.~Acke \inst{8}
\and B.~Merin\inst{9}
\and G.~J.~Herczeg \inst{6,10} 
\and The DIGIT team
          }

\institute{Max Planck Institute for Astronomy, K\"onigstuhl 17, D-69117 
 Heidelberg, Germany
\and  The University of Texas at Austin, Department of Astronomy, 
 2515 Speedway, Stop C1400 Austin, TX 78712-1205, USA
\and Astronomical Institute ``Anton Pannekoek'', University of Amsterdam, 
 PO Box 94249, 1090 GE Amsterdam, The Netherlands
\and Leiden Observatory, Leiden University, PO Box 9513, 2300 RA Leiden, 
 The Netherlands
\and Dep. de F\'{i}sica Te\'{o}rica, Fac. de Ciencias Universidad 
 Aut\'{o}noma de Madrid, Campus Cantoblanco 28049 Madrid, Spain
\and Max Planck Institute for extraterrestrial Physics, Garching, Germany
\and UJF-Grenoble 1 / CNRS-INSU, Institut de Plan\'{e}tologie et d'Astrophysique de Grenoble (IPAG), UMR 5274, Grenoble, F-38041, France
\and Instituut voor Sterrenkunde, Katholieke Universiteit Leuven, 
 Celestijnenlaan 200D, 3001 Heverlee, Belgium
\and ESAC, Madrid, Spain
\and Kavli Institute for Astronomy and Astrophysics, Ye He Yuan Lu 5, Beijing, 100871, P.R. China
\and SRON Netherlands Institute for Space Research, P.O. Box 800, 9700 AV Groningen, The Netherlands
             }
\offprints{B.~Sturm, \email{sturm@mpia.de}}

  \abstract
{We have analysed far--infrared spectra of 32 circumstellar disks around Herbig Ae/Be and T~Tauri stars obtained within the \emph{Herschel} key programme Dust, Ice and Gas in Time (DIGIT). The spectra were taken with the Photodetector Array Camera and Spectrometer (PACS) on board the \emph{Herschel Space Observatory}. In this paper we focus on the detection and analysis of the 69\,$\mu$m emission band of the crystalline silicate forsterite.
}
{This work aims at providing an overview of the 69\,$\mu$m forsterite bands present in the DIGIT sample. We use characteristics of the emission band (peak position and FWHM) to derive the dust temperature and to constrain the iron content of the crystalline silicates. With this information, constraints can be placed on the spatial distribution of the forsterite in the disk and the formation history of the crystalline grains.
}
{The 69\,$\mu$m forsterite emission feature is analysed in terms of position and shape to derive the temperature and composition of the dust by comparison to laboratory spectra of that band. The PACS spectra are combined with existing \emph{Spitzer} IRS spectra and we compare the presence and strength of the 69~$\mu$m band to the forsterite bands at shorter wavelengths.
}
{A total of 32 disk sources have been observed. Out of these 32, 8 sources show a 69\,$\mu$m emission feature that can be attributed to forsterite. With the exception of the T~Tauri star AS~205,  all of the detections are for disks associated with Herbig Ae/Be stars. Most of the forsterite grains that give rise to the 69\,$\mu$m bands are found to be warm ($\sim$100--200\,K) and iron-poor (less than $\sim$2\% iron). AB~Aur is the only source where the emission cannot be fitted with iron-free forsterite requiring approximately 3--4\% of iron. 
}
{Our findings support the hypothesis that the forsterite grains form through an equilibrium condensation process at high temperatures. The large width of the emission band in some sources may indicate the presence of forsterite reservoirs at different temperatures. The connection between the strength of the 69 and 33~$\mu$m bands shows that at least part of the emission in these two bands originates fom the same dust grains. We further find that any model that can explain the PACS and the \emph{Spitzer} IRS observations must take the effects of a wavelength dependent optical depth into account. \textnormal{We find weak indications of a correlation of the detection rate of the 69~$\mu$m band with the spectral type of the host stars in our sample. However, the sample size is too small to obtain a definitive result.}
}

 \titlerunning{The Forsterite 69$\mu$m band in the \emph{Herschel} DIGIT programme}
  \authorrunning{B. Sturm et al}
 
   \maketitle
%
%
\section{Introduction}
%
The diversity of planetary system architectures and planet properties is certainly related to a range of physical and chemical conditions in protoplanetary disks \citep[e.g ][]{Mordasini2012}. Dust particles in disks undergo a wide range of physical and chemical processes, ranging from dust growth through cohesive coagulation to annealing, sublimation and re-condensation \citep[e.g. ][]{Natta2007, HenningMeeus2011, Gail2010, GailAstromineralogy2010}. Infrared (IR) spectroscopy turns out to be a versatile tool to characterise the physical and chemical structure of the dust particles through the study of emission features arising from small dust grains.

Near- and mid-IR radiation comes from the upper optically thin layer of the disk atmosphere heated by stellar radiation \citep{Menshchikov1997, ChiangGoldreich1997}. The infrared spectra of these regions in the disks have been intensively studied, first by ISO \citep[e.g. ][]{Bouwman2003, Meeus2001} for intermediate-mass Herbig Ae/Be stars and later with much improved sensitivity provided by \emph{Spitzer} for disks around brown dwarfs/low-mass stars \citep[e.g. ][]{Pascucci2009}, T~Tauri stars \citep{Bouwman2008, KesslerSilacci2006, Olofsson2009, Furlan2009, Watson2009} and Herbig Ae/Be stars \citep{Juhasz2010}. Regions closer to the mid-plane can only be probed by observations at (sub)-millimetre wavelengths where the disks become optically thin.

These deeper layers of protoplanetary disks are optically thick to instruments in the near and mid--IR range. At around 70\,$\mu$m, the emission of regions closer to the midplane is detectable even in  relatively massive disks \citep[see, e.g.][]{Mulders2011}. An overview of the spectral features at these wavelengths and the information they carry is given by \citet{vanDishoeck2004}. Studies in this regime have been conducted with ISO, for example by \citet{Malfait1998}, \citet{Malfait1999b} and \citet{Lorenzetti2002}.

The emission band at $\sim$69~$\mu$m, associated with crystalline olivine grains, is much more sensitive to the temperature and the iron content of the crystalline dust particles than any of the olivine bands at shorter wavelengths \citep[e.g. ][]{Bowey2002,Koike2003,Suto2006}. Prior to the \emph{Herschel} observations, the 69~$\mu$m band has only been detected by ISO in one protoplanetary disk  \citep[HD~100546, ][]{Malfait1998}. However, that detection had a low signal--to--noise ratio (S/N) and the shape of the emission band was never analysed in terms of temperature or iron content. \emph{Herschel} PACS spectra of that particular source were discussed by \citet{Sturm2010}, who presented the 69~$\mu$m band at unprecedented S/N and resolution and who found the emission to emerge from very iron-poor dust at a temperature of about $\sim$150--200~K.

In this paper we present an overview and analysis on the 69\,$\mu$m forsterite emission band in the disk sample observed within the \emph{Herschel} key programme DIGIT. The sources in the DIGIT sample cover a wide range of disk properties including age, luminosity and effective temperature. Through the chosen set of disk properties the sample includes examples of different evolutionary phases. All sources in the brightness--limited sample have been drawn from previous studies with a focus on those objects for which high--quality \emph{Spitzer} IRS 5--35~$\mu$m spectra exist. This allows for a comparison of the hot, geometrically thin surface layer to the interior of the disks. All \emph{Herschel} observations in this work were taken with the PACS instrument.

In this paper we search for the 69\,$\mu$m emission band of forsterite (see section \ref{sec:forsterite-detection}). In section \ref{sec:forsterite-analysis} we analyse the position and shape of the detected emission bands in terms of temperature and iron content. A brief discussion on the possible influence of grain size is included. Furthermore, we compare the 69\,$\mu$m band to the forsterite emission bands at mid-IR wavelengths (e.g. 16 and 33\,$\mu$m) from \emph{Spitzer} observations. We search for a relation in the peak over continuum ratio in the different wavelength regimes, which would give further information about the spatial distribution and the formation history of the forsterite in the disk.

\section{Observation and data reduction}
\label{sec:datareduction_observation}
%

%
\begin{table*}
\begin{center}

\caption{Disk sources observed as part of the \emph{Herschel} key programme DIGIT.}
\label{tab:sources01}
\small
\begin{tabular}{@{}lrrlcrcrc@{}}
\toprule
        &   \multicolumn{2}{c}{Coordinates (J 2000)}   & &   \\
Star    &   RA [h m s] & DEC [d m s]                   &  spectral type & ref. & log($L_{\star}/L_{\astrosun}$) & ref  & age (Myr) & ref\\
\cmidrule{1-9}
HD 150193    &   16 40 17.92     &   -23 53 45.2    & A1$-$3Ve  & (1)& 1.47 &  (11)    & $>$2.0 &(11) \\
HD 97048     &   11 08 03.32     &   -77 39 17.5    & B9.5Ve+sh & (1)& 1.61 &  (11)    & $>$2.0 &(11) \\
HD 169142    &   18 24 29.78     &   -29 46 49.4    & A5V       & (5)& 1.58 &  (5)     & 7.7$\pm$2.0 &(10) \\
HD 98922     &   11 22 31.67     &   -53 22 11.5    & B9 Ve     & (11)& $>$2.96 &  (11)& $<$0.01 &(15) \\
HD 100453    &   11 33 05.58     &   -54 19 28.5    & A9 Ve     & (1)& 0.95 &  (12)    & $>$10  &(10) \\
HD 135344    &   15 15 48.95     &   -37 08 56.1    & F4Ve      & (1)& 1.02 &  (7)     & 9$\pm$2& (7) \\
HD 179218    &   19 11 11.25     &   +15 47 15.6    & B9e       & (1)& 2.50 &  (11)    & 1.1$\pm$0.7&(14) \\
HD 203024    &   21 16 03.02     &   +68 54 52.1    & A/B8.5 V  & (19)& $\sim$2.0 &  (19)      & ---- & \\
IRS 48       &   16 27 37.19     &   -24 30 35.0    & A?        & (2)& 1.16 &  (8)     & $\sim$15&(8) \\
SR 21        &   16 27 10.28     &   -24 19 12.5    & G~2.5     & (3)& 1.45 &  (3)     & $\sim$1.0&(3) \\
HD 38120     &   5 43 11.89      &   -4 59 49.9     & Be*       & (2)& 1.74 &  (9)     & $<$1.0 &(13) \\
HT Lup       &   15 45 12.87     &   -34 17 30.6    & K3 Ve     & (2)& ---- &        & 0.5$\pm$0.1 &(16) \\
HD 35187     &   5 24 01.17      &   +24 57 37.6    & A2Ve+A7Ve & (1)& 1.24 &  (10)    & 9.0$\pm$2 &(10) \\
S CrA        &   19 01 08.60     &   -36 57 20.0    & K6e       & (3)& 0.36 &  (3)     & $\sim$3.0 &(3) \\
HD 104237    &   12 00 05.08     &   -78 11 34.6    & A7.5Ve + K3 & (4)& 1.55&  (4)    & 5.5$\pm$0.5 &(10) \\
RY Lup       &   15 59 28.39     &   -40 21 51.2    & G0V       & (2)&  ---- &       & 29.7$\pm$7.5 &(16) \\
HD 144432    &   16 06 57.96     &   -27 43 09.8    & A9IVev    & (1)& $>$1.48 &  (11) & 5.5$\pm$2.0 &(14) \\
AS 205       &   16 11 31.35     &   -18 38 26.1    & K5        & (3)&  0.85 &  (3)    & $\sim$0.1 &(3) \\
HD 144668    &   16 08 34.29     &   -39 06 18.3    & A5$-$7III/IVe+sh  & (1)& 1.71 & (10) & 2.8$\pm$1.0 &(10) \\
RU Lup       &   15 56 42.31     &   -37 49 15.5    & G5 Ve     & (2)& ---- &        & 55.3$\pm$10.4 &(16) \\
HD 141569    &   15 49 57.75     &   -3 55 16.4     & A0 Ve     & (1)& 1.35 &  (11)    & 4.7$\pm$0.3 &(10) \\
EC 82        &   18 29 56.89     &   +1 14 46.5     & K8 D      & (18)& -0.179 & (18)  & $\sim$1.5 &(18) \\
HD 50138     &   6 51 33.40      &   -6 57 59.4     & B9        & (2)& 2.85 &  (11)    & 0.5$\pm$0.2 &(16) \\ 
HD 142666    &   15 56 40.02     &   -22 01 40.0    & A8Ve      & (1)& 1.13 &  (10)    & 9.0$\pm$2.0 &(10) \\
DG Tau       &   4 27 04.70      &   +26 06 16.3    & G Ve      & (2)& -0.05 &  (17)      & $\sim$0.6 &(17) \\
RNO 90       &   16 34 09.17     &   -15 48 16.8    & G5 D      & (2)& ---- &        & ---- & \\
HD 100546    &   11 33 25.44     &   -70 11 41.2    & B9Vne     & (1)& 1.51 &  (11)    & $>$10 &(11) \\
AB Aur       &   4 55 45.84      &   +30 33 04.3    & A0Ve+sh   & (1)& 1.68 &  (11)    & 5.0$\pm$1.0 &(10) \\
HD 163296    &   17 56 21.29     &   -21 57 21.9    & A3Ve      & (1)& 1.48 &  (11)    & 5.5$\pm$0.5 &(10) \\
HD 142527    &   15 56 41.89     &   -42 19 23.3    & F7IIIe    & (1)& 1.84 &  (11)    & 2.0$\pm$0.5 &(10) \\
HD 36112     &   5 30 27.53      &   +25 19 57.1    & A5 Ve     & (2)& 1.35 &  (11)    & 3.7$\pm$2.0 &(10) \\
HD~139614    &   15 40 46.38     &   -42 29 53.5    & A7Ve      & (1)& 0.88 &  (10)    & 10$\pm$2.0 & (10)\\
                                                      
\bottomrule                                           
\end{tabular}
\normalsize
\end{center}
\textbf{The references for spectral classifications and ages are:} (1) \citet{AckevdAncker2004}, (2) The Simbad database, (3) \citet{Prato2003}, (4) \citet{Boehm2004}, (5) \citet{Tilling2012}, (6) \citet{Meeus2002}, (7) \citet{Mueller2011}, (8) \citet{Brown2012}, (9) \citet{Hernandez2005}, (10) \citet{Meeus2012}, (11) \citet{vdAncker1998}, (12) \citet{Dominik2003}, (13) \citet{Sartori2010}, (14) \citet{Folsom2012}, (15) \citet{Manoj2006}, (16) \citet{Tetzlaff2011}, (17) \citet{Palla2002}, (18) \citet{Winston2010}, (19) \citet{Miroshnichenko1999}.\\
\textbf{Note:} Luminosities and ages are derived by different methods.

\end{table*}

In this work we present Photodetector Array Camera and Spectrometer \citep[PACS;][]{Poglitsch2010} spectra of protoplanetary disks in a sample of Herbig~Ae/Be and T~Tauri systems. These observations were taken as part of the Dust, Ice and Gas in Time (DIGIT) Herschel key programme. The list of sources we observed is provided in Table \ref{tab:sources01}.

The PACS instrument consists of a 5$\times$5 array of 9.4\arcsec$\times$9.4\arcsec\ spatial pixels (here after referred to as spaxels) covering the spectral range from 51-210~$\mu$m with $\lambda$/$\Delta\lambda$ $\sim$ 1000-3000. The spectrum is divided into four segments, covering $\lambda \sim$ 50-75, 70-105, 100-145, and 140-210~$\mu$m. The spatial resolution of PACS ranges from $\sim$9\arcsec\ at 50~$\mu$m to $\sim$18\arcsec\ at 210~$\mu$m. All of our targets were observed in the standard ``rangescan'' spectroscopy mode with a grating stepsize corresponding to Nyquist sampling \cite[see further ][]{Poglitsch2010}. In this paper we focus on the spectral range of 67--72~$\mu$m observed in the second spectral order with the ``blue'' detector. 

We processed our data using the Herschel Interactive Processing Environment \citep[HIPE,][]{Ott2010} using calibration version 42 and standard pipeline scripts. The infrared background emission was removed using two chop-nod positions 6\arcmin\ from the source in opposite directions. Absolute flux calibration was achieved by normalising our spectra to the emission from the telescope mirror itself as measured by the off-source positions, and a detailed model of the telescope emission available in HIPE. The sources were usually well-centred on the central spaxel which we used to extract the spectra as this provided the highest signal to noise ratio (S/N) in the spectra. However, small pointing errors and drifts of the telescope can lead to flux losses and spectral artifacts. To mitigate this we scaled the spectra derived from the central spaxel to the integrated flux over the entire array. This approach guarantees the best absolute flux calibration with the highest S/N spectra. Only for one case, HD~142666 , the 
telescope had such a large mispointing that the central spaxel did not contain most of the source flux. Here we opted to use the spectra integrated over the central 3$\times$3 spaxels to recover the total flux. Spectral rebinning was done with an oversampling of a factor of two and an upscaling of a factor of one corresponding to Nyquist sampling. For further details on the data reduction procedure for the entire DIGIT dataset we refer to Green et al. (subm.).

\section{Detection of forsterite}
\label{sec:forsterite-detection}
\subsection{Characteristics of the 69~$\mu$m emission band}
\label{subsect:detection1}
Olivine, a nesosilicate with an orthorombic crystal structure, forms a complete solid solution series from 
forsterite ($\text{Mg}_{2}\text{SiO}_4$) to fayalite ($\text{Fe}_{2}\text{SiO}_4$). The general chemical composition
is given by Mg$_{2(1-x)}$Fe$_{2x}$SiO$_4$, where $x$ is the fraction of iron cations relative to magnesium.
Magnesium--rich olivines give rise to a characteristic emission band at $\sim$69~$\mu$m. The most pronounced feature is found in forsterite --the iron-free end member of the olivines \citep{Henning2010}. The position and shape of the band depends on, in order of importance, the iron content, the temperature and the shape and size of the dust grains.

In this paper we will use the term ``forsterite'' or ``pure forsterite'' for completely iron-free olivines. ``Iron-poor forsterite'' is used for olivines with less than 5\% of iron cations relative to magnesium. When using the term ``iron-poor olivines'' we refer to olivines with no more than 10\% of iron. These terms are used to refer to the dust particles; the emission band is always labelled ``forsterite emission band'' or ``69~$\mu$m band'' as it is strongest in pure forsterite and vanishes completely in iron--rich olivines.

\subsubsection{Effects of iron content}
\label{subsubsect:ironcontent}
\begin{figure}[ht!]
\centering
\includegraphics[scale=.65, angle=90]{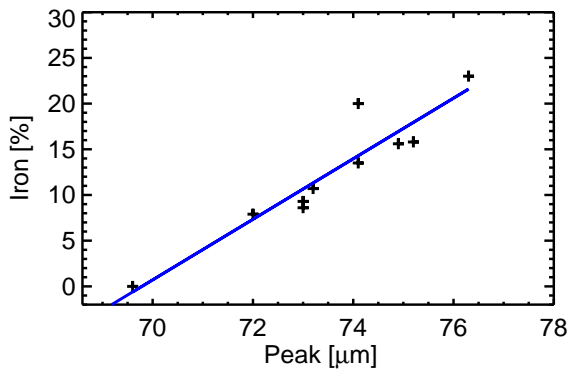}
\caption{The influence of the iron cation fraction relative to magnesium in the olivine material on the peak wavelength of the 69~$\mu$m band. The data (black crosses) are from \citet{Koike2003}, taken at 295~K and the blue line is a linear fit.}
\label{fig:peakIronLab}
\end{figure}

For olivines in general, the position and shape of the emission peak strongly depends on the iron content. The 69~$\mu$m band is shifted linearly toward longer wavelengths with the amount of iron in the crystals \citep[e.g.][]{Koike2003}. Even a small fraction of iron (x $\sim 0.05$) leads to a band position at $\lambda > 70$~$\mu$m (see Fig. \ref{fig:peakIronLab}).

%

Laboratory measurements are only available for pure forsterite and for olivines with more than  $\sim$10~\% of iron content. From the available data (presented in Fig. \ref{fig:peakIronLab}) a linear connection between iron content in the olivine dust grains and the peak position of the 69~$\mu$m band can be derived. Most of the available laboratory measurements were taken at room temperature. However, corrections for temperature effects can be applied if necessary, based on the temperature dependence of the pure forsterite and a sample with 9.3~\% iron \citep{Koike2006}.

The FWHM of the emission band is also increasing with the iron content in the grains. However, the available data is sparse, introducing additional uncertainties in the quantitative analysis of our detections. Therefore our estimated upper limits on the iron content in the crystalline silicates from which the 69~$\mu$m band emerges are mostly based on the position of the peak.

\subsubsection{Effects of temperature}
\label{subsubsect:temperature}
\begin{figure}[ht!]
\centering
\includegraphics[scale=.65, angle=90]{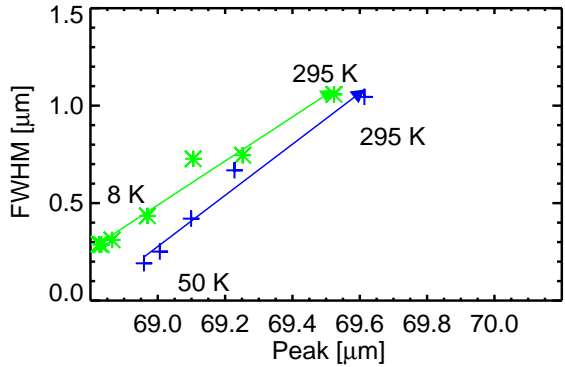}
\caption{The effect of temperature on the peak position and FWHM of the 69~$\mu$m band. The blue crosses are based on optical constants from \citet{Suto2006} using the distribution of hollow spheres by \citet{Min2003} with a grain size of 0.1~$\mu$m. The green asterisks represent mass absorption coefficients from \citet{Koike2006}. Both datasets were taken on pure forsterite.}
\label{fig:peakFWHMLab}
\end{figure}

Next to the strong dependency on the iron content, the position and width of the 69~$\mu$m forsterite band also depends on the grain temperature. This temperature dependence has been characterised in several laboratory experiments \citep[e.g.][]{Bowey2002,Koike2006,Suto2006}, covering a range from 8 to 295\,K. In Fig. \ref{fig:peakFWHMLab} we show the FWHM and peak position of the 69~$\mu$m band at different temperatures. The mass absorption coefficients from the optical constants by \citet{Suto2006} were computed using ``distribution of hollow spheres''  scattering theory \citep[DHS, ][]{Min2003}. We used a filling factor $f_{\mathrm{max}}$ of 1 and a grain size of 0.1~$\mu$m. From 8 to 295~K the peak position moved from 68.8~$\mu$m to 69.6~$\mu$m  and the FWHM of the emission band increases from $\sim$0.2 to 1.1~$\mu$m as can be seen from Fig.~\ref{fig:peakFWHMLab}.  Note that all of these measurements were carried out on pure forsterite. The temperature dependence of the 69~$\mu$m band at 9.3\% iron 
content is shown in \citet{Koike2006}. The effect is qualitatively similar to that in pure forsterite.

%

The offset between the different datasets shown in Fig. \ref{fig:peakFWHMLab} can be explained by the effects of grain size, grain shape, amount of lattice distortions, possible effects from the scattering theory \citep[see e.g., ][]{Mutschke2009} as well as eventual environmental contamination during laboratory experiments \citep{HenningMutschke2010}. Such differences can be the effect of different methods used in sample preparation. In the following section we will discuss, and take into account, the effects of grain size and shape but not that of lattice distortions. As noted by \citet{Koike2010}, the number of lattice distortions affects the shape (and to a minor extent the position) of the emission band. The broadening of the IR emission bands observed by \citet{Koike2010} and attributed to lattice distortions, however, is too small to be detected in the presence of the other effects we discuss in this paper. Therefore, we will will not take it into account in our further analysis.

\subsubsection{Effects of grain shape}
\label{subsubset:grainshape}

\begin{figure}[ht!]
\centering
\includegraphics[scale=.65, angle=90]{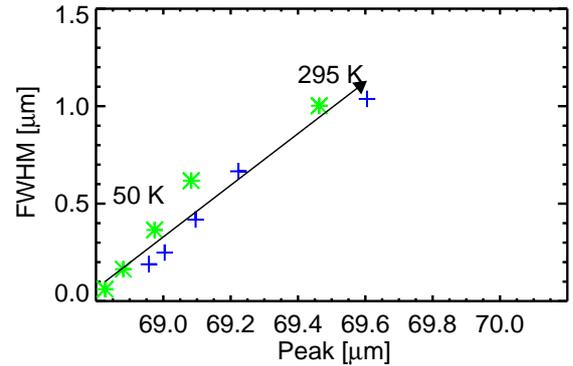}
\caption{The effect of the shape model used for the forsterite grains on the peak position and FWHM. Shown as blue crosses is the distribution of hollow spheres \citep{Min2003} while green asterisks stand for Mie scattering on spherical grains. Both examples are based on the data from \citet{Suto2006} and cover the temperature range from 50 to 295~K. The grain size is 0.1~$\mu$m and the black arrow indicates the temperature dependency.}
\label{fig:grainshapeLab}
\end{figure}

Another factor that has to be taken into account is the shape model which is needed to compute the mass absorption coefficients of dust grains from the refractive index of the material \citep{HenningMutschke2010}.  In Fig.~\ref{fig:grainshapeLab} we compare the calculated peak position and FWHM for compact spherical grains (Mie scattering) and the distribution of hollow spheres \citep[DHS; ][]{Min2003} model with a filling factor $f_{\mathrm{max}} = 1$, which was found to be a good model for the crystalline silicate grains producing features in the \emph{Spitzer} IRS wavelength regime (5--35~$\mu$m) \citep{Juhasz2010}. The temperature range covered is 50--295~K and the data is based on the results by \citet{Suto2006}. 

%

On average, both the peak position and the FWHM computed with the DHS model are shifted to larger values by about 0.1$\mu$m compared to Mie scattering. However, as can be seen from Fig.~\ref{fig:grainshapeLab}, the grain shape has little influence on the relation between the grain temperature and the peak position and FWHM of the forsterite band, which shows the far stronger effect. The same holds for the dependency on the iron content. From mid-IR observations we know that purely spherical grains do not reproduce the observed spectral features well. In this work, therefore, we adopt the DHS model with a maximum filling factor $f_{\mathrm{max}} = 1$ as shown in Fig. \ref{fig:grainshapeLab}. Results from a third widely used theory, the continuous distribution of ellipsoids \citep[CDE;][]{Min2003} are almost identical to those obtained from the DHS model. The difference is too small to lead to significant changes in the results for temperature or iron content. Also, as no grain size effects can be 
taken into account in this model, we do not consider CDE in the remainder of this paper.

\subsubsection{Effects of grain size}
\label{subsubsect:grainsize}
%
\begin{figure}
\centering
\includegraphics[scale=.65,angle=90]{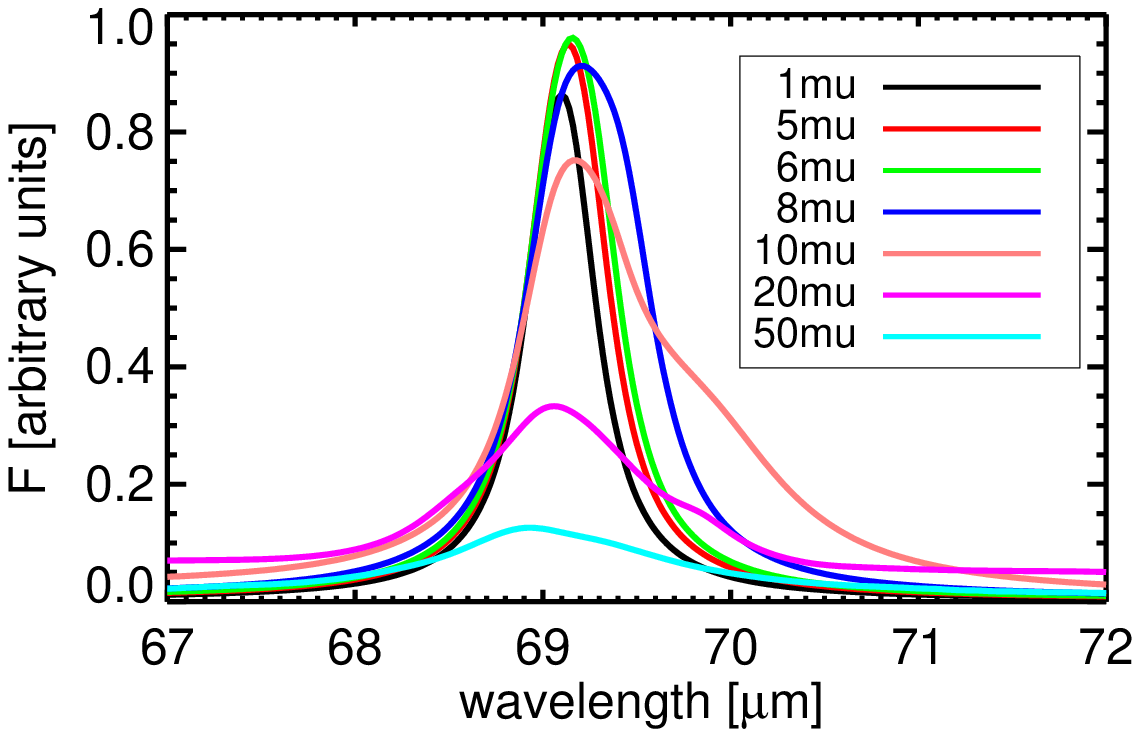}
\includegraphics[scale=.65,angle=90]{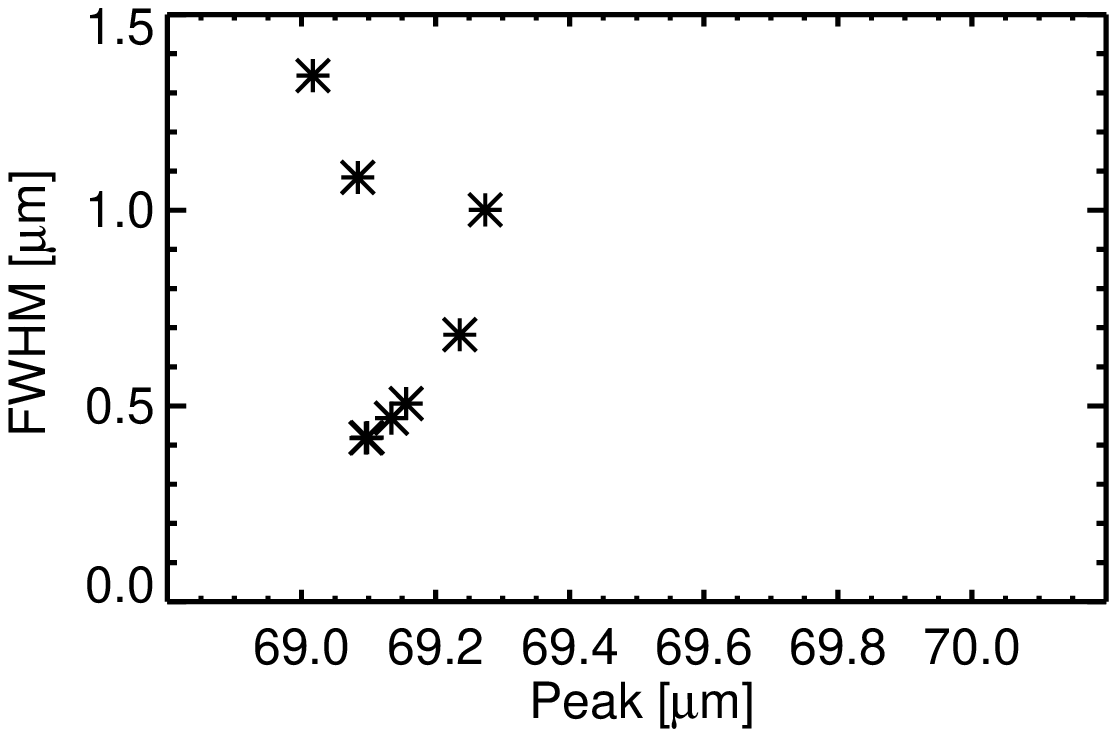}
\caption{The effect of grain  size on band shape. The top panel shows the profile of the 69~$\mu$m forsterite band at 150~K for different grain sizes. The profiles were computed using the optical constants from \citet{Suto2006} and the distribution of hollow spheres \citep{Min2003}. The lower panel shows the relation between peak position and FWHM of the profiles shown in the top panel. The asterisks in the lower panel correspond to the grain sizes in the upper panel, starting with 1~$\mu$m at the bottom.}
\label{fig:grainsize}
\end{figure}

A final parameter we discuss here is the size of the dust particles which also has an effect on the shape and peak position of the 69~$\mu$m band. Grain size effects could be important as we can expect to observe dust that is deeply embedded in the disk where grain growth and settling will likely play a role. We computed mass absorption coefficients for 7 different grain sizes, between 1 and 50\,$\mu$m, using the DHS scattering theory and optical constants from \citet{Suto2006}, and the results are displayed in Fig.\,4. The changes in the band can be clearly seen for grains larger than 1~$\mu$m, while the band profile is almost indistinguishable for grains with radii between 0.1 and 1.0~$\mu$m.  The maximum shift in the peak position we expect due to changes in the grain size is in the order of 0.2~$\mu$m.  The effects seen in the FWHM of the forsterite band is larger, however, the peak over continuum ratio decreases significantly for grains larger than 10~$\mu$m. With the S/N we achieve in our sample it 
would be impossible to detect the emission of grains larger than about 15--20~$\mu$m in radius, so we do not expect to detect the very broad emission bands which can be associated with large grains. The typical grain sizes we consider in this study are 0.1--1.0~$\mu$m. However, in our analysis (see section \ref{subsect:grainsize}) we test the conformity of the detected bands with larger grains.

\subsubsection{Summary of effects}
\label{subsubsect:summarygraineffects}
In summary, the iron content and the grain temperature have the largest effect on the 69~$\mu$m feature. However, for small measured shifts of the order of 0.2~$\mu$m in the peak position and FWHM, a degeneracy between the parameters exists and no clear distinction can be made between small changes in iron content, temperature, or grain size and shape. In our analysis we will discuss the constraints on iron content and grain temperature, and discuss if we see any indications for substantial grain growth beyond the typical grain size of 0.1--1~$\mu$m, the grain sizes typically observed in the mid-IR, taking into account these possible degeneracies.

\subsection{Searching for the 69~$\mu$m band in the PACS spectra}
\label{subsect:detection2}
To search for and analyse the 69\,$\mu$m forsterite band in our sample, we examine the 67--71.5\,$\mu$m region in the spectra, shown in the left column of Figures \ref{fig:overview01} to \ref{fig:overview01_h} in the Appendix. Aside from a few sources with very strong emission bands (e.g. HD~100546) noise reduction is important to separate forsterite bands from narrow features such as gas lines or noise peaks. 

In order to reduce the noise and to concentrate on wide features, all spectra underwent noise--filtering in their 67--71.5\,$\mu$m range. We compute the Fourier--transform (FT) of each spectrum and remove the contribution of all frequencies outside of the low frequency peak. After inverse FT the noise--filtered spectrum is overplotted on the unmodified data (see left column of Figures ~\ref{fig:overview01}--\ref{fig:overview01_h}). In general the filtering greatly improved the visibility of wider features such as the forsterite bands while gas lines, noise and other narrow spikes in the spectra are suppressed. However, some very strong gas lines (e.g. the 71~$\mu$m OH feature in DG~Tau) do not completely vanish through the filtering. Therefore, we carefully compared all forsterite detections with the unmodified spectra to avoid confusion with emission bands other than the forsterite 69~$\mu$m feature.

We use the noise--filtered spectra to search for possible 69~$\mu$m forsterite emission bands. In some cases the noise--filtered spectra suffer from the Gibbs--phenomenon at the outer edges, introducing artificial features that could be confused with emission bands. To avoid complications we restrict the position for the peak of possible emission bands to the 68--71\,$\mu$m range. Visual inspection did not reveal any forsterite peaks outside this range. Some of the spectral resolution is lost due to the noise--filtering process. To make sure that no false positives were created through the filtering we compare the filtered to the unmodified spectrum. The comparison is done by overplotting the unmodified with the filtered version of the spectra (see leftmost column in Figures \ref{fig:overview01}--\ref{fig:overview01_h}).

The 67--71.5\,$\mu$m region of each noise--filtered spectrum is fitted with a 2$^{nd}$ or 3$^{rd}$ order polynomial to account for the continuum emission, and a superimposed Lorentz profile to describe one possible emission band. The Lorentz profile is restricted to a peak position in the range of 68--71\,$\mu$m, a FWHM of less than 1.5\,$\mu$m.  We considered only positive values for the integrated flux to avoid fitting absorption bands. These restrictions prevent the fit from converging on gas lines or features that are extended outside the fitted spectral range. Any fit solution where one or more parameters reached these limits was not used to claim a detection.

As the data reduction does not provide reliable uncertainties, the first fit is done using equal weights (1) for all points. This leads to improbably low reduced $\chi^2$ ($\chi^2_r$) values, so the fit is repeated with assumed error bars of $\sqrt{\chi^2_r}$ for each datapoint. If the parameter values remain the same, this will result in a new $\chi^2_r = 1$. If the parameters change, the procedure is iterated until $0.95 < \chi^2_r < 1.05$ is reached. A proper estimate of the uncertainties in the spectrum is important to obtain reliable estimates for the uncertainties of the model parameters.

The best-fit models overplotted on the noise--filtered spectra are shown in the middle column of Figures \ref{fig:overview01}--\ref{fig:overview01_h}. The parameter values found for the Lorentzian are listed in Table \ref{tab:overview01}. The table also includes the formal uncertainties of the peak position and the FWHM derived by the IDL implementation of the least squares minimisation \texttt{mpfit} \citep{Markwardt2009} from the covariance matrix. The uncertainty of the integrated flux is computed from the standard deviation of the residuals.


Finally we check how likely our best-fit Lorentz curve is to be explained by a combination of residual noise. The best-fit model is subtracted from the noise--filtered data. We then take the standard deviation of the residuals as an estimate for error bars (in which case the residuals could be fitted with F($\lambda$) = 0) and determine the probability that the best-fit Lorentzian is a result of that noise. We compute
\begin{equation}
    D = \sum_i \left(1/(\sigma_{\text{res}}) L(\lambda_i,p)\right)^2
\end{equation}
where $i$ is the index of the wavelength points in the spectrum, $L$ the Lorentz function and $p$ the parameters from the fit.

Assuming $D$ follows a $\chi^2$ distribution with $N$ (= number of wavelength points minus number of parameters) degrees of freedom, we compute the probability $P$ to find a value of $D$ as high as measured or higher. The values of $P$ are listed in Table \ref{tab:overview01}. The residuals, overplotted with the F($\lambda)$ = 0 and the error bars as well as the best-fit Lorentzian is shown in the right column of Figures \ref{fig:overview01}--\ref{fig:overview01_h}.


The criteria for a detection of the forsterite 69\,$\mu$m band are aj integrated flux/error ratio $>$~3 in the best-fit model, a low probability for the fitted band to be a result of residual noise (we do not have to specify a precise number as the probability is in all cases either almost unity or almost zero), and that none of the fit parameters have reached their boundaries. Furthermore, all band models that meet these criteria are checked against the unmodified spectrum to make sure that the band was not significantly altered through the noise--filtering. These checks were done through over-plotting the best-fit model on the unmodified spectrum. A more rigorous solution would be to fit the model also to the unmodified spectrum and check if the parameter values are the same as with the filtered spectrum. As the emission band is very weak in some cases (e.g. AS~205, HD~104237) this would require some fine tuning of start values for the parameters. Therefore, we only visually checked the model overplotted on 
the unmodified spectrum.

The sources with firm detections following the analysis presented in this paper are AB~Aur, HD~100546, HD~104237, HD~141569, HD~179218, HD~144668, AS~205 and IRS~48. Continuum--subtracted 67--72$\mu$m spectra of these objects are shown in Figure \ref{fig:fo-details01}. The detections in AS~205 and HD~144668 are weak and could be labelled as marginal. However, all the formal criteria described above are fulfilled and the visual inspection of the spectrum in the 67--71~$\mu$m range (see Figures \ref{fig:overview01}--\ref{fig:overview01_h} in the appendix) shows that the fitted peak is stronger than any other feature in its vicinity. Several other sources also show signs of an emission band in the region of 69~$\mu$m but did not allow for a definitive identification. We will discuss these cases below.

The upper limits for the sources without detectable bands are computed by inserting a band of the same shape and position as the one found in HD~100546 and scale the integrated flux until 3--3.2 $\sigma$ over the residuals is reached.

%
\begin{table*}
\begin{center}
\caption{Properties of detected forsterite bands and upper limits for nondetections}
\label{tab:overview01}
\small
\begin{tabular}{@{}lrrrrr@{}}
\toprule
Star    & \multicolumn{1}{c}{Peak}   & \multicolumn{1}{c}{FWHM}   & \multicolumn{1}{c}{Flux} & \multicolumn{1}{c}{$\sigma$} &  \multicolumn{1}{c}{probability} \\
        &      \multicolumn{1}{c}{[$\mu\textnormal{m}$]}  &  \multicolumn{1}{c}{[$\mu$m]} & \multicolumn{1}{c}{$10^{-16}$ [W/m$^2$]} & & \multicolumn{1}{c}{for $\chi^2$ [\%]}\\
\cmidrule{1-6}
         \multicolumn{6}{c}{Detections}\\
\cmidrule{1-6}

AB~Aur       &  69.957 $\pm$ 0.018  &  0.509 $\pm$ 0.063  &      10.14  $\pm$  0.90   &  11.2  &  0 \\
HD~100546    &  69.194 $\pm$ 0.004  &  0.681 $\pm$ 0.015  &      94.60  $\pm$  1.27   &  74.6  &  0  \\
HD~104237    &  69.224 $\pm$ 0.012  &  0.351 $\pm$ 0.038  &       2.14  $\pm$  0.21   &   9.8  &  0 \\
HD~141569    &  69.303 $\pm$ 0.021  &  0.600 $\pm$ 0.092  &       6.32  $\pm$  0.49   &  12.9  &  0 \\
HD~179218    &  69.196 $\pm$ 0.005  &  0.502 $\pm$ 0.020  &       7.33  $\pm$  0.18   &  39.7  &  0 \\
HD~144668    &  69.088 $\pm$ 0.014  &  0.599 $\pm$ 0.057  &       2.60  $\pm$  0.13   &  19.8  &  0 \\
IRS~48       &  69.168 $\pm$ 0.007  &  0.521 $\pm$ 0.025  &       6.11  $\pm$  0.21   &  28.9  &  0 \\
AS~205       &  68.745 $\pm$ 0.017  &  0.595 $\pm$ 0.085  &       4.59  $\pm$  0.29   &  15.6  &  0 \\

\cmidrule{1-6}
         \multicolumn{6}{c}{False positives}\\
\cmidrule{1-6}

DG~Tau       &  69.187 $\pm$ 0.038  &  0.506 $\pm$ 0.153  &       2.21  $\pm$  0.39   &   5.6  &  100 \\
HD~100453    &  69.427 $\pm$ 0.043  &  0.749 $\pm$ 0.157  &       2.93  $\pm$  0.34   &   8.6  &  100 \\
HD~203024    &  69.466 $\pm$ 0.043  &  0.712 $\pm$ 0.204  &       3.26  $\pm$  0.41   &   8.0  &  100 \\
HD~35187     &  68.605 $\pm$ 0.039  &  0.695 $\pm$ 0.153  &       1.98  $\pm$  0.23   &   8.5  &  100 \\
HT~Lup       &  69.723 $\pm$ 0.016  &  0.285 $\pm$ 0.056  &       3.33  $\pm$  0.61   &   5.5  &  100 \\
SR~21        &  69.390 $\pm$ 0.025  &  0.510 $\pm$ 0.101  &       2.60  $\pm$  0.31   &   8.4  &  94 \\

\cmidrule{1-6}
         \multicolumn{6}{c}{Non--detections (upper limits)}\\
\cmidrule{1-6}

HD~144432    &    &    &       0.84  $\pm$  0.27  &    3.1  &  100 \\
HD~97048     &    &    &       1.96  $\pm$  0.63  &    3.1  &  100 \\
HD~36112     &    &    &       0.72  $\pm$  0.23  &    3.1  &  100 \\
HD~142666    &    &    &       6.53  $\pm$  2.11  &    3.1  &  100 \\
HD~163296    &    &    &       0.82  $\pm$  0.26  &    3.1  &  100 \\
HD~50138     &    &    &       2.08  $\pm$  0.67  &    3.1  &  100  \\
HD~150193    &    &    &       0.67  $\pm$  0.22  &    3.1  &  100 \\
HD~135344B   &    &    &       1.12  $\pm$  0.36  &    3.1  &  100 \\
HD~169142    &    &    &       1.73  $\pm$  0.56  &    3.1  &  100 \\
S~CrA        &    &    &       1.21  $\pm$  0.39  &    3.1  &  100 \\
HD~139614    &    &    &       0.75  $\pm$  0.24  &    3.1  &  100 \\
HD~98922     &    &    &       1.15  $\pm$  0.37  &    3.1  &  100 \\
RNO~90       &    &    &       1.06  $\pm$  0.34  &    3.1  &  100 \\
RU~Lup       &    &    &       0.90  $\pm$  0.29  &    3.1  &  100 \\
RY~Lup       &    &    &       0.98  $\pm$  0.32  &    3.1  &  100 \\
EC~82        &    &    &       5.09  $\pm$  1.64  &    3.1  &  100 \\
HD~38120     &    &    &       0.60  $\pm$  0.19   &   3.1  &  100 \\
HD~142527    &    &    &       2.67  $\pm$  0.86   &   3.1  &  100 \\

\bottomrule
\end{tabular}
\normalsize
\end{center}
\textbf{Description:} Fitted--band centre position, FWHM, flux and flux/uncertainty as well as the probability of finding a value in a $\chi^2$ distribution greater than or equal to the one found for comparing the Lorentzian to the noise in the residuals (see description in the text). The limits for the parameters of the Lorentzian were as follows: 68~$\mu$m $<$ Peak $<$ 71~$\mu$m, FWHM $<$ 1.5~$\mu$m and 10$^{-30}$W/m$^2$ $<$ flux. The first part of the table lists the firm detections. In the second part all sources with an emission band that has a formal significance of more than three $\sigma$ but are rejected for various reasons (see text for a discussion) are shown. The third part contains all sources where no band could be fitted and we present upper limits instead. 

\end{table*}

\subsection{False positives}
\label{detection3}

For six systems, namely DG~Tau, HD~100453, HD~203024, HD~35187, HT~Lup and
SR~21, our Lorentz curve fitting formally fits a peak in the studied wavelength
range. However, the $\sigma$ value of the fits is low and the probability to
find a value in a $\chi^2$ distribution greater than or equal to the one found
for comparing the Lorentzian to the noise in the residuals is for these six
systems (near) 100~\%. This means that the probability for the central peak to
be the result of residual noise is very high and we, therefore, classify these
as false positives. By visually inspecting the spectra of these sources,  one can
clearly see multiple ``peaks'' not associated with forsterite emission and which
are most likely low level spectral artifacts.  In the following analysis we will
treat the values quoted in Table~\ref{tab:overview01} for these six sources as
upper limits.

\section{Analysis of forsterite emission bands}
\label{sec:forsterite-analysis}

\subsection{The iron content and temperature of the crystalline silicates}
\label{subsect:ironcontent_and_temperature}
%
\begin{figure}[ht!]
\centering
\includegraphics[scale=.65, angle=90]{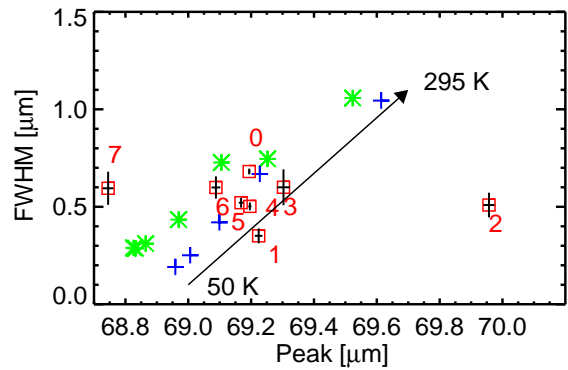}
\caption{FWHM of the forsterite `69\,$\mu$m band' plotted over the peak position, fitted with a Lorentz profile. The red squares with error bars represent measurements taken from our PACS observations. The numbers refer to the stars as follows: (0) HD\,100546, (1) HD\,104237, (2) AB\,Aur, (3) HD\,141569, (4) HD\,179218, (5) IRS~48, (6) HD~144668, (7) AS~205. The green asterisks are taken from \citet{Koike2003} and the blue crosses are based on \citet{Suto2006}, computed with the DHS model \citep{Min2003} and a grain size of 0.1~$\mu$m. The black arrow indicates the temperature dependency.}
\label{fig:peakFWHM}
\end{figure}

As discussed in Sections~\ref{subsubsect:ironcontent} and
\ref{subsubsect:temperature}, the two parameters most strongly influencing the forsterite band are the iron-to-magnesium ratio and the grain
temperature. In Figure~\ref{fig:peakFWHM} we plot the measured peak position and
FWHM of the detected 69~$\mu$m features. Comparing these observed values to the
laboratory measurements shown in Figures~\ref{fig:peakIronLab} and
\ref{fig:peakFWHMLab} it is immediately clear that the crystalline olivine
grains can not contain much more than a few percent iron and have a typical
temperature of about 150~K.  Both AS~205 and AB~Aur show deviating values
compared to the other sources. Where the observed 69~$\mu$m band of AS~205 seems to be consistent only with pure
forsterite grains, the band position observed for AB~Aur can only be explained if
the olivine grains contain some fraction of iron.

Ideally, one could have used extensive laboratory studies, exploring in detail
the combined effects of iron contend and grain temperature on the forsterite
band. Unfortunately, almost all of the laboratory measurements of the position
of the forsterite band as a function of iron content were taken at room
temperature \citep{Koike2003}. The only temperature-dependent laboratory
measurements available are for pure forsterite (Mg$_2$SiO$_4$) crystals \citep[e.g.
][]{Suto2006,Koike2006} and for iron-poor olivine with a relative iron mass
fraction of $\sim$10\% \citep{Koike2006}, for which the peak of the emission
band has already shifted to $\sim$72.2~$\mu$m, far beyond any of the observed
peak positions. 

To make a quantitative statement about the iron content and temperature of the
olivine grains, we assume that the peak position and FWHM of the
69~$\mu$m feature scale linearly with the two main parameters. These assumptions are consistent with the data shown in Figures \ref{fig:peakIronLab} and \ref{fig:peakFWHMLab}. We fitted Lorenzians to the laboratory measurements of olivine bands by \citet{Koike2003} and linearly
interpolated between those, creating a fine grid of feature shapes as a function
of iron fraction (between 0--6~\%). We also fitted the position and FWHM of the 69~$\mu$m band as function of grain temperature to the data by \citet{Koike2006,Suto2006} in the range of $\sim$10--300~K and linearly interpolated between them. 

The linear combination of both the iron and the temperature dependence provides us with a model grid for the band characteristics. Since this model grid is based on both mass absorption coefficients \citep{Koike2003,Koike2006} of several different samples and a DHS model with a grain size of 0.1~$\mu$m using the optical constants from \citet{Suto2006}, we can give neither a shape model nor a fixed grain size for the individual models in the interpolated grid. 

\begin{figure*}[ht!]
\centering
\subfigure{
    \includegraphics[scale=0.65]{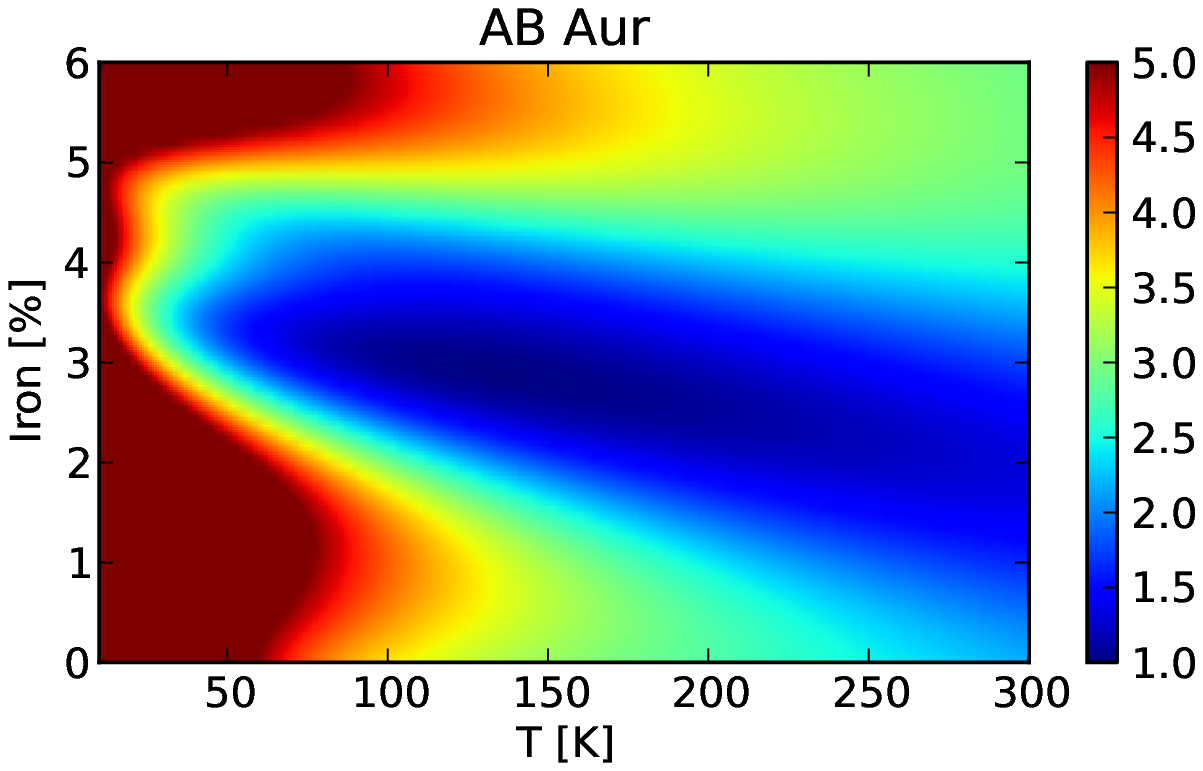}
    \label{fig:AB-Aur-chisqr}}
\subfigure{
    \includegraphics[scale=0.65]{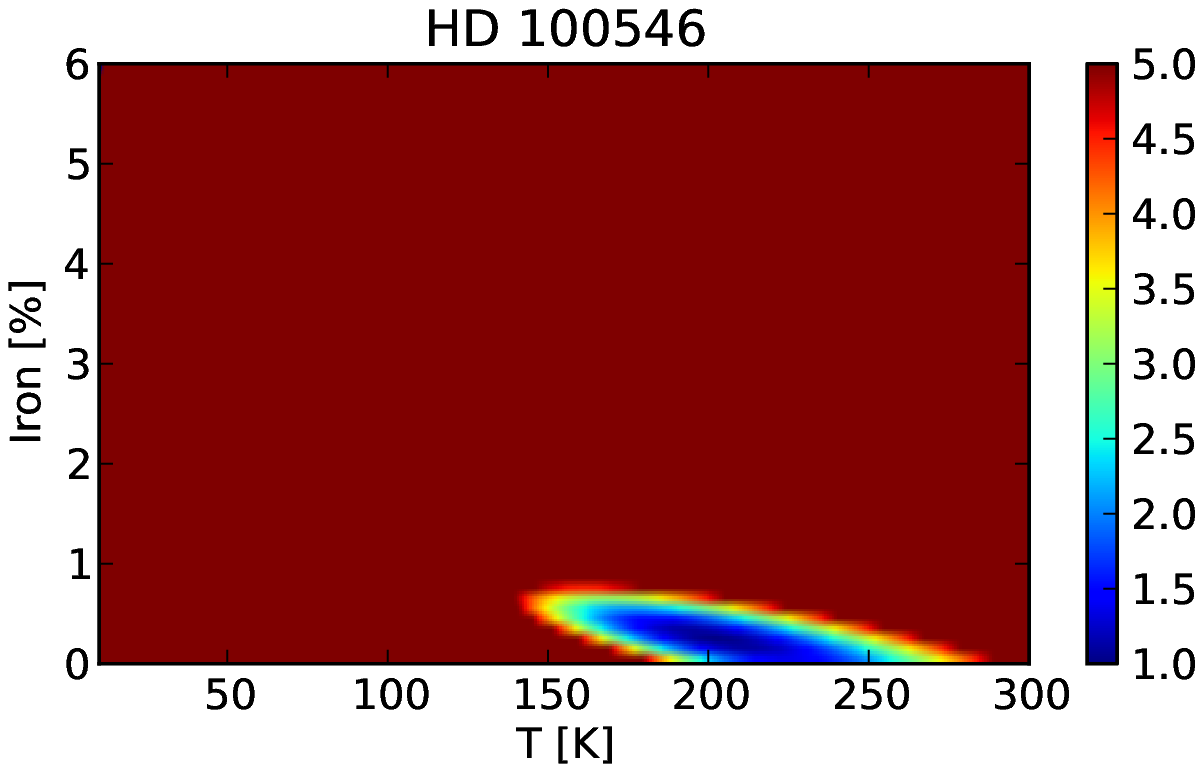}
    \label{fig:HD100546-chisqr}}
\subfigure{
    \includegraphics[scale=0.65]{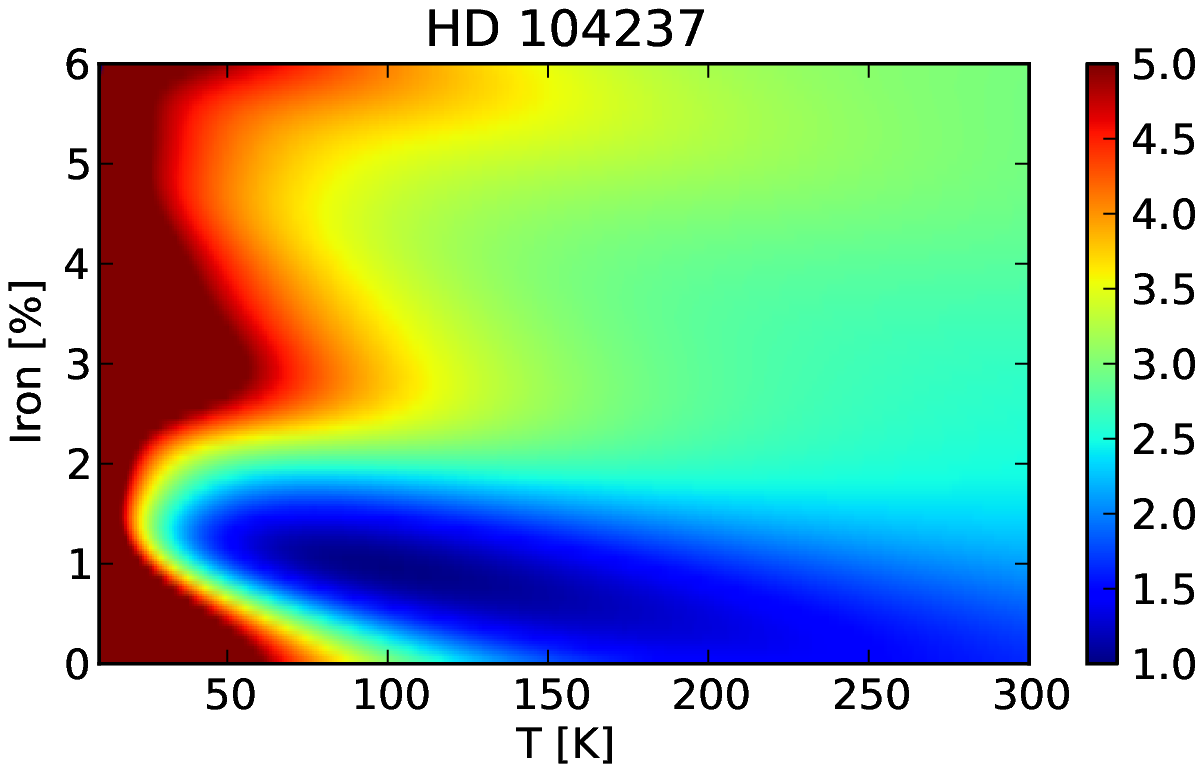}
    \label{fig:HD104237-chisqr}}
\subfigure{
    \includegraphics[scale=0.65]{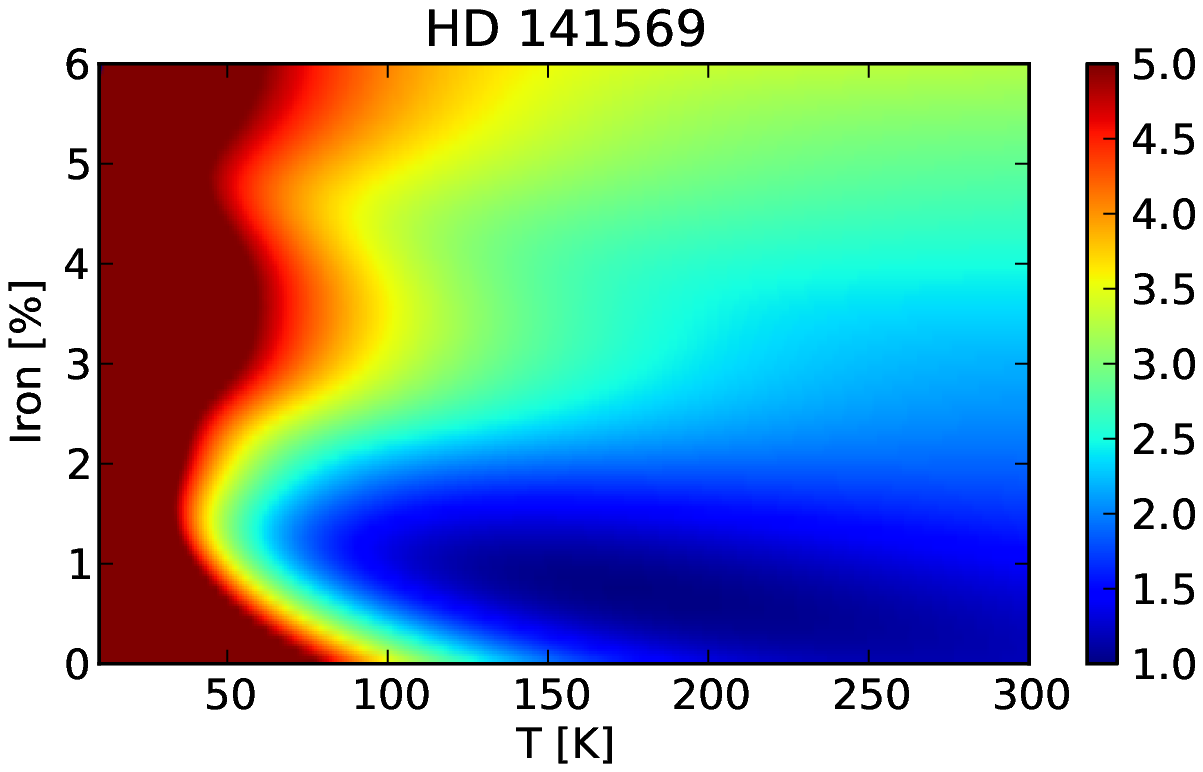}
    \label{fig:HD141569-chisqr}}
\subfigure{
    \includegraphics[scale=0.65]{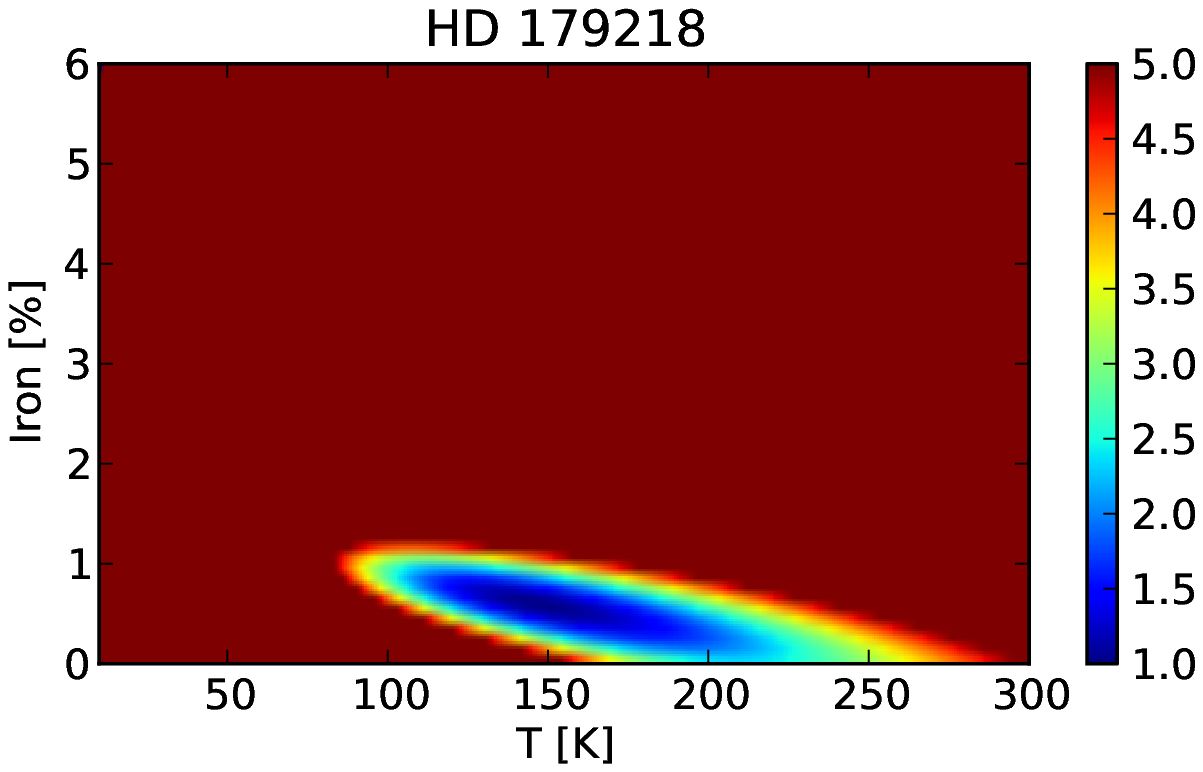}
    \label{fig:HD179218-chisqr}}
\subfigure{
    \includegraphics[scale=0.65]{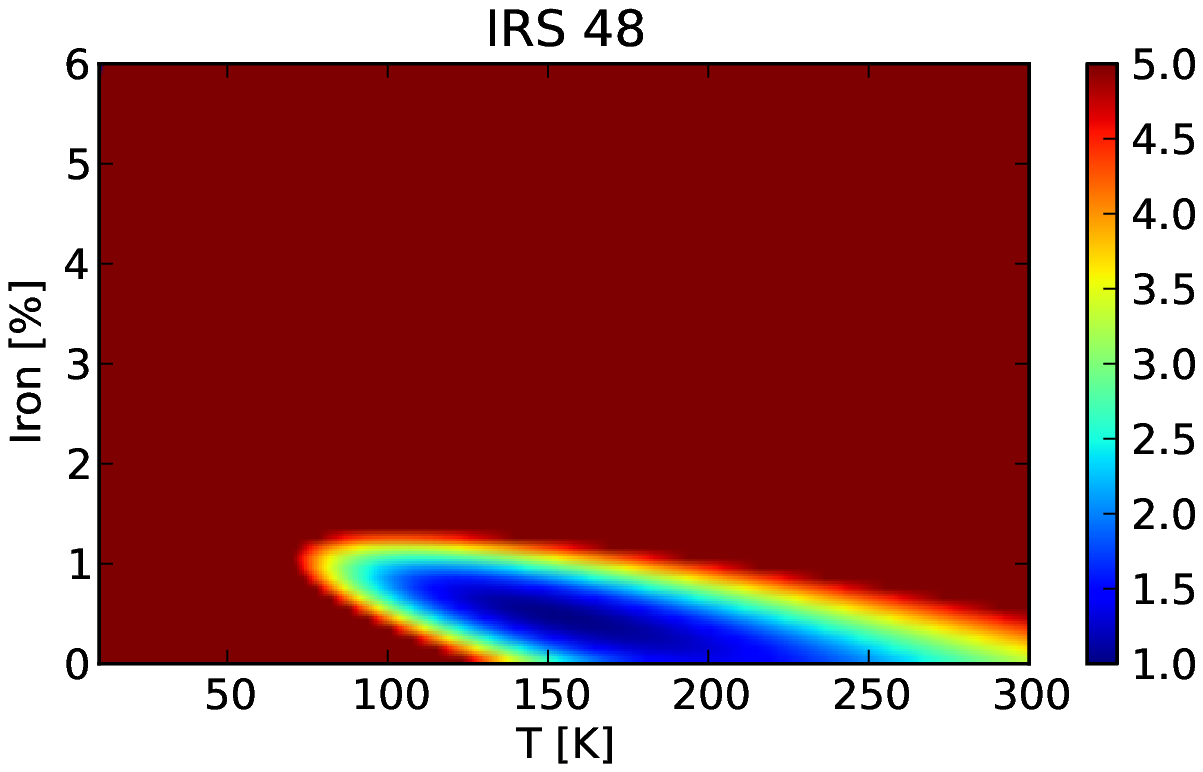}
    \label{fig:IRS48-chisqr}}
\subfigure{
    \includegraphics[scale=0.65]{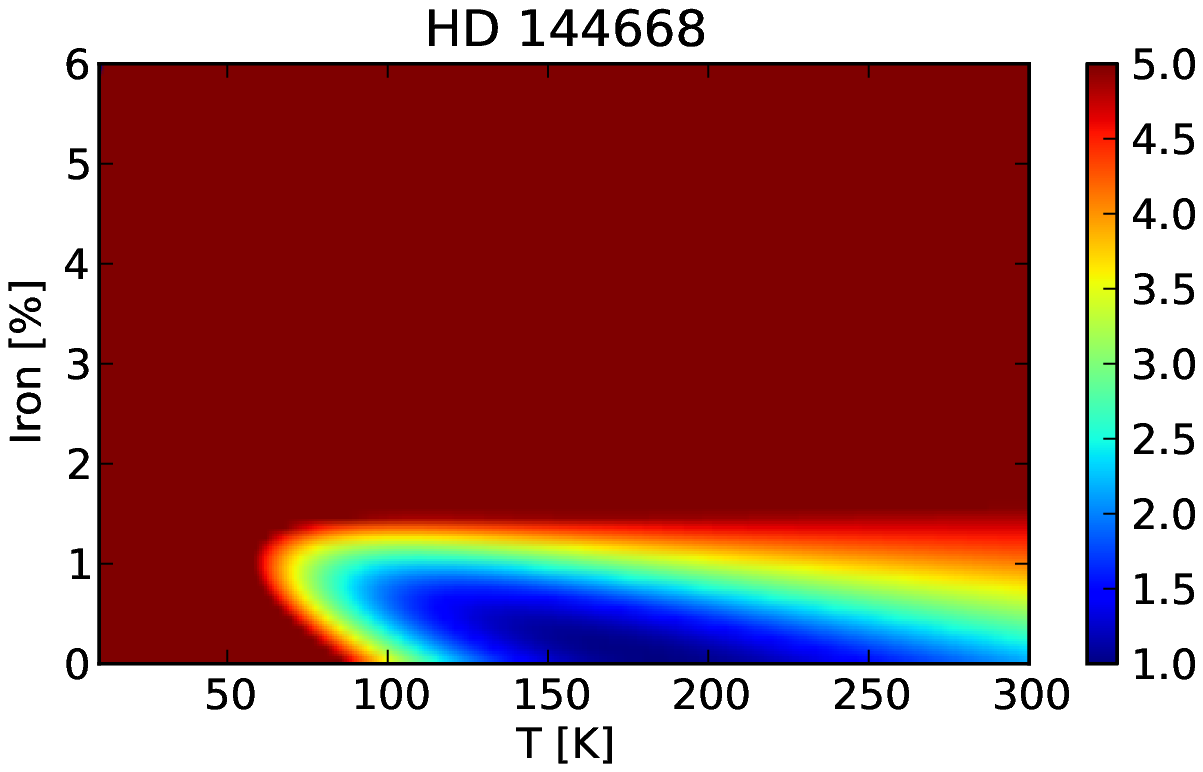}
    \label{fig:HD144668-chisqr}}
\subfigure{
    \includegraphics[scale=0.65]{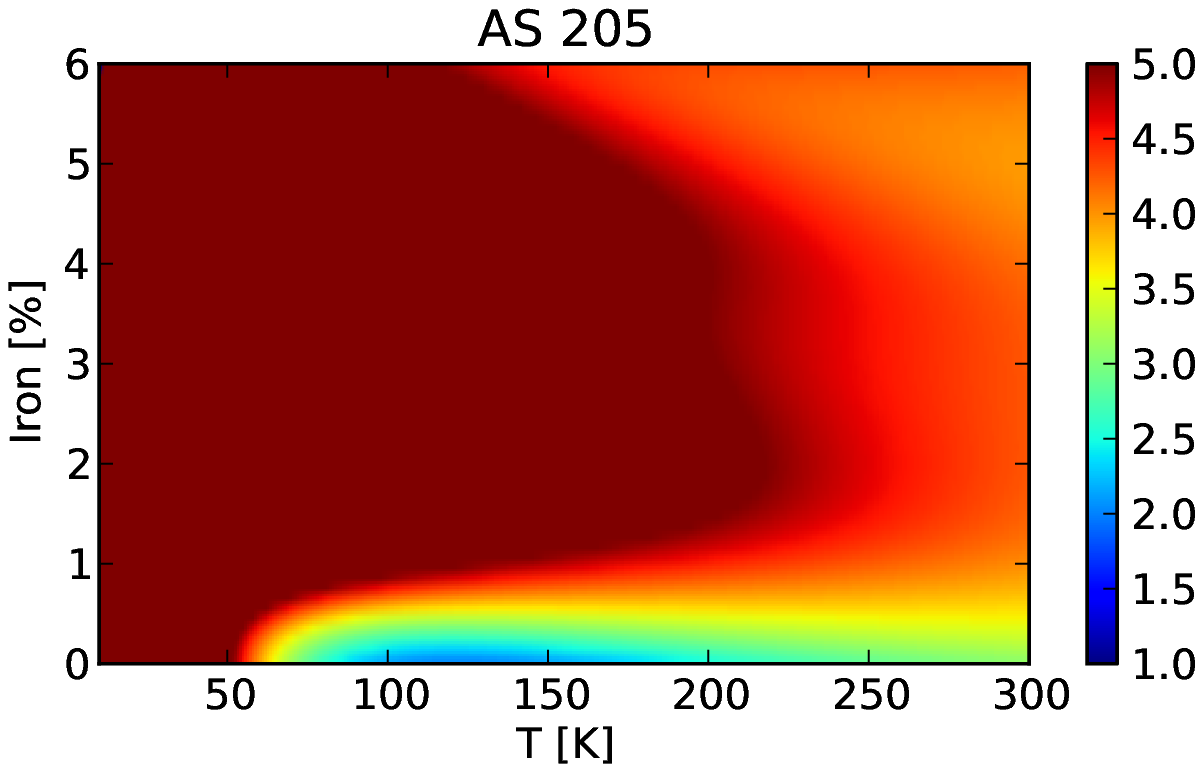}
    \label{fig:AS205-chisqr}}
\caption{Resulting reduced $\chi^2$ distributions from a comparison between our model grid for the 69~$\mu$m forsterite feature as a function of iron fraction and grain temperature, and the  \emph{Herschel}-observed bands in eight of our targets. Our model grid is an interpolation based on several different measurements of both optical constants with the DHS shape model and with absorption coefficients. Thus no fixed grain size can be given. Where applicable a grain size of 0.1~$\mu$m was used.}
\label{fig:chisqr}
\end{figure*}

\begin{figure*}[ht!]
\centering
\subfigure{
    \includegraphics[scale=0.65]{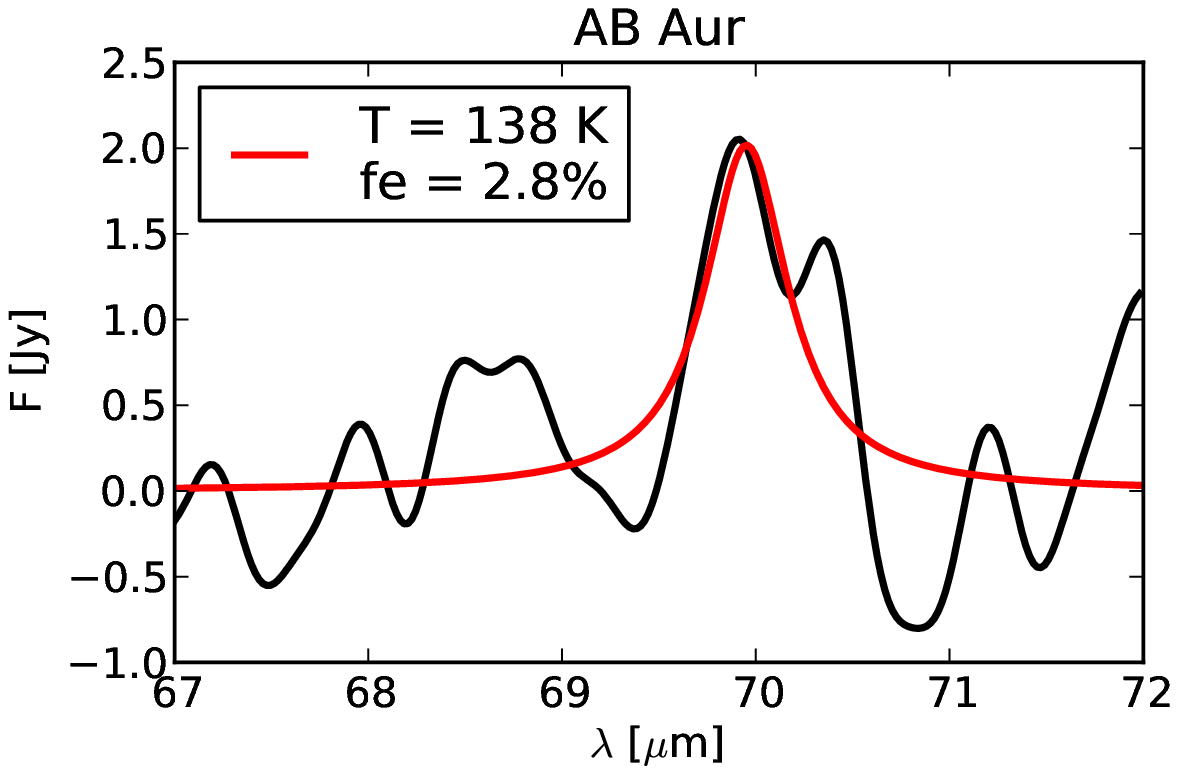}
    \label{fig:AB-Aur-chi_opt}}
\subfigure{
    \includegraphics[scale=0.65]{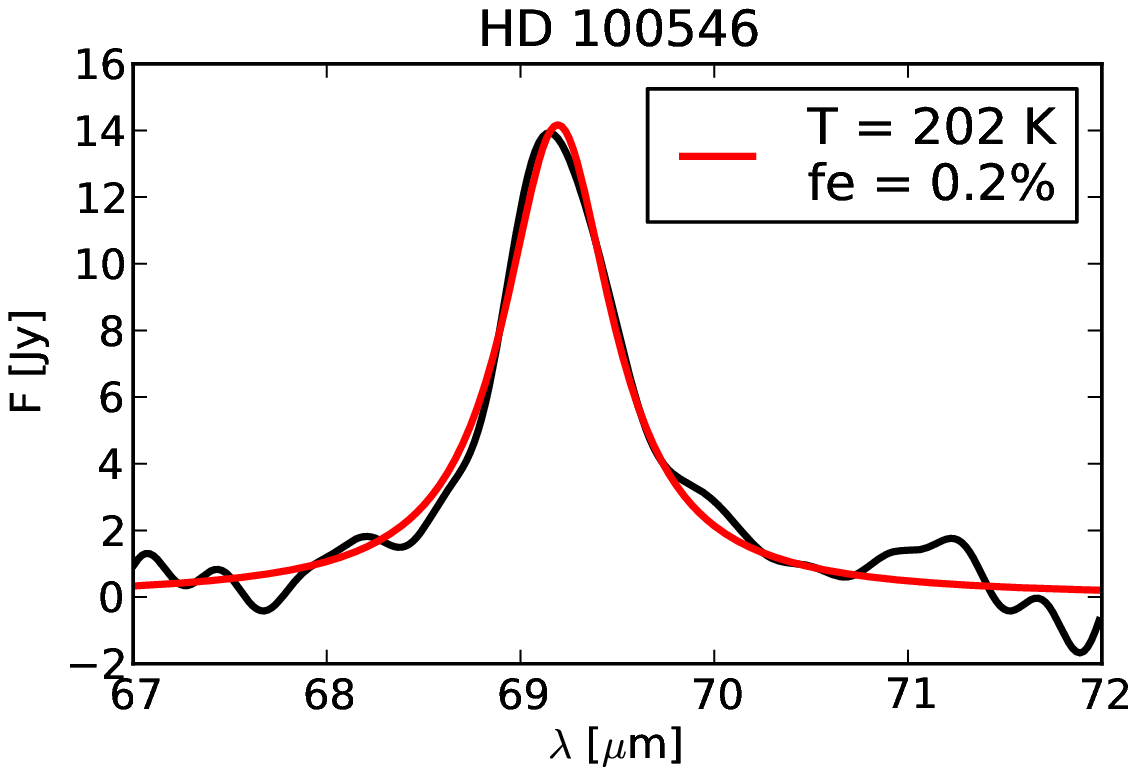}
    \label{fig:HD100546-chi_opt}}
\subfigure{
    \includegraphics[scale=0.65]{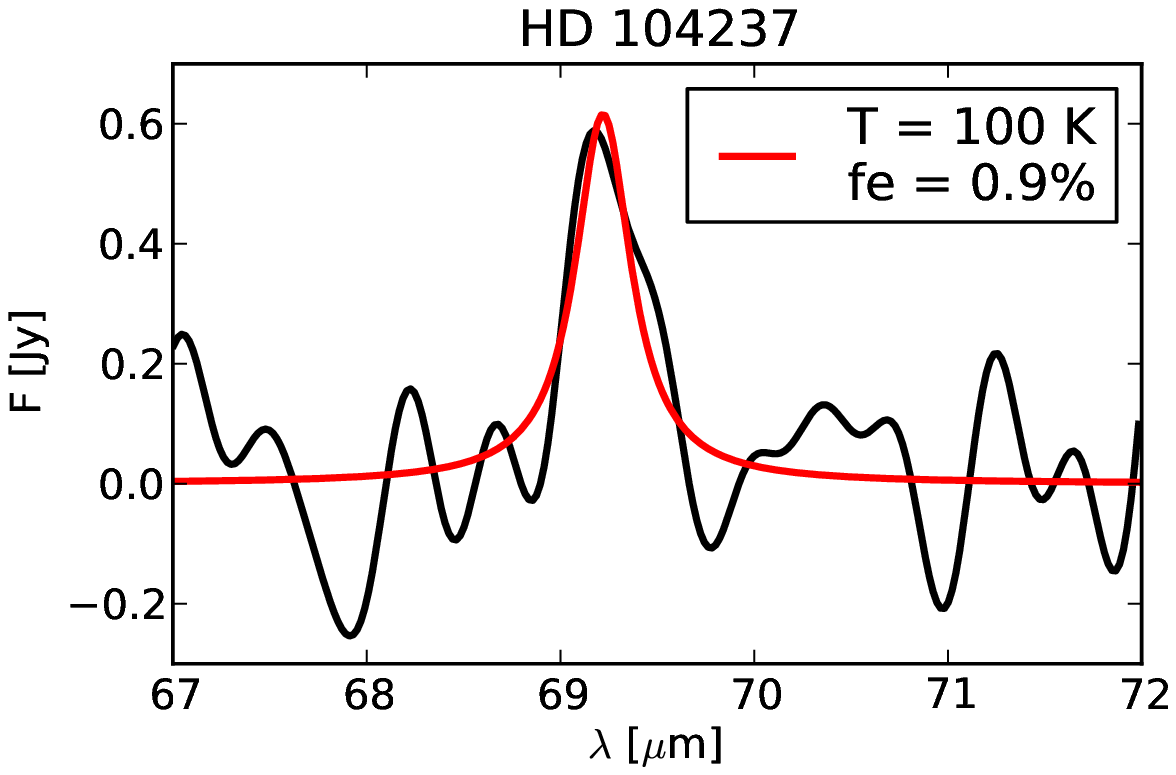}
    \label{fig:HD104237-chi_opt}}
\subfigure{
    \includegraphics[scale=0.65]{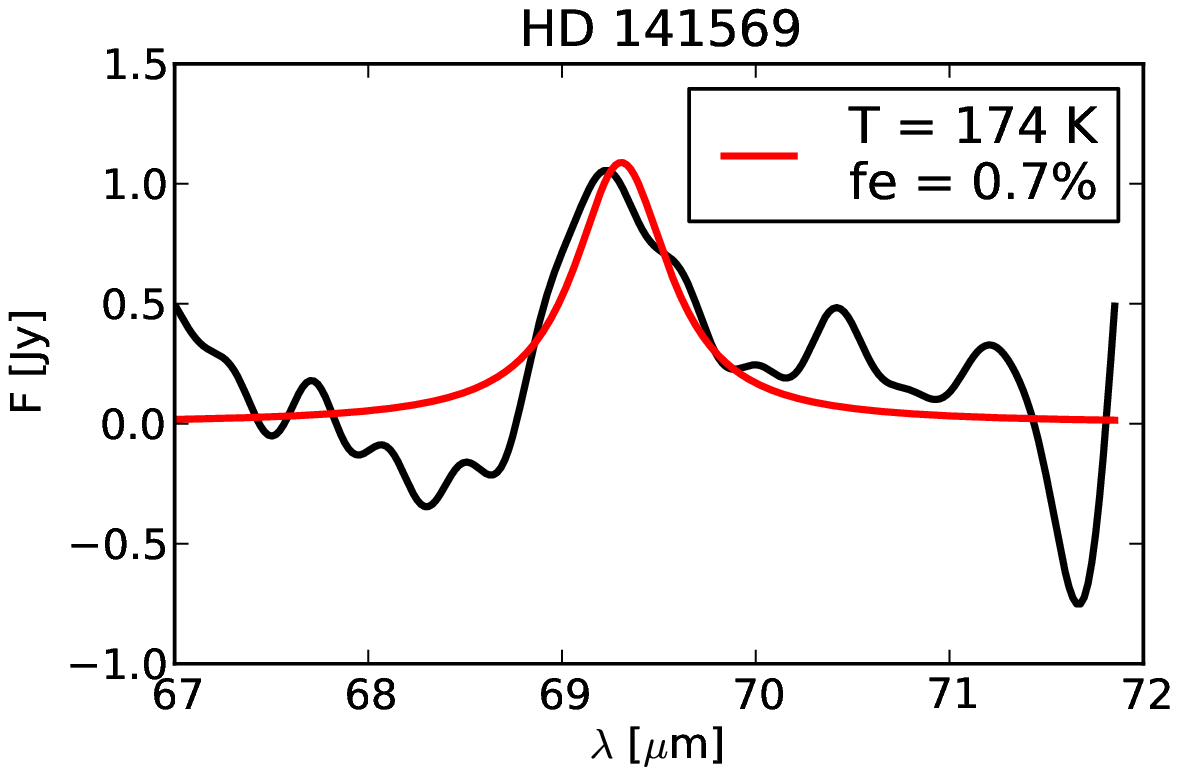}
    \label{fig:HD141569-chi_opt}}
\subfigure{
    \includegraphics[scale=0.65]{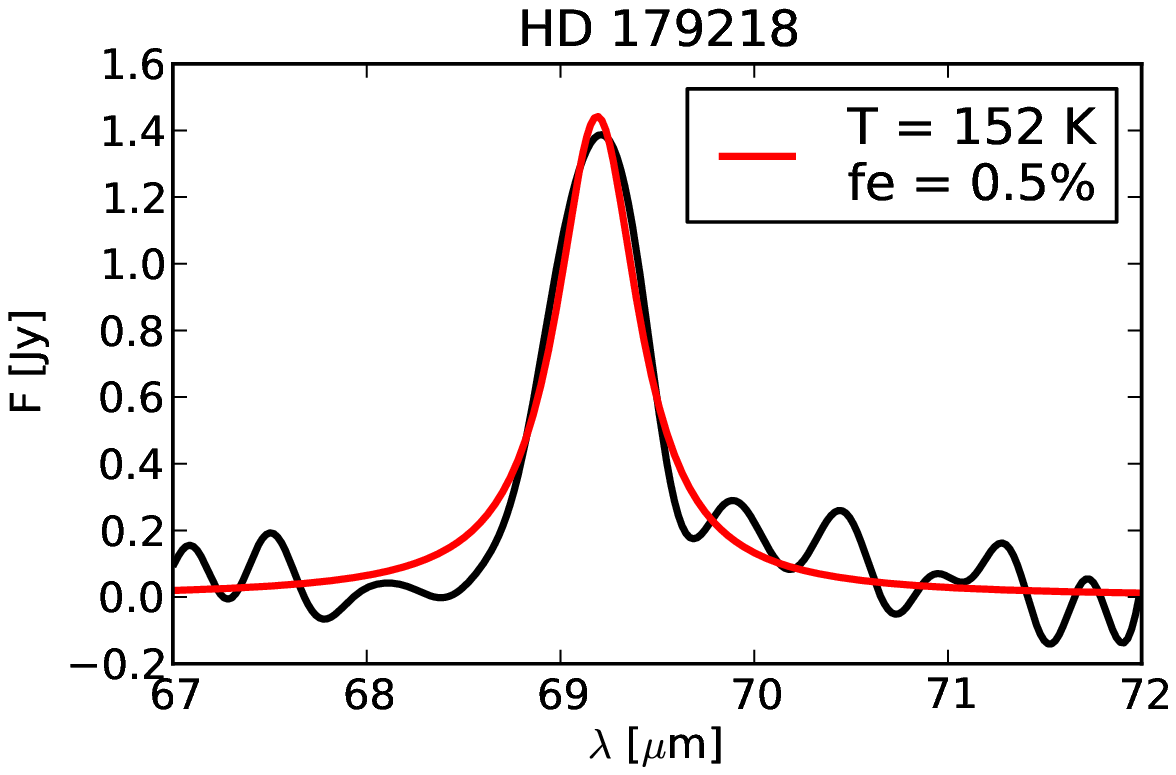}
    \label{fig:HD179218-chi_opt}}
\subfigure{
    \includegraphics[scale=0.65]{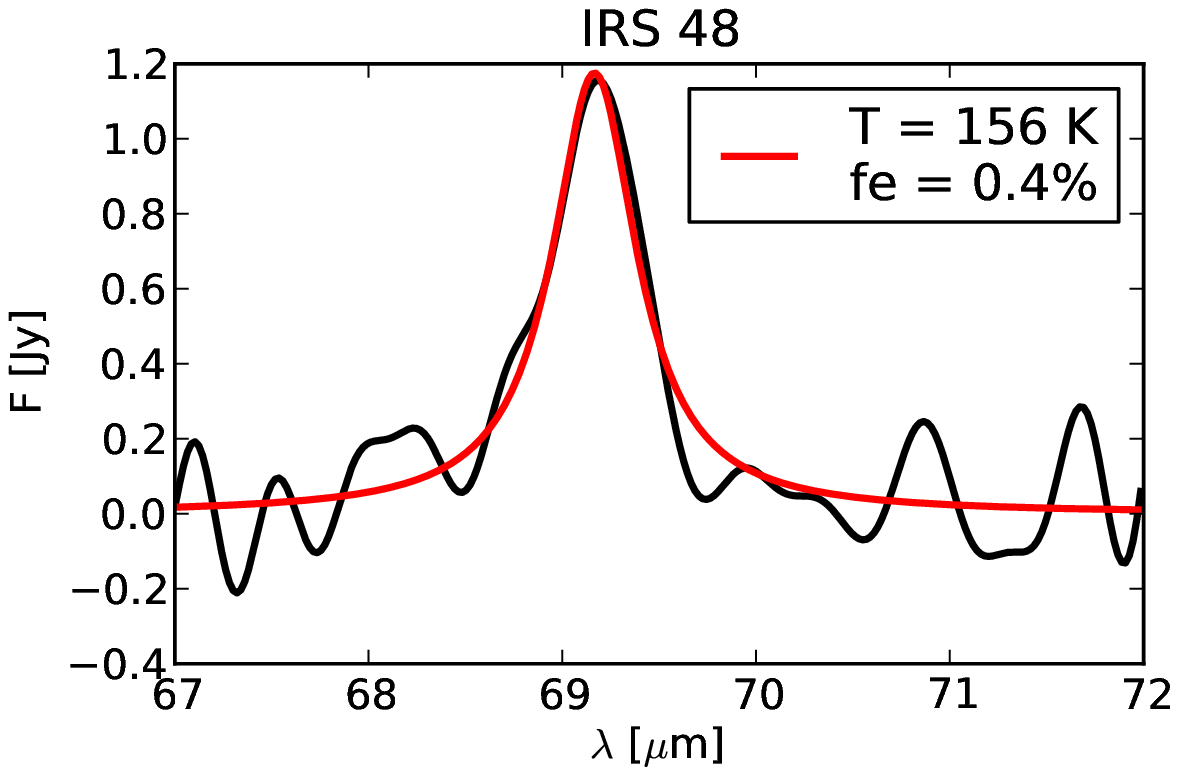}
    \label{fig:IRS48-chi_opt}}
\subfigure{
    \includegraphics[scale=0.65]{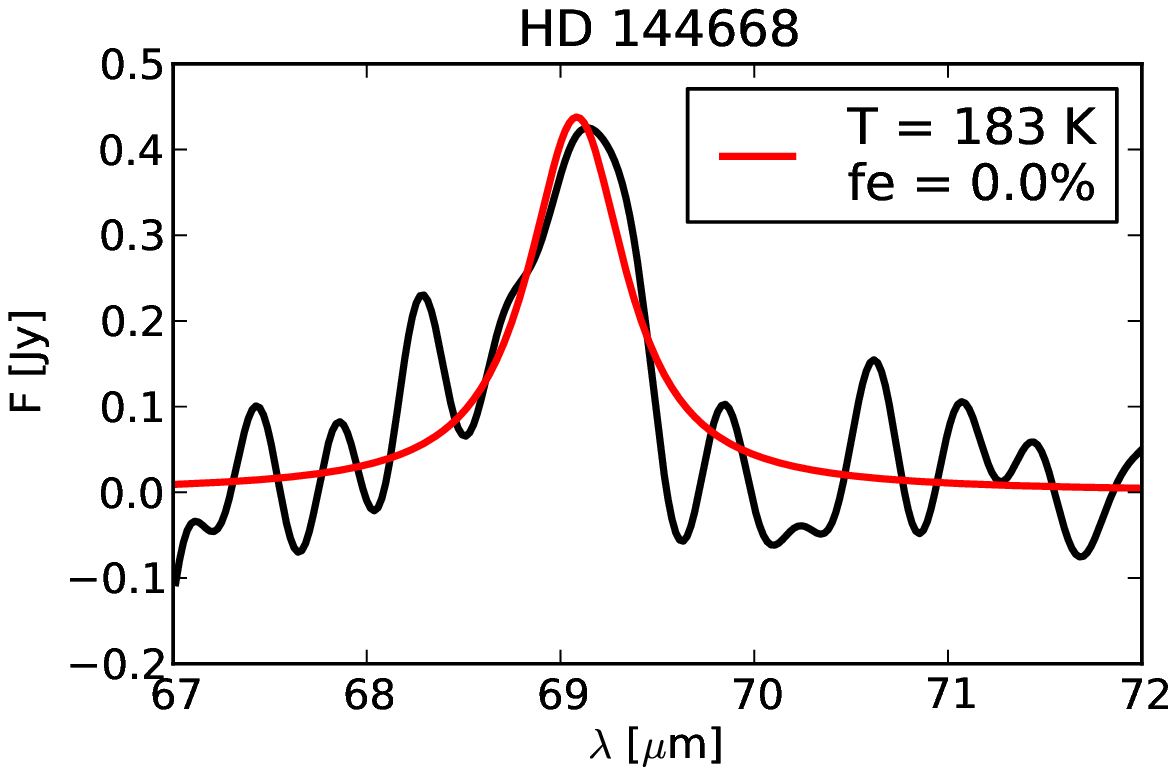}
    \label{fig:HD144668-chi_opt}}
\subfigure{
    \includegraphics[scale=0.65]{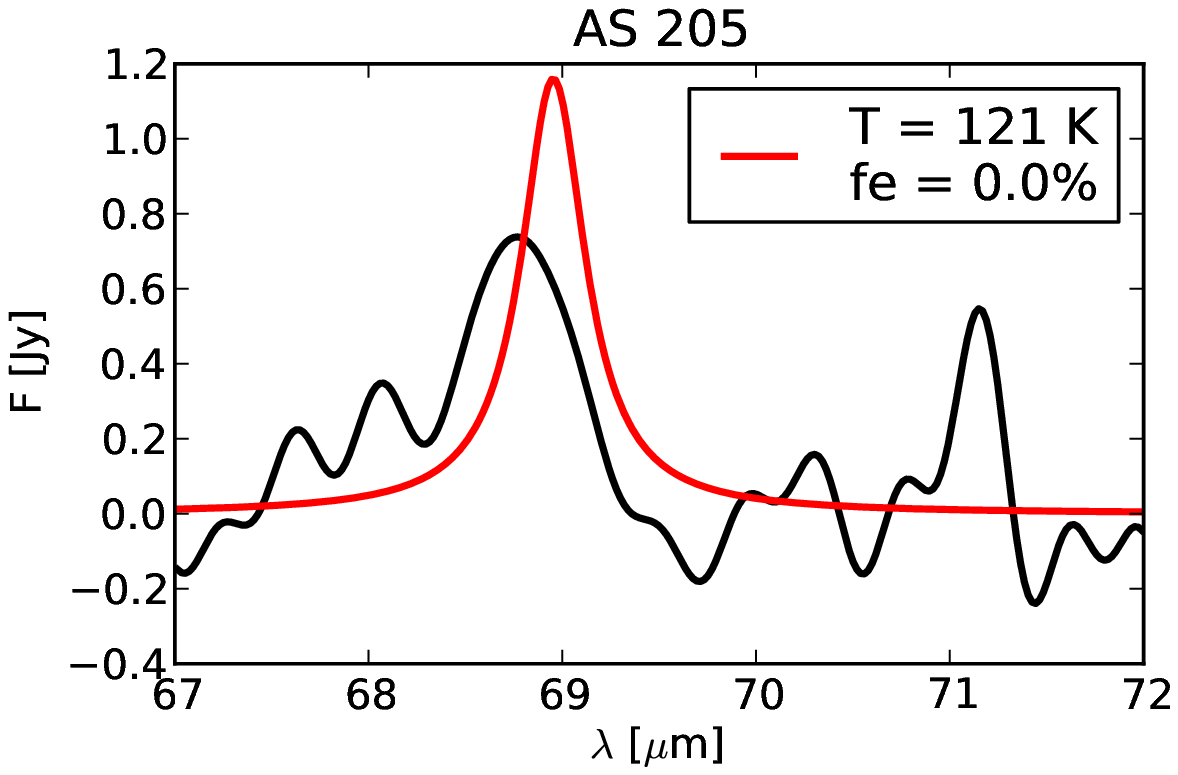}
    \label{fig:AS205-chi_opt}}
\caption{Model fits to the observed 69~$\mu$m forsterite feature. The models correspond to the minimum in the $\chi^2$ distributions shown in Fig.~\ref{fig:chisqr}. Our model grid is an interpolation based on several different measurements of both, optical constants with DHS shape model and absorption coefficients. Thus no fixed grain size can be given. Where applicable a grain size of 0.1~$\mu$m was used.}
\label{fig:chisqr_opt}
\end{figure*}

We performed a minimum $\chi^2$ analysis, comparing the interpolated laboratory measurements to the observed forsterite features. The resulting $\chi^2$ maps and best-fit curves are plotted in Figs~\ref{fig:chisqr} and \ref{fig:chisqr_opt}. 

From the minimum $\chi^2$ analysis we computed confidence intervals for the iron content and the dust temperature. First, we integrated the reduced $\chi^2$ distribution with $\nu$ degrees of freedom from $x = \chi^2/\nu$ to infinity. The new distribution, $Q(\chi^2/\nu,\nu)$, describes the probability that a value of $\chi^2/\nu$ or larger is produced by random noise. Consequently we searched for a value of $\chi^2/\nu$ so that $1 - Q(\chi^2/\nu,\nu) = 0.997$. In Table~\ref{tab:chsqrfitresults} we list the highest and lowest temperature and iron content fractions within the range of models below the 3~$\sigma$ threshold. In the case of AS~205, however, the minimum $\chi^2/\nu$ is already larger than the computed 3~$\sigma$ limit as the fit is dominated by systematic errors due to a deviation of our band model from the measured profiles at very low temperatures. We therefore  cannot give a proper confidence interval for the dust temperature or iron content in that system. We discuss this result in more detail below.

Our detailed analysis confirms our initial estimate that the observed 69~$\mu$m
features are consistent with iron-poor olivine. From Fig.~\ref{fig:chisqr}
it is clear that the olivine dust grains in all sources but AB Aur contain at
most 1--2\% of iron. Though the olivine grains in AB~Aur can not be iron-free,
with a minimum fraction of 2\%, the iron fraction can not be more than 5\%. The
dust temperatures are less well constrained as the iron fraction though most
fitted temperatures seem to be between 100--200~K. We will return to
this in the following sections as we discuss the effects of temperature and grain
size distributions.

The poorest match to the observed band profile is achieved for AS~205. Our best
fit model has a band profile which is too narrow, and peaks at slightly longer
wavelengths compared to the observed band, as can be seen from
Fig.~\ref{fig:chisqr_opt}. We believe this is a result of the deviation of the linear model of the FWHM as function of temperature from the laboratory data by \citet{Koike2006} for low temperatures ($\le$30~K). Lattice distortions in
the laboratory samples are known to lead to an increase in FWHM, independent of
temperature \citep{Imai2009}. If the grains are more distorted than those used
in the laboratory experiments a  higher FWHM would be expected. Also the grain
shape can influence the resulting spectrum \citep[see Section~ 3.1.3;
][]{Imai2009}. If the olivine grains in AS~205 are more spherical than the
sample used by \citet{Koike2006}, they would peak at slightly shorter
wavelengths (see also Fig.~\ref{fig:grainshapeLab}). Independent of these
uncertainties, it is clear that the forsterite grains in AS~205 must 
be extremely iron-poor. Even small fractions of iron would shift the
forsterite feature to such long wavelengths that effects like lattice distortions or
grain shape would not explain the observed band position any longer.

%
%
%
%
\begin{table}
\begin{center}
\caption{Confidence intervals for iron fraction, grain temperature and dust distance to the host star.}
\label{tab:chsqrfitresults}
\small
\begin{tabular}{@{}lllllrr@{}}
\toprule
Star        &  \multicolumn{2}{c}{Iron fraction [\%]}   & \multicolumn{2}{c}{Temperature [K]} & \multicolumn{2}{c}{distance [AU]}  \\
            & min   & max     &  min  & max    &   min  & max    \\
\cmidrule{1-7}
AB~Aur      &   1.9 & 3.5     &   74  & ~~273    &   16   & 221 \\
HD~100546   &   0.1 & 0.3     &   184 & ~~223    &   20   &  29 \\
HD~104237   &   0.4 & 1.2     &    60 & ~~184    &   31   & 289 \\
HD~141569   &   0.0 & 1.2     &   107 & $>$300 &   $<$9 &  72 \\
HD~179218   &   0.4 & 0.7     &   126 & ~~173    &  104   & 196 \\
HD~144668   &   0.0 & 0.4     &   130 & ~~224    &   25   &  74 \\
IRS~48      &   0.1 & 0.6     &   124 & ~~195    &   17   &  43 \\
\cmidrule{2-7}
AS~205      &  \multicolumn{2}{c}{0.0}   &   \multicolumn{2}{c}{121}  &   \multicolumn{2}{c}{32} \\
\bottomrule                                           
\end{tabular}
\normalsize
\end{center}
\textbf{Note:} Iron fraction and dust temperature are fitted while the distance of the dust to the host stars is estimated based on the temperature (see text) The best-fit models are shown in Fig.~\ref{fig:chisqr_opt}.
\end{table}

\subsection{Spectral decomposition: the effect of a temperature distribution}
\label{subsect:spectral-decomposition}
\begin{figure*}
\centering
\subfigure{
    \includegraphics[scale=0.60]{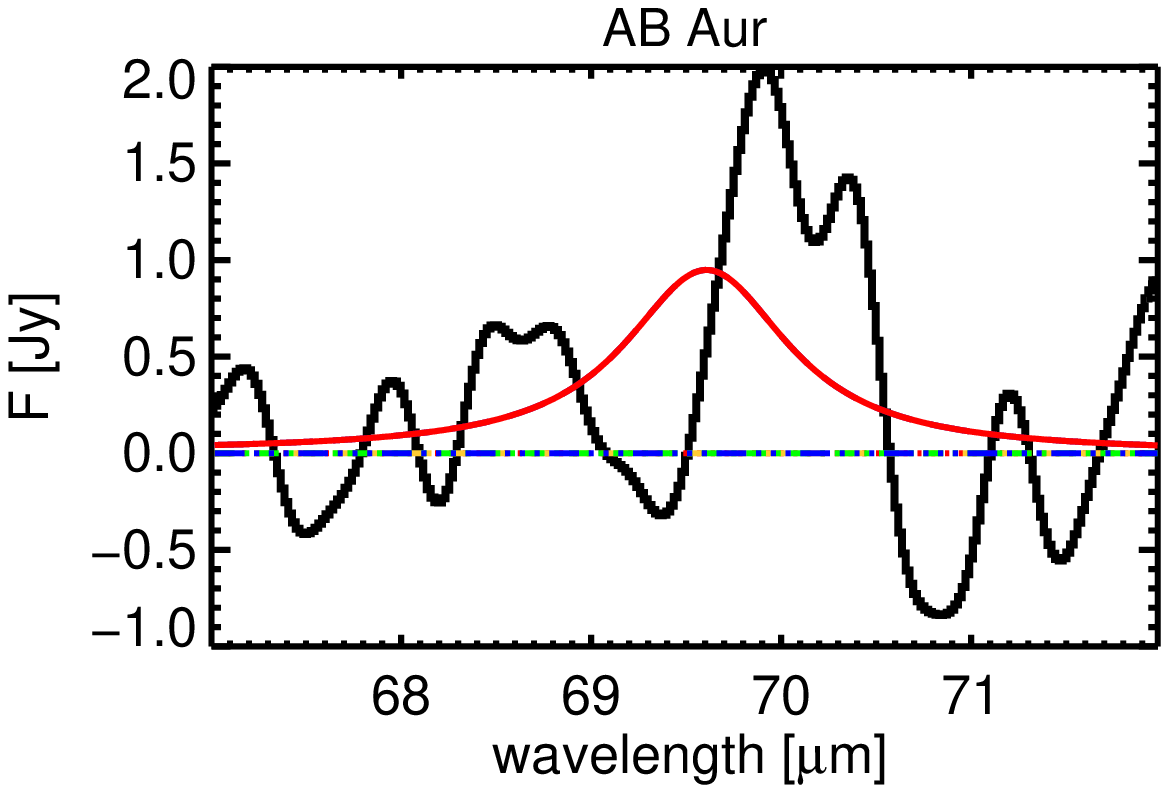}
    \label{fig:AB-Aur-filtered-detail}}
\subfigure{
    \includegraphics[scale=0.60]{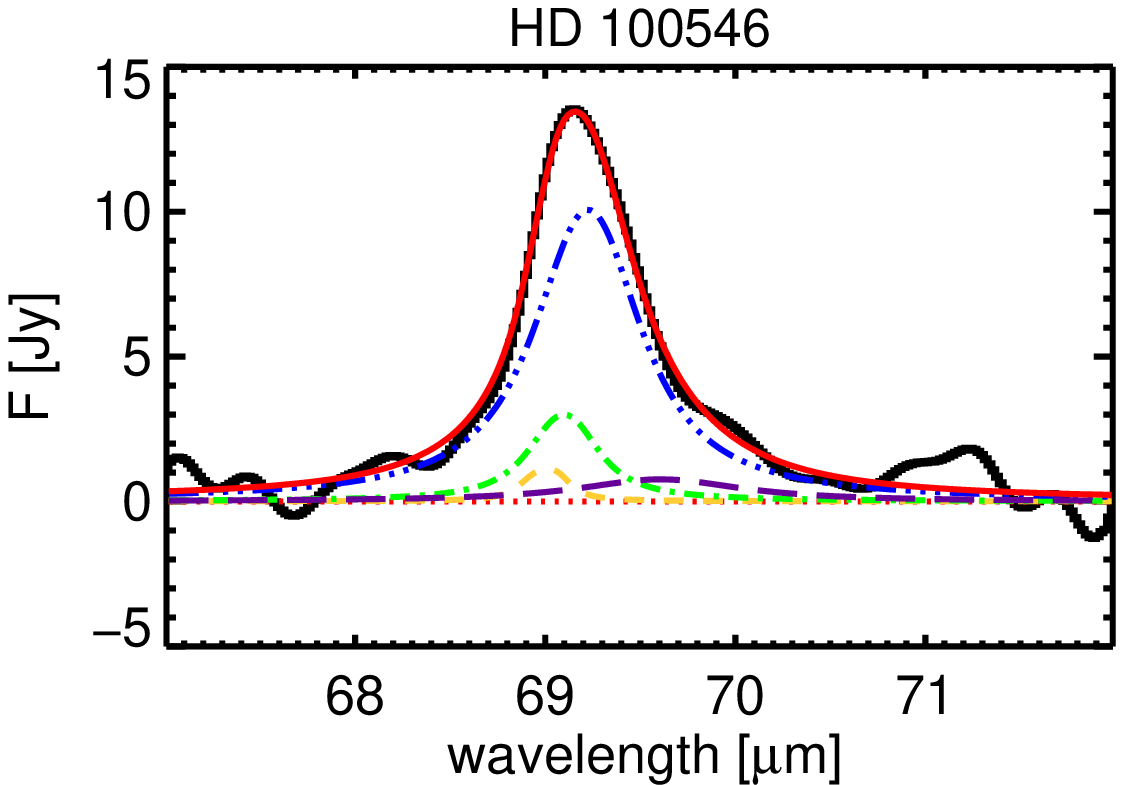}
    \label{fig:HD100546-filtered-detail}}
\subfigure{
    \includegraphics[scale=0.60]{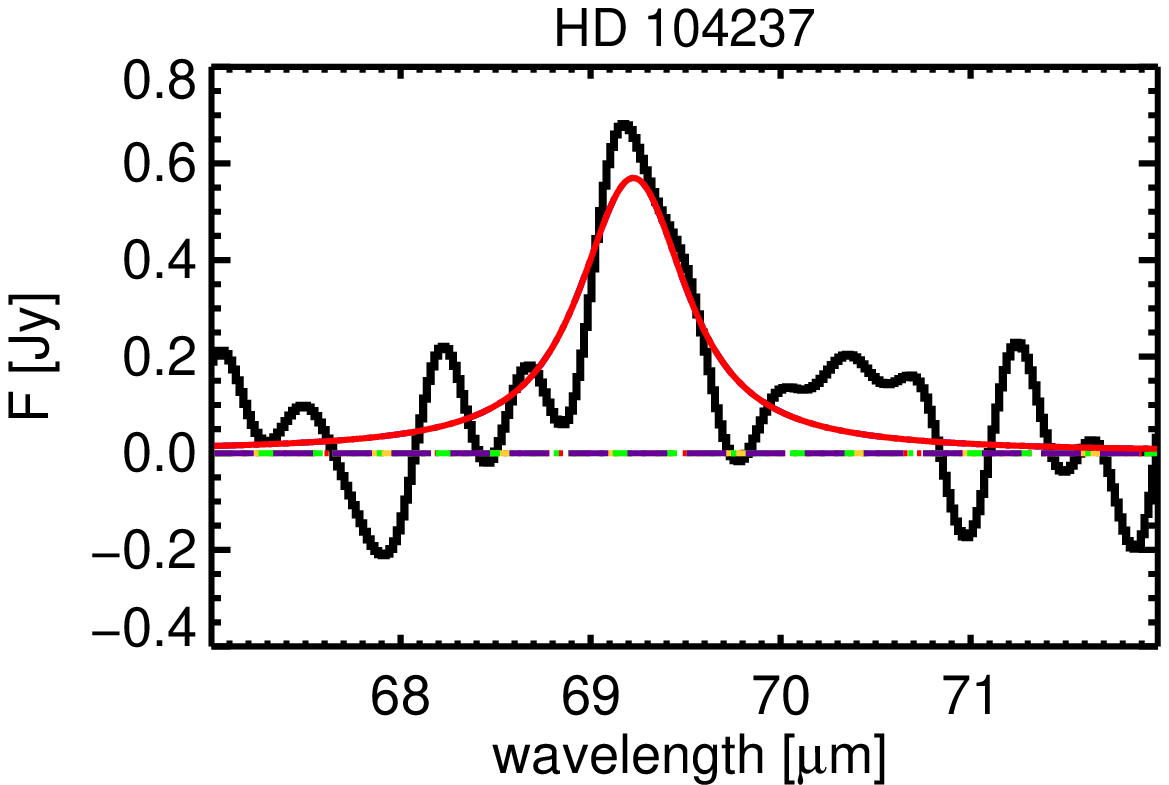}
    \label{fig:HD104237-filtered-detail}}
\subfigure{
    \includegraphics[scale=0.60]{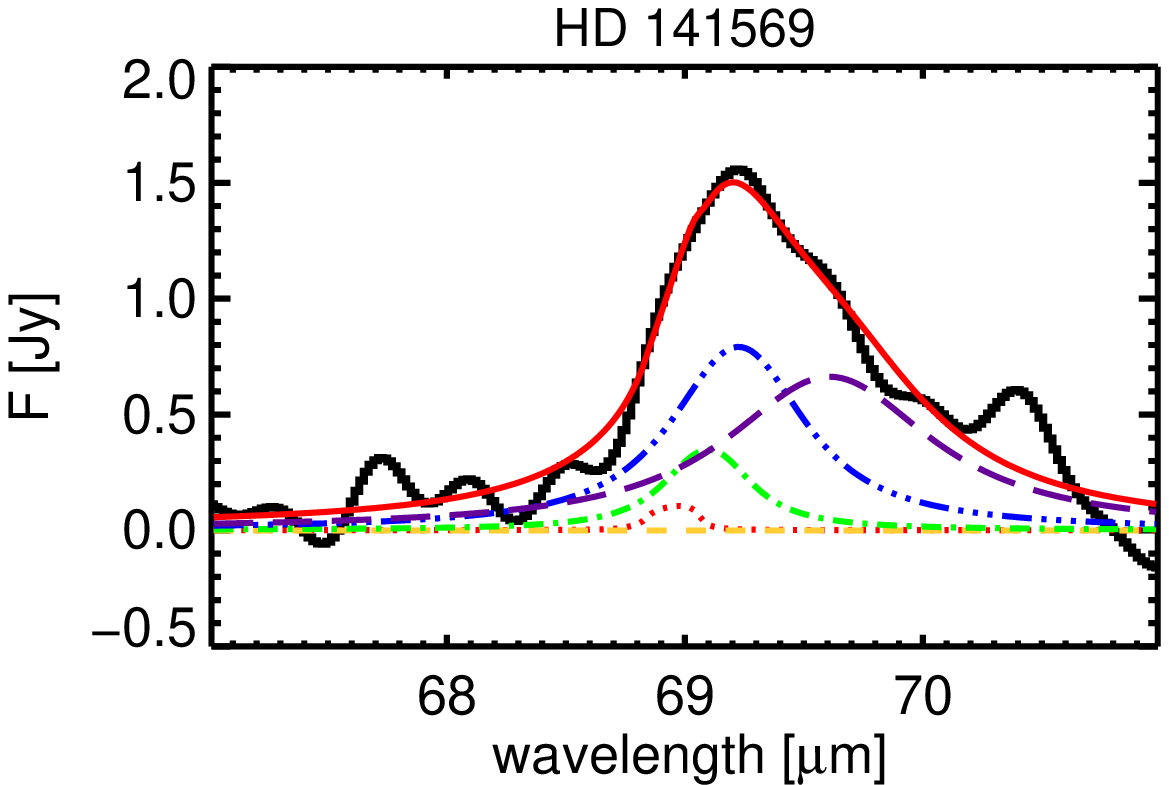}
    \label{fig:HD141569-filtered-detail}}
\subfigure{
    \includegraphics[scale=0.60]{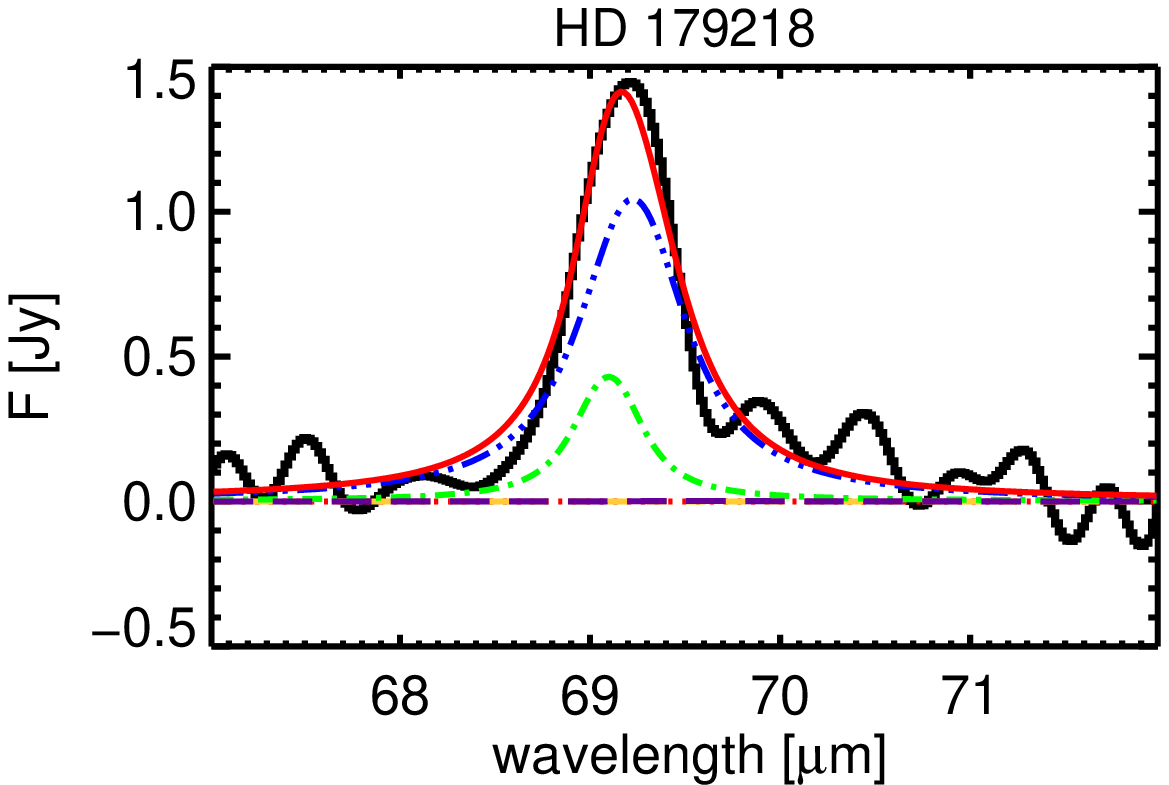}
    \label{fig:HD179218-filtered-detail}}
\subfigure{
    \includegraphics[scale=0.60]{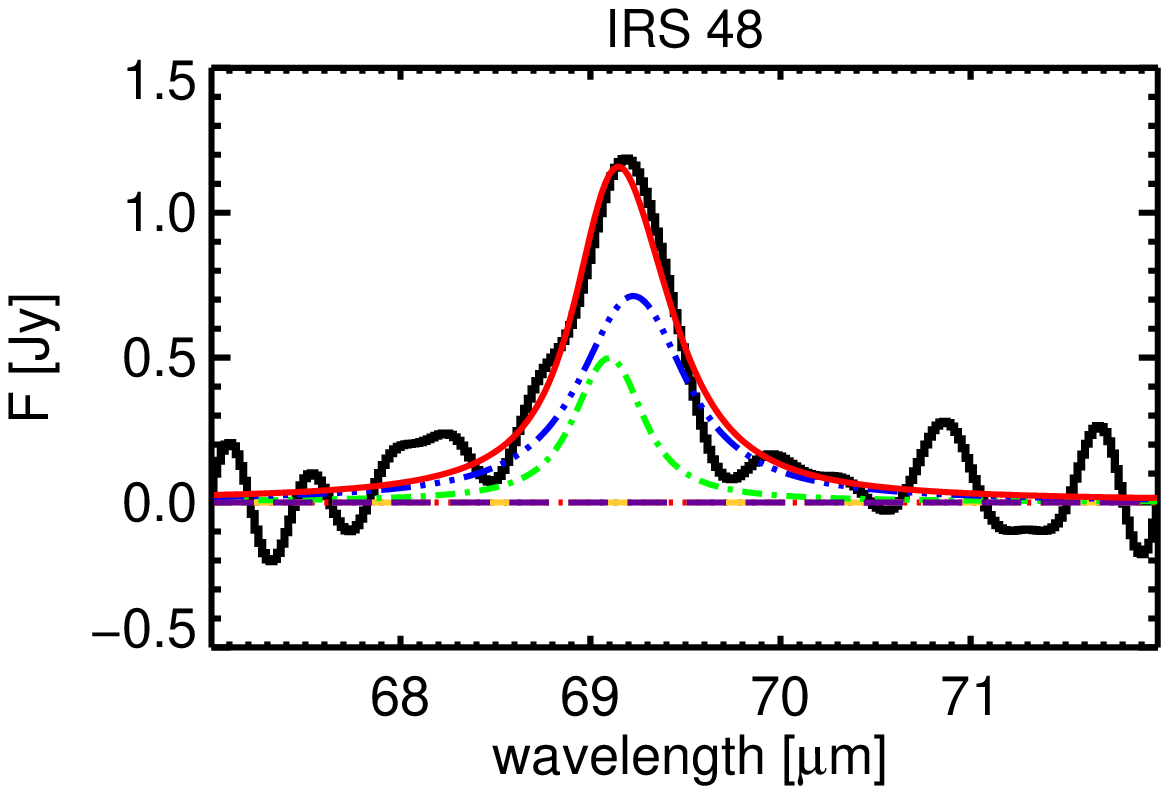}
    \label{fig:IRS48-filtered-detail}}
\subfigure{
    \includegraphics[scale=0.60]{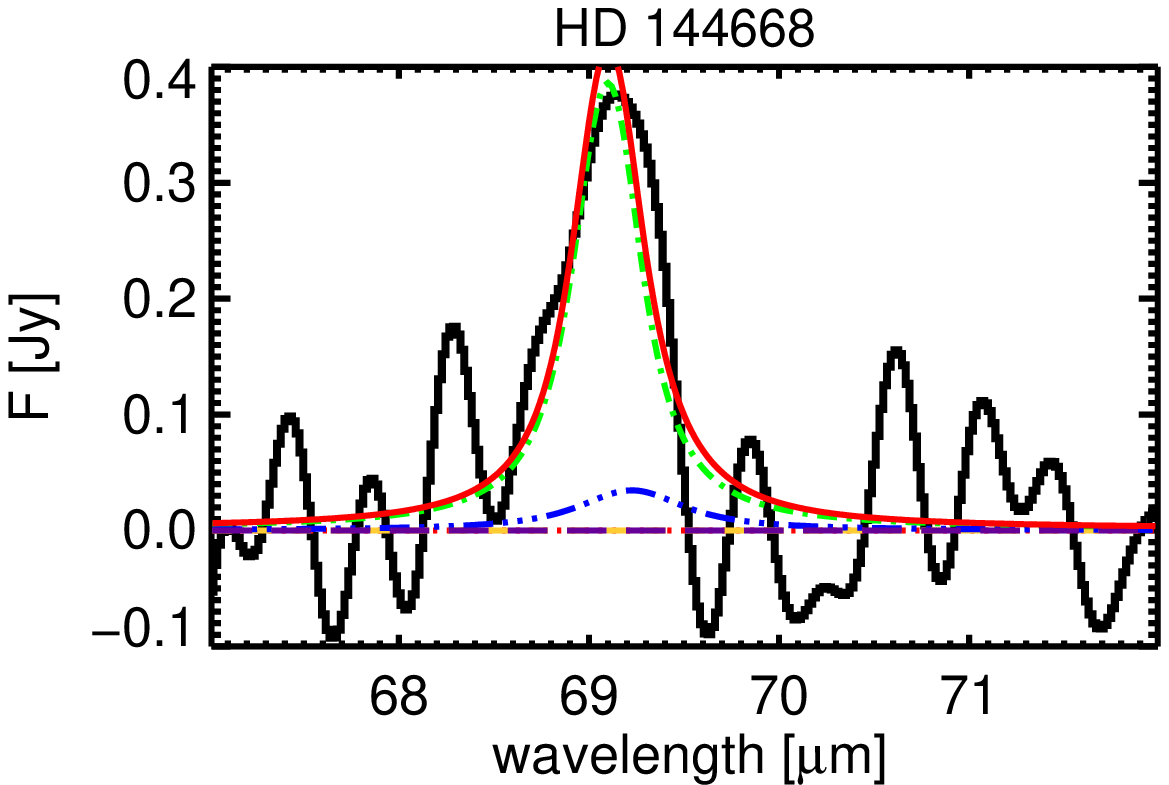}
    \label{fig:HD144668-filtered-detail}}
\subfigure{
    \includegraphics[scale=0.60]{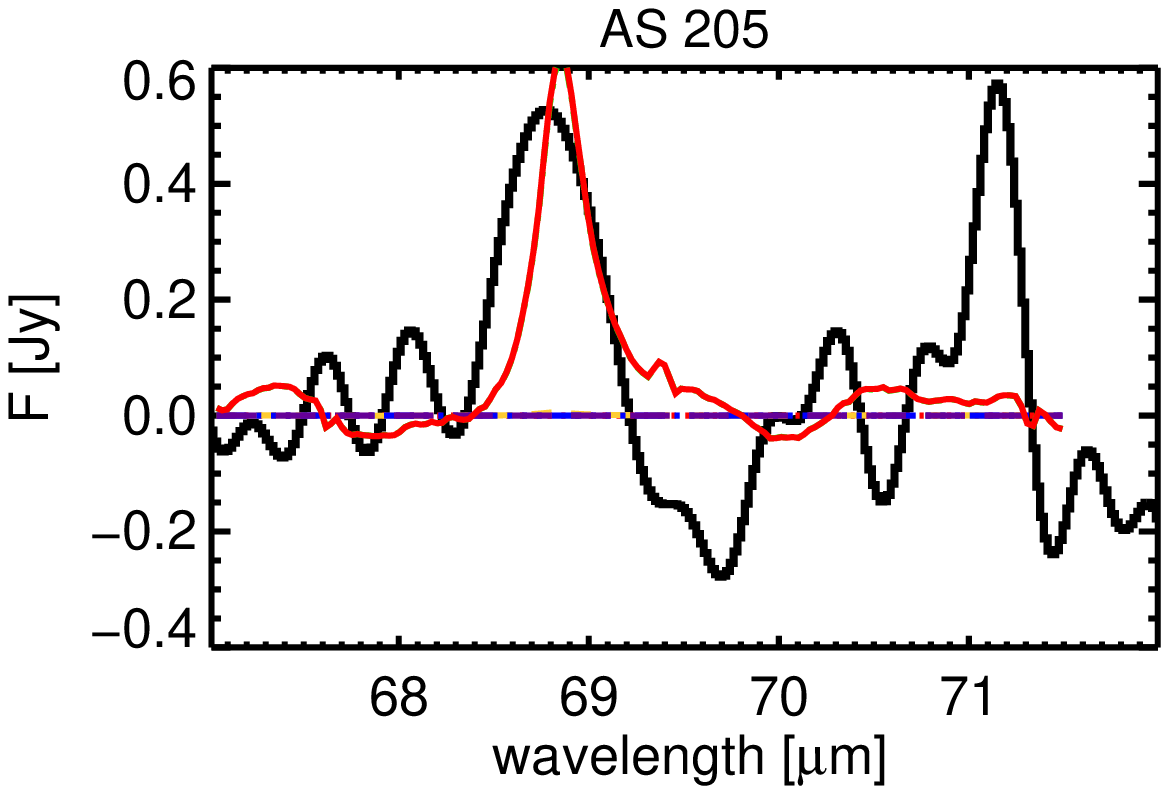}
    \label{fig:AS205-filtered-detail-detail}}
\caption{Results of fitting eq. \ref{eq:dustmodel} with pure forsterite to the detected 69\,$\mu$m forsterite emission bands. The measured spectrum is plotted as a black line, over-plotted with the best-fit model (continuous red line). The contribution of crystals at different temperatures is shown as follows: The dotted red line is the 50\,K component, dashed yellow represents 100\,K, dash--dotted green stands for 150\, while the dashed--three--dot blue curve represents the 200\,K sample and dashed purple the 295\,K data. In all fits except for AS~205 we use the data from \citet{Suto2006} with the DHS shape model and a grain size of 0.1~$\mu$m. In AS~205 the position of the peak indicates very low temperatures so we used the laboratory data from \citep{Koike2006} which include 8-- (red, dotted) and 20~K (dashed, yellow) samples. The profile of AB~Aur cannot be fitted with pure forsterite. In this case the admixture of some percent iron greatly improves the fit (see fig. \ref{fig:AB-Aur_Fe})
.}
\label{fig:fo-details01}
\end{figure*}

In addition to the analysis of the previous section we also investigated the influence of a temperature distribution on the derived dust properties. For a more precise analysis of the temperature of the forsterite, we fitted the observed profiles with a weighted sum of laboratory data \citep{Suto2006,Koike2003} at different temperatures (50, 100, 150, 200 and 295~K)\footnote{The peak of the emission band in AS~205 is at less than 69~$\mu$m and therefore can only be fitted with laboratory data in the 8--150~K range} (Fig. \ref{fig:fo-details01}). The measured flux $F_{\rm {67-72}\mu m}$ is modeled by
\begin{equation}
F_{\text{67-72}\mu\text{m}}(\lambda) = F_{\textnormal{cont}}(\lambda) + \sum_{i=1}^{5}{w_i \cdot \kappa(T_{\textnormal{dust}}^i, \lambda) \cdot B_{\lambda}(\lambda, T_{\textnormal{dust}}^i)}.
\label{eq:dustmodel}
\end{equation}
Here $F_{\rm{cont}}$ is the local continuum (a 2$^{\text{nd}}$ or 3$^{\text{rd}}$ order polynomial), and the $\kappa$ values are the mass absorption coefficients, computed from the optical constants of the \citet{Suto2006} or directly taken from \citet{Koike2006} pure forsterite sample and $B_{\lambda}$ are the Planck function at the dust  temperatures $T^i_{\mathrm{dust}}$. Finally, the $w_i$ values are the relative weights of the flux at the different temperatures. In addition, we had to specify a dust shape distribution in order to calculate the mass absorption coefficients from the optical constants given by \citet{Suto2006}. We choose the DHS model \citep{Min2003} with a grain size of $0.1\,\mu$m and a filling factor of $f_{\mathrm{max}}$ of 1, which has been found to be a good representation of observed silicate profiles in \emph{Spitzer} data \citep[e.g.,][]{Juhasz2010}. The parameters are fitted with the IDL implementation of the least squares minimisation \texttt{mpfit} \citep{Markwardt2009}. 

%

%
\begin{figure}
\centering
\includegraphics[scale=.65]{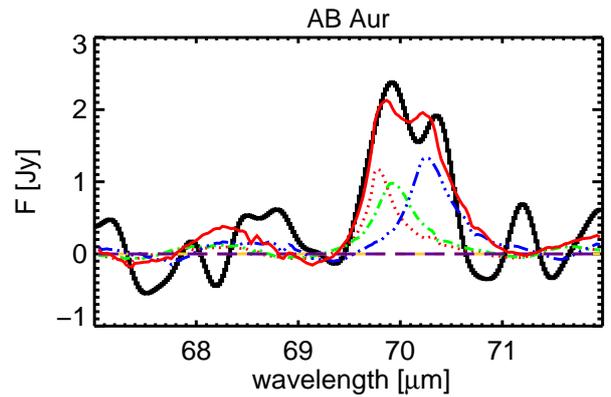}
\caption{AB~Aur: The continuum--subtracted forsterite emission at 69\,$\mu$m (histogram) over-plotted with the best-fit model (red line) which is given by eq. \ref{eq:dustmodel}. This model is a sum of the following components: 50\,K 3\% iron (red dotted line), 50\,K with 4\% iron (yellow dashed line), 100\,K 3\% iron (green dash-dotted line) and 100\,K 4\% iron (blue dash-three-dots line) and 200\,K 4\% (purple dashed line). Each of the components is multiplied by a fitted weighting factor.}
\label{fig:AB-Aur_Fe}
\end{figure}

In a first attempt we only use the pure forsterite data of \citet{Suto2006} in the fit, except for AS~205 where the laboratory data from \citet{Koike2003} was taken to account for the position of the peak (see Figure \ref{fig:fo-details01} for results). All detected bands with the exception of AB~Aur are very well described under this assumption by the model from Eq. \ref{eq:dustmodel}. The most dominant components for almost all sources are at 100--200\,K, a result which is consistent with the analysis presented in Section \ref{subsect:ironcontent_and_temperature}. The major contribution in AS~205 stems from dust at 20~K. In one case (HD\,141569) a large spread of temperatures (50--295~K) is found which may suggest two separate reservoirs of forsterite in the disk. As we show in section \ref{subsect:grainsize} the shape of the emission band in HD~141569 can alternatively be explained by large ($\sim$10~$\mu$m) grains at a single temperature. The bump in the data at around 70.4~$\mu$m is not a significant 
feature and consequently we do not investigate it further.

The peak of the emission band in AB~Aur is found beyond 70~$\mu$m and can only be fitted if we allow for an admixture of some ($\sim$\,3--4\%) iron. The best fit to the AB~Aur spectrum using eq. \ref{eq:dustmodel} that includes iron is shown in fig. \ref{fig:AB-Aur_Fe}. The mass absorption coefficients used for these models were interpolated from the \citet{Suto2006} and the \citet{Koike2003} data. For the interpolation we assumed that the band is shifted linearly with the iron content and exhibits the same temperature dependency for all different iron fractions up to 10\%. Even for AB~Aur we find that no more than 3--4\% iron is needed to obtain a good fit.

For each source we computed a mass-averaged temperature from our fit results. The confidence intervals for the dust temperatures were derived from a $\chi^2$ analysis similar to the procedure we discussed in section \ref{subsect:ironcontent_and_temperature}. Since we only have laboratory data at five different temperatures, and the effects of nonzero iron fractions $<$ 1\%  cannot properly be taken into account as all laboratory measurements were made using pure forsterite, the values derived here may suffer from larger systematic effects than the ones given in Section \ref{subsect:ironcontent_and_temperature}. For AS~205 and AB~Aur we cannot compute confidence intervals as even the best-fit value already has too large of a reduced $\chi^2$. In the case of AS~205, the mismatch between the observed band profile and the best-fit model can be explained as a consequence of insufficient laboratory data concerning the optical properties at temperatures of $\le$50~K. For AB~Aur we have to manually select the iron content in the data used to 
fit the temperature distribution. Since the iron content is not a parameter in the fit in this model (see eq. \ref{eq:dustmodel}) the minimal $\chi^2$ achieved in the fit is too large for the computation of a 3~$\sigma$ confidence interval.

We summarise our fit results in Table~\ref{tab:fitresults}. All detected 69~$\mu$m bands except the ones in AS~205 and AB~Aur can be fitted with pure or iron-poor and warm (150--200~K) forsterite. In AS~205 the most dominant component is found at $\sim$20~K, while AB~Aur requires some percent of iron at a temperature of $\sim$50~K.

The temperatures, and therefore the radial distance estimates based on them, are consistent with those derived from the $\chi^2$ maps in section \ref{subsect:ironcontent_and_temperature}. Only AS~205 shows strong deviations between the results of the two different methods. This is most likely attributed to the poor quality of the fit in either of the two scenarios. In fact, the minimum $\chi^2$ in the fit to the observed profile in AS~205 is too large to derive a confidence interval with the method described in section \ref{subsect:ironcontent_and_temperature}.

\begin{table*}

\begin{center}
\caption{Fit results of the spectral decomposition of the detected forsterite bands.}
\label{tab:fitresults}
\small
\begin{tabular}{@{}lllllllr@{}}
\toprule
            &   \multicolumn{5}{c}{Relative mass fractions}         &  M.A.T  & distance \\
Star        &    @ 50~K   & @ 100~K  & @ 150~K & @ 200~K& @ 295~K   &   \multicolumn{1}{c}{[K]}   & \multicolumn{1}{c}{AU} \\
\cmidrule{1-8}
HD~100546   &    0.00     &    0.11  &   0.22  &   0.63 &    0.04   &   166 -- 204  & 24 -- ~~36 \\
IRS~48      &    0.00     &    0.00  &   0.45  &   0.55 &    0.00   &   156 -- 189  & 19 -- ~~27 \\
HD~144668   &    0.00     &    0.00  &   0.93  &   0.07 &    0.00   &   135 -- 183  & 37 -- ~~69 \\
HD~179218   &    0.00     &    0.00  &   0.32  &   0.68 &    0.00   &   174 -- 186  & 90 -- 103 \\
HD~141569   &    0.00     &    0.00  &   0.37  &   0.23 &    0.40   &   148 -- 248  & 13 -- ~~38 \\
HD~104237   &    0.00     &    0.00  &   0.00  &   1.00 &    0.00   &   180 -- 211  & 23 -- ~~32 \\
  \multicolumn{8}{c}{}\\
\cmidrule{2-6}
            &   @ 8.K     &  @ 20.K  & @ 50.K  & @ 100.K&   @ 150.K &    \\
\cmidrule{2-6}
AS~205      &    0.00     &    0.78  &   0.22  &   0.00 &    0.00   &    $\sim30$  & $\sim516$ \\
  \multicolumn{8}{c}{}\\
\cmidrule{2-6}
            & @ 50~K, 3\% Fe&@ 50~K, 4\% Fe&@ 100~K, 3\% Fe& @ 100~K, 4\% Fe&@ 200~K, 3\% Fe &    \\
\cmidrule{2-6}
AB~Aur      &   0.75      &    0.00  &   0.10  &   0.15 &     0.00  &    $\sim63$  & $\sim304$\\

\bottomrule                                           
\end{tabular}
\normalsize
\end{center}
\textbf{Note:} Listed are the relative mass fractions of the forsterite components at different temperature or with different iron mass fractions. Also listed is the confidence intervals for the mass-average-temperature (M.A.T.) of the fitted dust components. The distance intervals are computed based on the ranges of dust temperature (see text). 
\end{table*}

\subsection{Distance of the forsterite grains from their host stars}
\label{subsect:distance}

With the temperature of the forsterite grains derived from the profiles of the 69~$\mu$m emission band (see Tab. \ref{tab:chsqrfitresults} and Tab. \ref{tab:fitresults}) we can estimate the distance of the dust particles from their host star. For this estimate we have to assume that the dust grains are located in an environment that is optically thin, in both optical wavelengths where the star emits and IR wavelengths where the dust grains re-emit, an assumption that is very likely invalid for almost all sources in our sample. In that case the estimated distances must be interpreted as upper limits. More reliable distances would require full radiative transfer modeling for each of the sources, which is beyond the scope of this paper.

The equilibrium temperature of a dust grain in an optically thin environment depends on the optical properties not only at the wavelengths where it emits but also at the wavelengths around the peak of the stellar emission, where it receives most of the radiation. Since the forsterite grains in protoplanetary disks are likely part of larger, porous aggregates mixed together with other materials, these properties will be very different from those of pure forsterite grains. However, the exact composition and structure of these dust aggregates cannot be determined from the infrared spectra in the \emph{Spitzer} IRS and \emph{Herschel} PACS wavelength ranges since most of the possible constituents dominating the opacities at optical wavelengths (e.g. carbon, iron) have no emission bands in this spectral region. \citet{Lisse2009} present an approach, where it is assumed that the dust aggregates in circumstellar disks have optical properties similar to those in cometary material in the solar system for which we 
know the temperature-distance relation. We therefore use the relation from \citet[][ section 2.2.6]{Lisse2009} and solve for the distance of the forsterite particles from their respective host star.


We estimate the location of the forsterite grains in the disks based on the dust temperatures derived in sections \ref{subsect:ironcontent_and_temperature} and \ref{subsect:spectral-decomposition} using the temperature-distance relation by \citet{Lisse2009}. The derived distances are listed in Tables \ref{tab:chsqrfitresults} and \ref{tab:fitresults}. The distance estimates based on the two different methods used to derive the dust temperature agree reasonably for most of the sources with minimum distances of 10--30~AU. The larger values in HD~179218 are easily explained by the larger luminosity of the star compared to the rest of our sample. In HD~104237 and AB~Aur, compositional effects play a more significant role, thus explaining the different results for these sources in Tables \ref{tab:chsqrfitresults} and \ref{tab:fitresults}. The emission profile of the 69~$\mu$m band in AS~205 is not well-fitted in either of our two approaches, leading to a large difference in the computed distance of the forsterite from the 
central star, depending on the method used to determine it.

The most notable difference between the results of the two methods we used in Sections  \ref{subsect:ironcontent_and_temperature} and \ref{subsect:spectral-decomposition}  to derive the temperature is the size of the confidence interval. The spectral decomposition method in Section \ref{subsect:spectral-decomposition} generally produces much narrower confidence intervals on the temperature and thus the radial distribution of the forsterite. The reason for this is that in the spectral decomposition we only use pure forsterite and even a small amount of iron could significantly decrease the temperature of the best-fit solution (see section \ref{subsect:detection1}). The effect is also illustrated in Figure \ref{fig:chisqr} where the $\chi^2$ contours extend to lower temperatures as we move to higher iron fractions.

In fact all the distance values given in Tables \ref{tab:chsqrfitresults} and \ref{tab:fitresults} are estimates and should be taken with caution. For the computation we have  assumed that the forsterite is located in an optically thin environment. This assumption may be violated in protoplanetary disks like the ones observed in our sample. As shown for example by \citet{Mulders2011} the temperature profile inside an optically thick disk is dramatically different from that of the optically thin disk atmosphere.

A more precise determination of the location of the forsterite in the protoplanetary disks requires detailed radiative transfer modeling which is beyond the scope of this paper. Such modeling would also greatly benefit from a more complete set of optical constants, covering olivines with low  iron content at different temperatures. Radiative transfer modeling is very likely to decrease the derived distance of the forsterite reservoirs from their central stars, unless the grains are located at the inner edge of a disk \citep[as in HD~100546, see ][]{Mulders2011}. The distances derived in this paper can be regarded as upper limits.

\subsection{Grain size effects}
\label{subsect:grainsize}
The effect of grain size of the emitting dust on the position and shape of the 69~$\mu$m band has been discussed in Section \ref{subsect:detection1} and illustrated in Figure \ref{fig:grainsize}. The largest grains that produce a 69~$\mu$m band with a profile that is compatible to our detections have radii of about 10~$\mu$m.

If we take large grains (1--10~$\mu$m) into account, the temperature estimate we derive from the peak position and FWHM (similar to Figure \ref{fig:peakFWHM}) is slightly lower than for small grains. Assuming that almost all the dust consists of grains with radii of 10~$\mu$m allows a reasonable fit to all detected bands with only one temperature component at around 100~K. However, for most of the objects the fit is significantly worse than the one presented in Figure \ref{fig:fo-details01} where we used smaller grains.

\begin{figure}
\centering
\includegraphics[scale=.65]{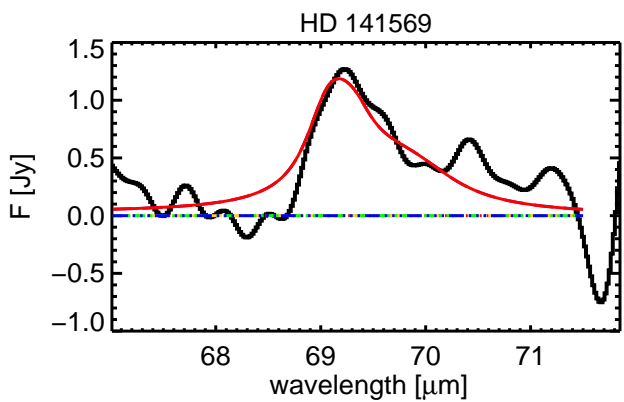}
\caption{The 69~$\mu$m band in HD~141569 fitted with a combination of the emission from different grain sizes at 150~K. Only the emission from grains of 10~$\mu$m diameter contributes to the best-fit model.}
\label{fig:HD141569-grainsize150K}
\end{figure}

A more detailed analysis by fitting a weighted sum of different grain sizes to the data (similar to Eq. \ref{eq:dustmodel}) is done for each of the temperatures at which \citet{Suto2006} published optical constants (50, 100, 150, 200 and 295~K). The best results are typically found for 100--150~K dust with no significant contributions (mass fraction $\le10^{-3}$) from 5--8~$\mu$m grains. AB~Aur cannot be fitted with pure forsterite and we have no laboratory data available from which we could derive the cumulative effect of iron and grain size. The band in AS~205 is at very short wavelength and very narrow. None of the fits with large grains could describe the observed emission band. Only for HD~141569 do we find indications for a significant contribution from 10~$\mu$m grains. For this system the 69~$\mu$m band can be modeled with emission from 10~$\mu$m grains only (Fig. \ref{fig:HD141569-grainsize150K}) at a temperature of 150~K without any significant change in fit quality compared to the model we discussed in the previous subsection.

\subsection{Comparison to \emph{Spitzer} IRS data}

Most of our targets were observed with the InfraRed Spectrograph (IRS) onboard the
\emph{Spitzer Space Telescope}. To compare the results of our analysis of the  \emph{Herschel}
spectra with those of the forsterite emission bands seen in the  \emph{Spitzer} observations,
we use the spectra published by \citet{Juhasz2010} and  \citet{Olofsson2009}. 
For the targets where we detected a 69~$\mu$m band, and those with marginal \emph{Spitzer} detections,
we carefully re-reduced the \emph{Spitzer} spectra, using the same data reduction approach as
\citet{Juhasz2010}, but with the latest calibration version (S18.18.0). Our re-processed spectra only marginally 
differ from those reported by \citet{Juhasz2010} and \citet{Olofsson2009}.  However, we report the detection of forsterite bands between $\sim$19 to 33~$\mu$m in the
\emph{Spitzer} spectrum of HD~141569, not reported by \citet{Juhasz2010}.
Interestingly, this source was classified by \citet{Meeus2001} as a Group~Ib
source, i.e. a strongly flaring disk with no silicate emission bands.

\begin{figure}
\centering
\includegraphics[scale=.65,angle=90]{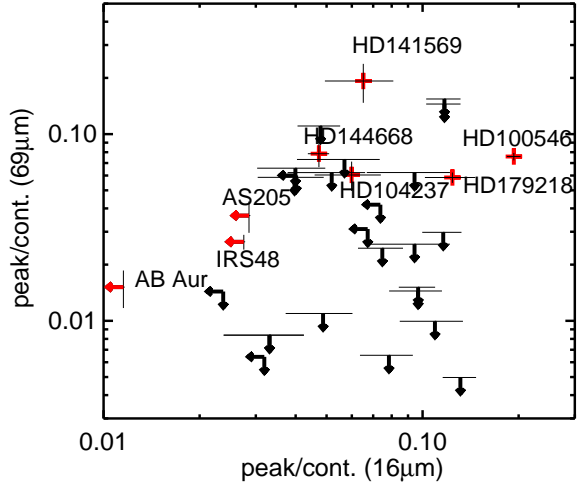}
\caption{Comparison of the peak/continuum of the 69\,$\mu$m to the 16\,$\mu$m bands in our sample. The sources with a firm detection of the 69\,$\mu$m band are labeled. }
\label{fig:peakContinuum16}
\end{figure}

\begin{figure}
\centering
\includegraphics[scale=.65,angle=90]{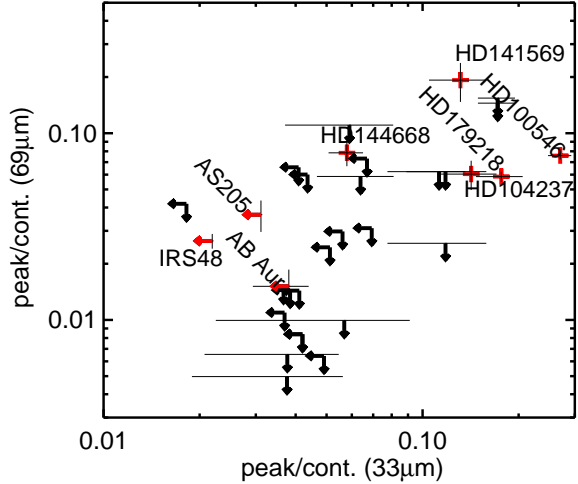}
\caption{Comparison of the peak/continuum of the 69~$\mu$m to the 33~$\mu$m bands. Sources with a 69~$\mu$m detection are labeled.}
\label{fig:peakContinuum33}
\end{figure}

We searched for correlations between the strength of the forsterite bands seen in the
\emph{Spitzer} data and the 69~$\mu$m band. For this, we analysed the forsterite features seen in the \emph{Spitzer} spectra in an identical fashion to our analysis of the 69~$\mu$m band.
For all the forsterite bands we
computed the ratio of the Lorentzian and the continuum polynomial at the
position of the peak of the best-fit Lorentzian. The comparisons between the 16~$\mu$m and
33~$\mu$m peak over continuum values derived from the \emph{Spitzer} data and the 
69~$\mu$m peak over continuum values are shown in Figures ~\ref{fig:peakContinuum16} 
and \ref{fig:peakContinuum33}. We compare the \emph{Herschel} results 
with the 16 and 33~$\mu$m bands as those have very little confusion with other 
emission bands, and probe the warmest and coldest dust detectable in the \emph{Spitzer} wavelength range, respectively.

%

\begin{figure}
\centering
\includegraphics[scale=.35]{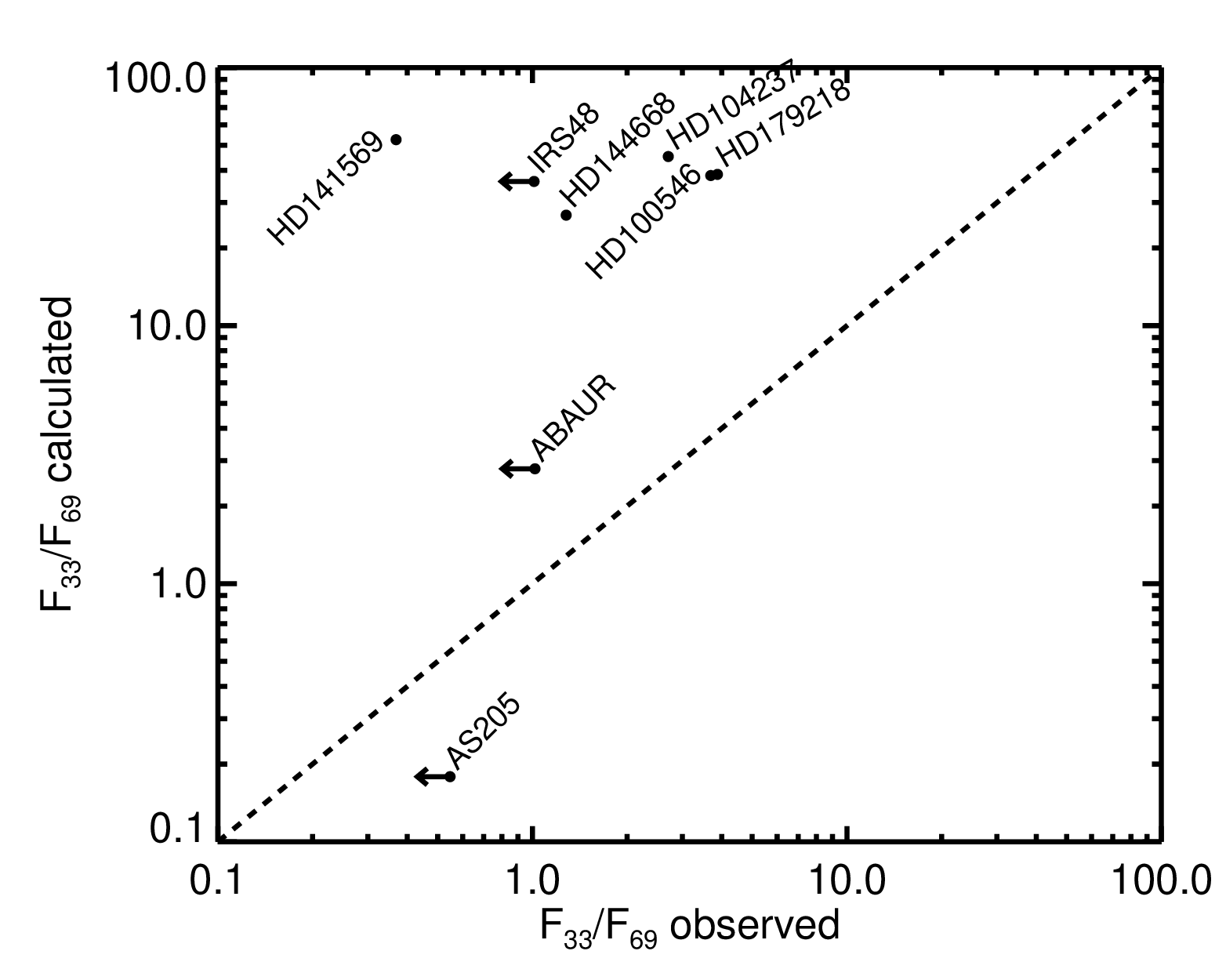}
\caption{Comparison between the measured 33~$\mu$m over 69\,$\mu$m peak fluxes in the forsterite bands and the expected ratio's based on the derived mass averaged temperatures of the forsterite grains as listed in Table~\ref{tab:fitresults}. The dashed line is the expected behaviour if no optical depth effects play a role.}
\label{fig:ratios_comparison}
\end{figure}

%
\begin{figure}
\centering
\includegraphics[scale=.65]{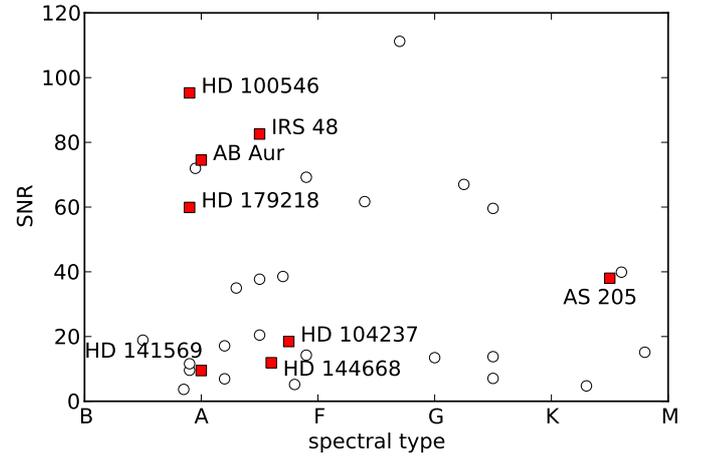}
\caption{Spectral type versus signal--to--noise ratio. Sources with a detected 69~$\mu$m forsterite band are shown as red squares. The S/N is computed as the average flux divided by the mean point--to--point difference in the spectra. }
\label{fig:SpTypeSNR}
\end{figure}

The comparison between the 16~$\mu$m and 69~$\mu$m bands does not show any
obvious correlation, implying that there is no straightforward observational
correlation between the emission from the warmest and the coldest forsterite grains as probed 
by the 16~$\mu$m and, respectively,  the 69~$\mu$m band.
There appears to be a slightly stronger correlation with the 33~$\mu$m band: those sources which show the 
strongest 69~$\mu$m band also show an on average stronger 33~$\mu$m band.
A correlation between the 33~$\mu$m and the 69~$\mu$m feature might be expected due to the relative warm temperatures of the forsterite grains ($\sim$150--200~K, see Table~\ref{tab:fitresults}). However, as discussed by \citet{Sturm2010} for HD~100546, the 33~$\mu$m over 69~$\mu$m band ratio is consistent with a $\sim$70~K grain temperature, rather than the $\sim200$~K derived from the 69~$\mu$m feature. The solution to this was proposed by \citet{Mulders2011}, who showed that the 33~$\mu$m band can be substantially suppressed relative to the 69~$\mu$m feature by optical depth effects. These effects could blur any straightforward correlation between the \emph{Spitzer} and \emph{Herschel} observations. 

To check whether any of the other disks could have similar optical depth effects as those seen in in the disk of HD~100546, we plotted in Fig.~\ref{fig:ratios_comparison} the observed 33~$\mu$m over 69~$\mu$m band ratio against the expected ratios based on the temperatures listed in Table~\ref{tab:fitresults}. Except for AS~205, all disks must be  
optically thick at mid-IR wavelengths to suppress the forsterite bands in the \emph{Spitzer} IRS wavelength range. For these disks, the emission seen with \emph{Spitzer} arises from the disk surface, while the bulk of the 69~$\mu$m emission originates from the disk interior. Most of the emission should come from a disk region where 
dust temperatures in the range of 150--200~K are reached, which for a typical Herbig~Ae system is $\sim$10~AU from the central star. Only for AS~205 are our findings are consistent with a complete lack of warm forsterite grains.  In that case, the emission originates from a reservoir of cold grains with very low iron content located either in a region far out in the disk, or in the cold interior, which is shielded from direct stellar radiation. Note that as this system is a T~Tauri star, low temperatures are reached closer to the central star than in the Herbig~Ae systems. The physical location of the forsterite grains might therefore not be that different from the $\sim$10~AU we expect 
for the intermediate-mass systems.

In contrast to the amorphous silicate grains, no evidence for substantial growth of the forsterite crystals
is found from \emph{Spitzer} observations \citep[e.g.][]{Juhasz2010}. The observed mid-IR emission bands are all consistent with emission from small ($\le$1~$\mu$m) forsterite grains. Also in our \emph{Herschel} observations no conclusive evidence is found for the presence of larger grains. The only source where indications of larger (10~$\mu$m) grains are found is HD~141569, though the 69~$\mu$m band can also be modeled with small grains only.
As the far-IR observations tend to probe closer to the disk midplane, if grain growth and gravitational settling of the larger grains occur in these disks, our \emph{Herschel} observations should be more sensitive to a population of larger grains. As we do not detect these larger grains, these results are consistent with the notion that the small forsterite grains are embedded into larger aggregates consisting of mainly amorphous silicates or carbonaceous material. As \citet{Min2008} show for aggregates, materials that have a very low abundance appear spectroscopically as if they were in very small grains, while more abundant materials appear spectroscopically to reside in larger grains.

We also investigated whether a correlation exists between the presence of the 69~$\mu$m band and the global disk shape. \citet{Meeus2001} divided the Herbig~Ae/Be systems into 2 groups: Group~I, having a large far-IR excess corresponding with a strongly flaring disk, and group~II, having a SED consistent with a flattened disk structure. HD~104237 is the only group~II source with a forsterite detection; all other sources are classified as group~I. This might, however, be an observational bias. Due to the SED shape of the group~II sources, the far-IR fluxes are systematically lower than those of the group~I sources and, therefore, the S/N of the spectra are typically lower. An example of this is HD~150193, a source with one of the strongest forsterite features at mid-IR wavelengths, but with no detected 69~$\mu$m band. Due to the low far-IR flux of this source, the upper limit we can place on the forsterite band would still be consistent with a relatively strong feature, like the one observed in HD~179218. 

\subsection{Correlation of detection rate with S/N and spectral type}
\label{subsect:correlations1}
We carefully looked for a correlation of the detection of a 69~$\mu$m band with the S/N. To account for the complex continua in our sample, we computed the S/N as the average flux divided by the mean point--to--point difference in the spectra. Among the 16 sources with the highest S/N, we find five detections, compared to three detections among the 16 sources with lower S/N. This indicates a weak correlation of the detection rate with S/N. 

\begin{figure}
\centering
\includegraphics[scale=.65]{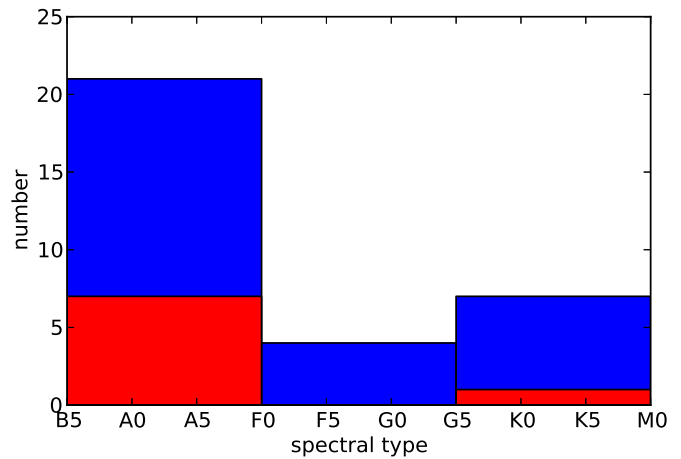}
\caption{The spectral type distribution in the observed sample (blue) and the distribution of the 69~$\mu$m detections (red).}
\label{fig:sptype_hist}
\end{figure}

Remarkably, almost all stars with a detection of the 69~$\mu$m forsterite feature have spectral types of B9--A9 (see fig. \ref{fig:SpTypeSNR}). The only exception is AS~205, which has a spectral type of K5. In total we have detected the 69~$\mu$m band in 7 out of 21 objects with spectral types earlier than F compared to only 1 detection among 11 objects of spectral types F and later. This seems to indicate a correlation of the detection rate with spectral type. However, the correlation is not significant. We show the spectral type distribution in our sample and the distribution of the detections of the 69~$\mu$m forsterite band in Fig. \ref{fig:sptype_hist}. The number of bins in the histogram was chosen following the strategy presented in \citet{Hogg2008}. Assuming the 8 detections of the 69~$\mu$m forsterite band were randomly distributed among the 32 sources in our sample, the probability to find the observed distribution would be $\sim$8\%.

We also carefully checked if there are significant deviations in the S/N distribution within our target sample for stars with spectral types earlier or later than F0. A non--parametric Mann--Whitney two--sample U--test indicates that the probability that the S/N distribution is the same for both early-- and late--type stars is $P = 0.38$, or formulated differently, the probability that our Herbig Ae/Be sample has a different S/N distribution than the T~Tauri sample has only 1~$\sigma$ significance. Thus, we have not detected a significant S/N bias favouring either early or late spectral types in our sample.

AS~205 may show up as the only 69~$\mu$m detection in a spectral type later than A because of the geometry and orientation of this particular disk. \citet{Andrews2009} find that the scale height of the protoplanetary disk in AS~205 is the largest in their sample, while the mass is at the lower end of the spectrum. This implies that the dust density is low compared to the other sources, allowing for an optically thin medium and therefore enabling us to detect all of the forsterite mass at 69~$\mu$m while in more massive/dense disks only a fraction would be visible. This is supported by our finding that we do not need to take optical depth effects into account when comparing the \emph{Herschel} and \emph{Spitzer} data (see Fig. \ref{fig:ratios_comparison}). Also, the large scale height indicates a strongly flaring disk, which will intercept a higher portion of the central star's radiation compared to flatter disks. Thus, the disk of AS~205 might be warmer than those of stars with comparable spectral 
types. Finally, the disk of AS~205 is seen nearly face--on \citep{Andrews2009}, maximising the projected surface which is seen by \emph{Herschel}.

\section{Conclusions}
\label{sec:conclusion}
We searched for the 69~$\mu$m forsterite emission band in 32 disk sources (see Table \ref{tab:sources01}) and identified it in only eight of the objects. Several more spectra show some emission features between 69 and 70~$\mu$m which can not clearly be attributed to forsterite. This detection rate (even including the features with uncertain identification) is much lower than the one for the forsterite bands in the \emph{Spitzer} IRS wavelength range of the same sample. 

This work provides the first overview of the presence and properties of the 69~$\mu$m band in a sample of disks that covers various ages, disk masses and stellar properties. Only one detection of the 69~$\mu$m band in a disk source was reported from an ISO observation \citep[HD~100546, ][]{Malfait1998}. Thanks to the unprecedented sensitivity and spectral resolution of \emph{Herschel} PACS we can for the first time derive the temperature and composition of the crystals from the shape and position of the emission band.

\subsection{Iron content, size, temperature and location of the forsterite grains}
\label{subsect:temperature_iron_and_location}
We compared the observed 69~$\mu$m forsterite emission band with laboratory measurements of this feature for different iron fraction  and grain temperature. In accordance with the ISO detection of the 69~$\mu$m band in HD~100546 we find that all but one of the detected bands in our sample are consistent with emission from iron-poor grains. For one source, AB~Aur,  a 3--4\%  iron fraction is needed in order to explain the position of the detected emission band. In all other objects pure forsterite can account for the 69~$\mu$m features although a small amount of iron ($\le$1\%) cannot be ruled out. Our results are consistent with the recent detection of the 69~$\mu$m band in the debris disk system $\beta$~Pictoris, which shape and position is consistent with olivine with an iron fraction of at most 1~\% \citep{deVries2012}.

Our analysis shows no conclusive evidence for larger (5--10~$\mu$m) forsterite grains. 
The observed 69~$\mu$m bands can be modeled with emission from small (1~$\mu$m or less) forsterite grains. These inferred small grain sizes are consistent with the grain sizes for the forsterite crystals derived from \emph{Spitzer} observations \citep[e.g.][]{Juhasz2010}. The only source where indications of larger (10~$\mu$m) grains are found is HD~141569, though the observed 69~$\mu$m feature can also be modeled with small grains only.

From the analysis of the position and shape of the emission bands we derive a best-fit grain temperature for pure forsterite grains of around 150--200~K for all sources except for AB~Aur and AS~205. The latter source requires very cold dust grains at temperatures of 30~K or less to explain the observed band position, though an exact temperature could not be determined due to the sparsity of laboratory measurements of the optical properties of forsterite at these very low temperatures. In the case of AB~Aur, no pure forsterite can explain the observed spectral feature. Determining grain temperatures using iron-containing olivine grains leads to similar results though the derived temperatures are less well-constrained due to systematic uncertainties (see Section~3.1) caused by the interchangeability of the effects of grain temperature and iron content on the band position and shape.

We used the derived grain temperatures to estimate the radial distance from the forsterite grains to their respective host stars. We find that the forsterite grains we observe with \emph{Herschel} are located at least 10~AU or farther from the central star. In case of pure forsterite grains, we find the forsterite for most of our detections (except AB~Aur and AS~205) to be located in a relatively narrow ring in the circumstellar disk.  Note that for our simple distance estimation we assume an optically thin environment and that all of the values will be modified when taking radiative transfer effects in the disks with wavelength-dependent optical depth into account. The distances estimates in this paper should, therefore, be regarded as upper limits.  In the case of HD~100546, an analysis of the forsterite grain location using a full radiative transfer model has been published \citep{Mulders2011}. The derived location in a narrow ring of around 13-20~AU radius by \citet{Mulders2011} is at least 
qualitatively in agreement with the narrow location we derive of $\sim$20-30~AU.

\subsection{Formation history of the forsterite grains}
\label{subsect:formation}
With the constraints on the grain temperature and composition we derived from the \emph{Herschel} PACS spectra, we can speculate on formation history of the forsterite grains. The very low iron content in the crystalline silicates supports a crystal formation scenario through condensation from the gas phase at high ($\sim$1500~K) temperatures \citep{Gail2010}. This is in apparent contradiction to the relatively low grain temperatures we derive for the forsterite, as was already noted from the analysis of \emph{ISO} and \emph{Spitzer} spectra \citep[e.g][]{Juhasz2010, Malfait1998}. Several scenarios have been proposed to explain this discrepancy, such as radial mixing from the hot inner disk to the outer disk regions \citep[e.g.][]{Bockelee-Morvan2000} or transient heating events like shocks \citep{HarkerDesch2002} temporarily heating up circumstellar material in disk regions otherwise too cold for forsterite formation to occur \citep[see also][for a review on dust formation and processing]{Henning2010}.

Given the large uncertainties in the temperature estimate and the fact that our assumption of an optically thin medium may be violated we can only speculate whether any radial mixing could have occurred. Since the temperature gradient in optically thick disks is much steeper than in an optically thin environment \citep[e.g.][]{Mulders2011}, not only will the best estimate for the dust distance become smaller but also the confidence interval narrower if we include radiative transfer effects. Thus it seems likely that the forsterite we detected is mostly located in narrow rings/tori. The distribution produced by large-scale radial mixing, however, is expected to be much more radially extended \citep[e.g.][]{Gail2001}. \citet[][]{Juhasz2010} also argue against radial mixing due to a lack of crystalline dust species other than forsterite at lower temperatures, which should have higher mass fractions than forsterite at the outer disk regions, according to radial mixing models \citep[][]{Gail2001}.

For one source in our sample, HD~100546, we seem to be able to rule out the radial-mixing scenario. For this system a full radiative transfer model \citep{Mulders2011} is available which shows that the forsterite is concentrated in a rather narrow ring. Since the temperatures in most of the sources are similar to HD~100546, it seems reasonable to assume that the distribution of the forsterite grains in those sources is also similar to that of HD~100546. Thus it seems unlikely that radial mixing can explain the narrow distribution of forsterite found in in the disks of our sample. As an alternative to radial mixing \citet{Bouwman2003} propose a connection between the formation of the crystalline silicates on site and the creation of a gap in the disk of HD~100546 through the gravitational interaction with a massive planet at around 10~AU. A planet opening a gap in the disk, would produce shock waves through tidal interaction with the disk \citep[e.g.][]{ LinPapaloizou1980,BossDurisen2005}, which might be 
enough to heat the circumstellar material to high enough temperatures for crystallisation to occur \citep[][]{Desch2005}. Planetary companions are not the only possible cause for shockwaves in protoplanetary disks, which could also be induced by disk instabilities \citep[e.g. ][]{HarkerDesch2002}. Given that all of our targets are still gas-rich \citep[][ and references therein]{Meeus2012,Fukagawa2010,Brown2012,Andrews2009}, small dust grains should be efficiently heated to high temperatures at disk radii of up to at least $\sim$10~AU.

Alternatively, collisions of large, possibly differentiated objects (planetesimals), could also produce crystalline olivines. \citet{Lisse2009} found that the \emph{Spitzer} spectrum of the debris disk around HD~172555 is dominated by silica emission and shows possible features arising from SiO gas, which the authors explained as evidence for hypervelocity collisions of planetesimals. However, contrary to the system of HD~172555, only very small amounts of silica are reported by \citet{Juhasz2010} for our targets where we detect the 69~$\mu$m forsterite emission band. Also no clear indication of SiO gas can be found in the residuals of the IRS spectra. Further, such collisions are less likely in the gas disks found in our sample \citep[][ and references therein]{Meeus2012,Fukagawa2010,Brown2012,Andrews2009} than in a gas-poor debris disk like HD~172555 since the interaction between disk and planetesimals will dramatically reduce the eccentricity of orbits \citep{BitschKley2010}.

Furthermore, the formation of olivines in larger, differentiated objects is a scenario that would result in a much higher iron content than we found \citep{Gail2010}. It is interesting to note that only for a few debris disk sources is there solid evidence for olivine with a high ($\sim$20\%) iron fraction, suggesting that the olivine material in these debris disks originated from larger differentiated bodies \citep{Olofsson2012}. This is also consistent with the recent detection of the 69~$\mu$m band in the young debris disk system $\beta$~Pictoris, which shape and position is consistent with olivine with an iron fraction of at most 1~\% \citep{deVries2012}. It seems that the bulk of the olivine dust in protoplanetary and young debris disks is formed through an equilibrium condensation process at high temperatures, and that only during the later debris disk phases a substantial amount of iron-rich crystalline silicates is produced through disruptive collisions of differentiated bodies.

From the evidence at hand we conclude that it is unlikely, although not impossible, that the crystalline silicates in our sample to be produced by large hypervelocity collisions of differentiated objects.


\subsection{Correlation with mid--IR and disk-- or host star properties}
\label{subsect:correlations}
No correlation between the detection or strength of the 69~$\mu$m feature and the 16~$\mu$m bands was found. A possible weak correlation between the 69~$\mu$m and the 33~$\mu$m bands could exist: sources with the strongest 69~$\mu$m band tend on average to be those sources with the strongest 33~$\mu$m bands. Note, however, that a detection at 69~$\mu$m does not automatically imply the presence of forsterite bands at the shorter wavelengths. Examples for this are AB~Aur, IRS~48 and AS~205 which have no detectable forsterite bands at mid--IR wavelengths and still showing a 69~$\mu$m band.

No conclusive correlation between the existence of the 69~$\mu$m band and the disk geometry has been found. Although only one of the eight forsterite detections was found in a flattened disk (HD~104237), the possibility exists that this is an observational bias due to the systematically lower fluxes of sources with flattened disks compared to those with strongly flaring disks. 

In Figures \ref{fig:SpTypeSNR} and \ref{fig:sptype_hist}, we show that the detection rate of the forsterite 69~$\mu$m band is highest among early spectral types with high S/N. The correlation between spectral type and detection rate indicated by this finding is not significant. The chance of finding the observed distribution of detections in our sample by chance is $\sim$8\%. The disk of AS~205 has geometric and compositional properties such as large scale height, low density, and near face-on orientation \citep{Andrews2009}, unusual among the disks of T~Tauri stars.  These properties increase the chance that \emph{Herschel} would be able to detect the 69~$\mu$m band. Therefore, we conclude that the indicated correlation of detection rate and spectral type is most likely an observational bias.

\bibliographystyle{aa}
\bibliography{DIGIT_Fo}

\Online

\begin{appendix}
\onecolumn
\section{}
%
\begin{figure*}
\centering
\subfigure{
\includegraphics[scale=0.75]{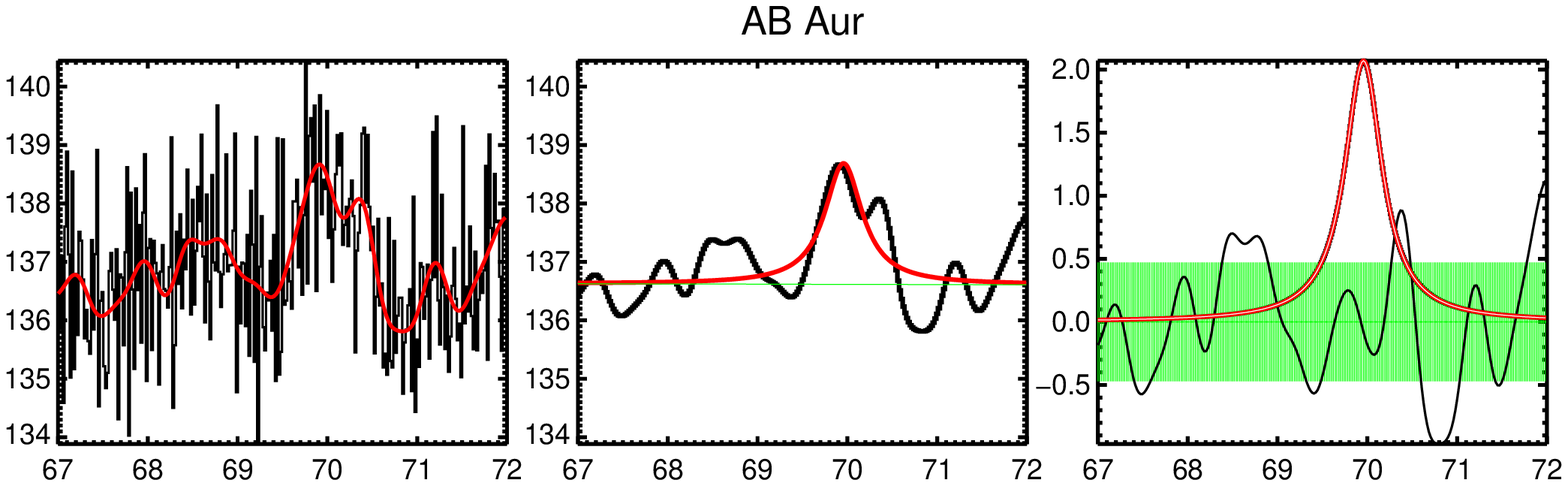}
\label{fig:abaur-overview01}}
\subfigure{
\includegraphics[scale=.75]{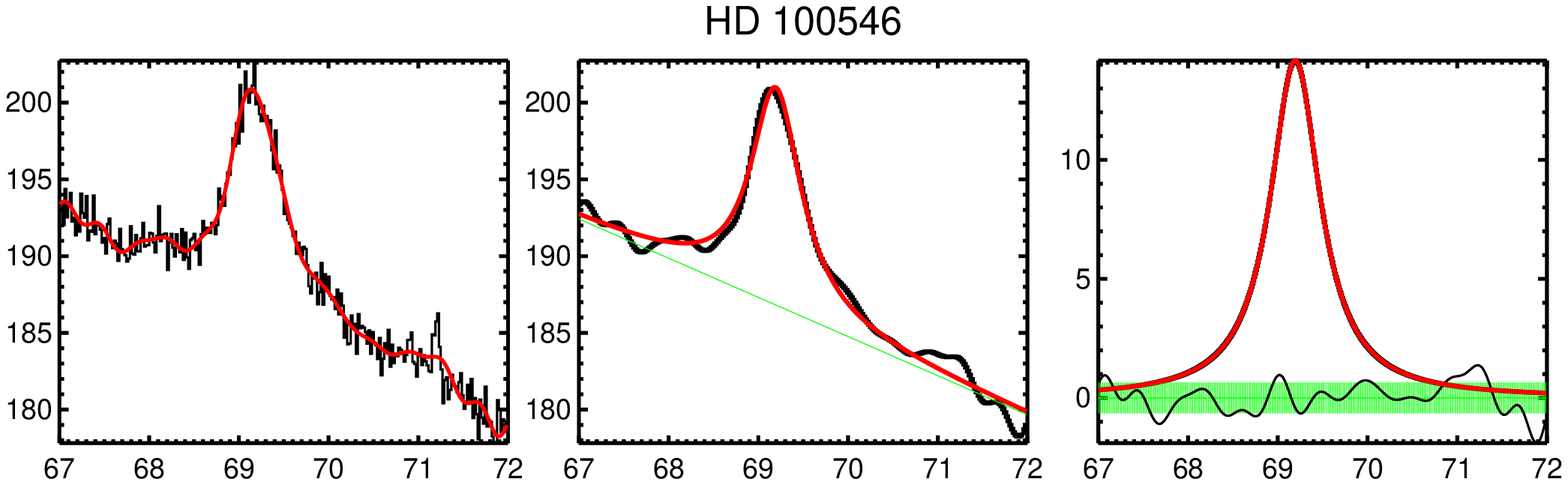}
\label{fig:hd100546-overview01}}
\subfigure{
\includegraphics[scale=.75]{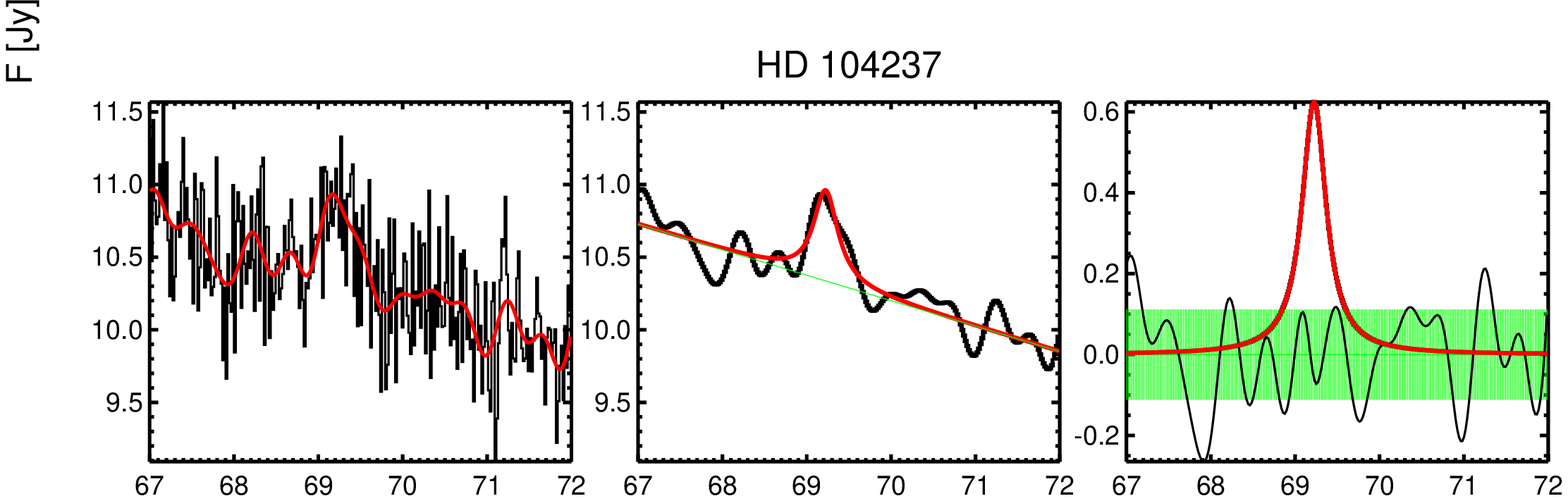}
\label{fig:hd104237-overview01}}
\subfigure{
\includegraphics[scale=.75]{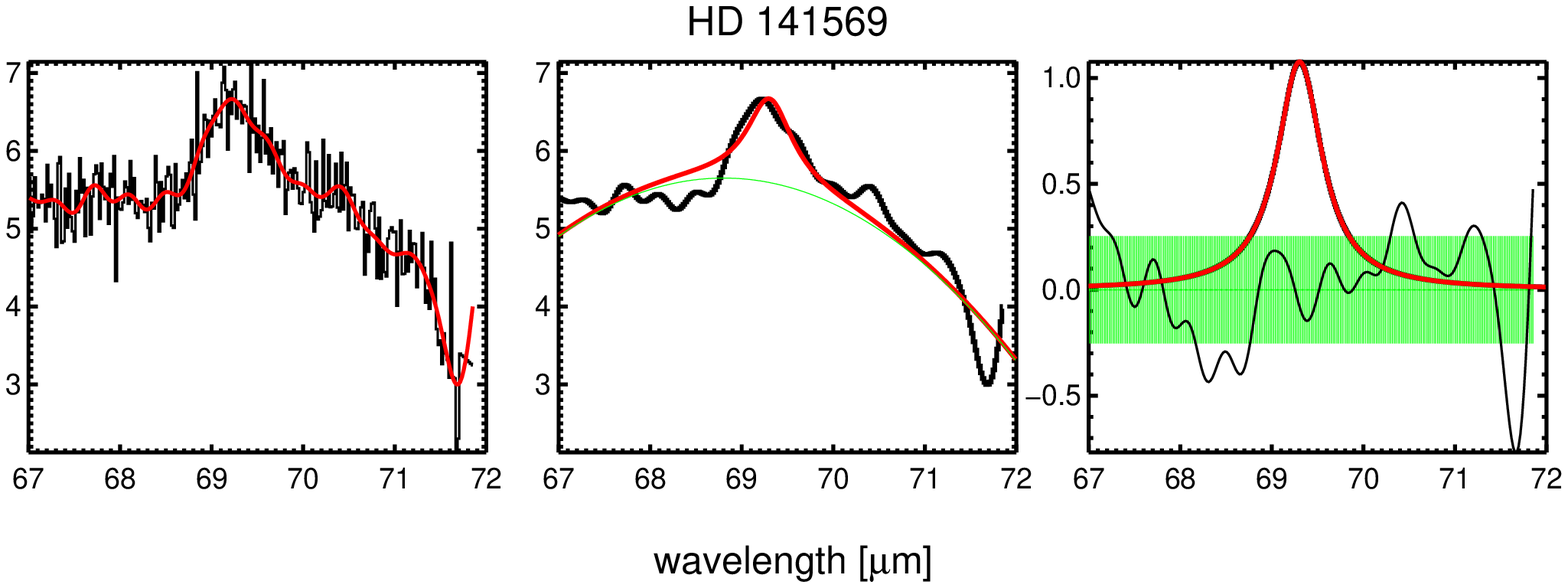}
\label{fig:hd141569-overview01}}
\caption{Comparison of the unmodified (black) and the noise--filtered (red) spectrum in the left column. The middle column shows the noise--filtered spectrum (black) overplotted with the best-fit model (red). The model is a sum of a 3$^{rd}$ order polynomial and a Lorentzian. In the right column we show a comparison of the residuals to the Lorentzian from the model. The error bars on the residuals are corresponding to the standard deviation of the residuals. In all sources shown in this figure the 69~$\mu$m forsterite band has been detected.}
\label{fig:overview01}
\end{figure*}

\begin{figure*}
\centering
\subfigure{
\includegraphics[scale=.75]{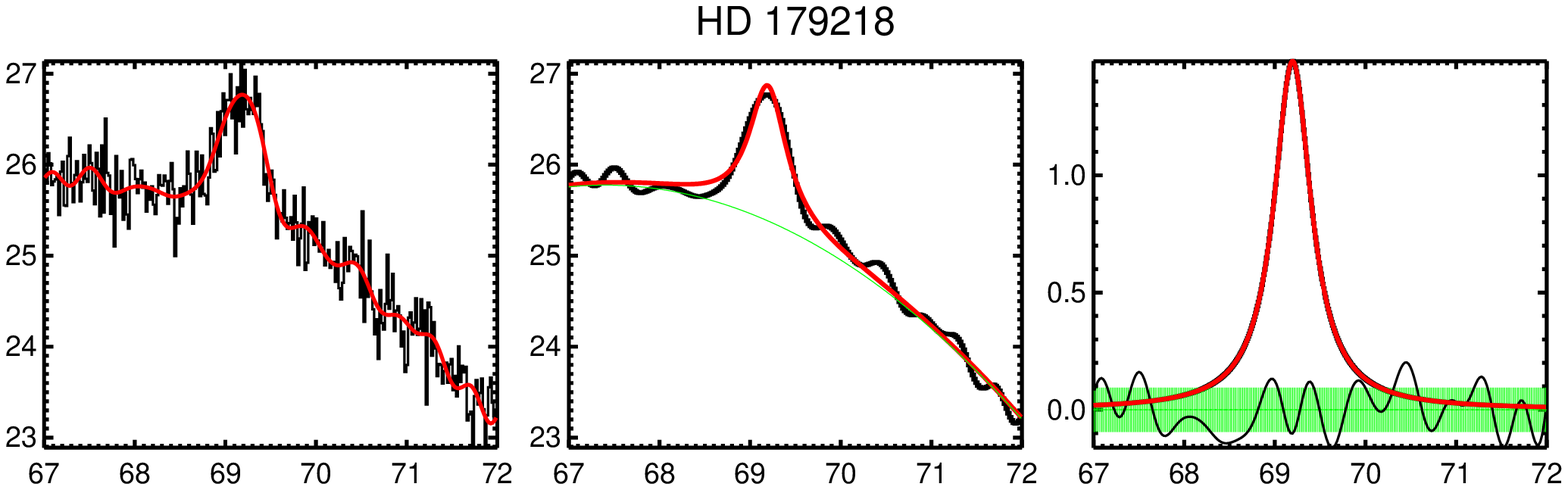}
\label{fig:hd179218-overview01}}
\subfigure{
\includegraphics[scale=.75]{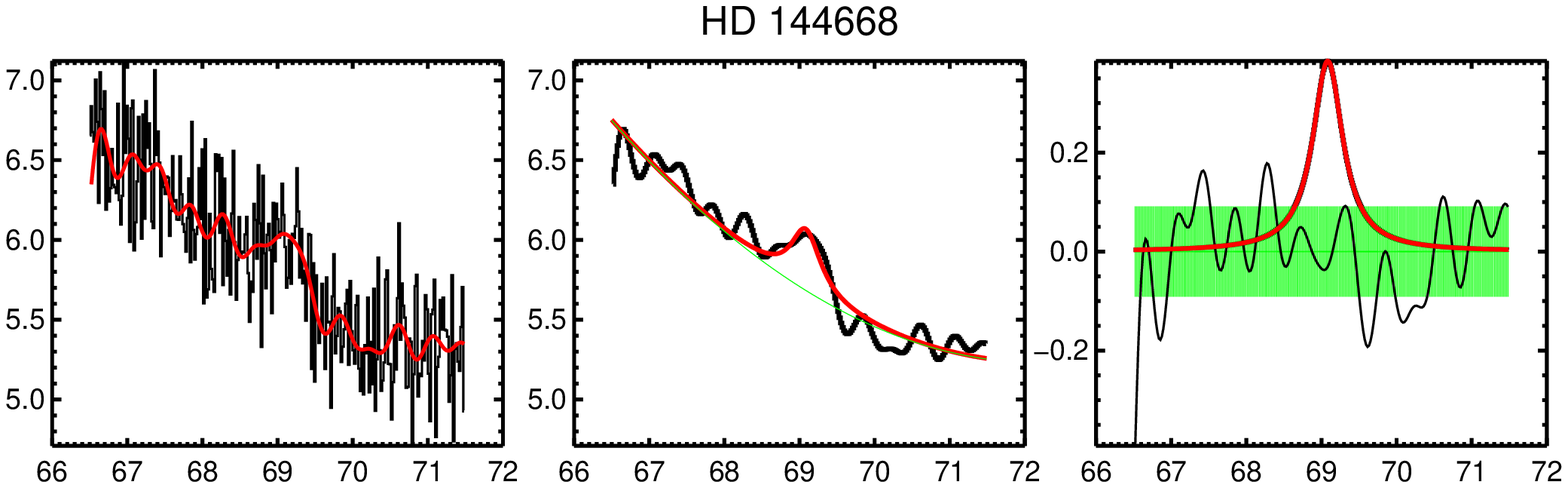}
\label{fig:hd144668-overview01}}
\subfigure{
\includegraphics[scale=.75]{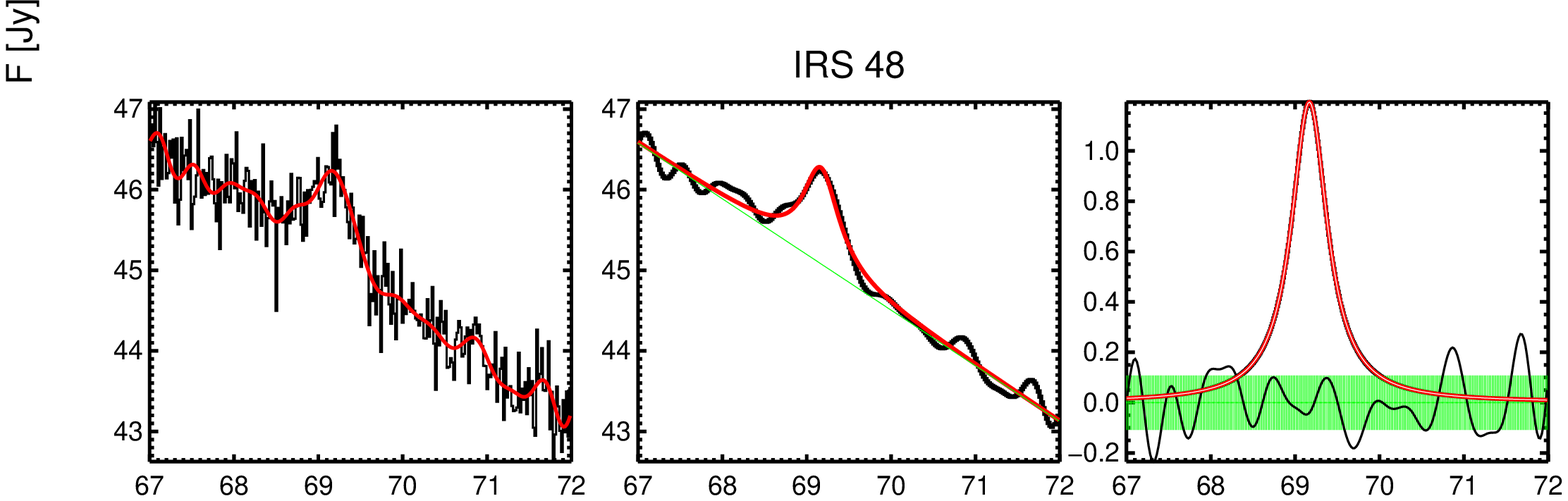}
\label{fig:irs48-overview01}}
\subfigure{
\includegraphics[scale=.75]{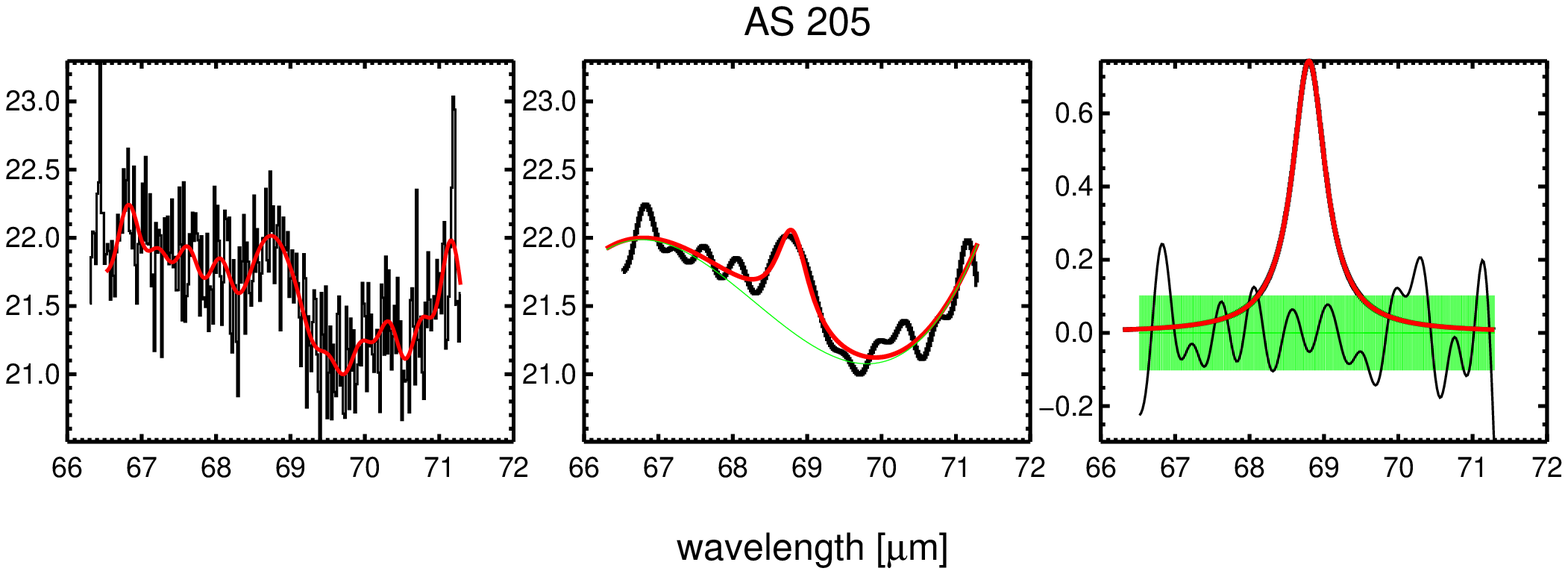}
\label{fig:as205-overview01}}
\caption{Same as fig. \ref{fig:overview01}. In all sources in this figure the 69~$\mu$m forsterite band has been detected.}
\label{fig:overview01_b}
\end{figure*}

\begin{figure*}
\centering
\subfigure{
\includegraphics[scale=.75]{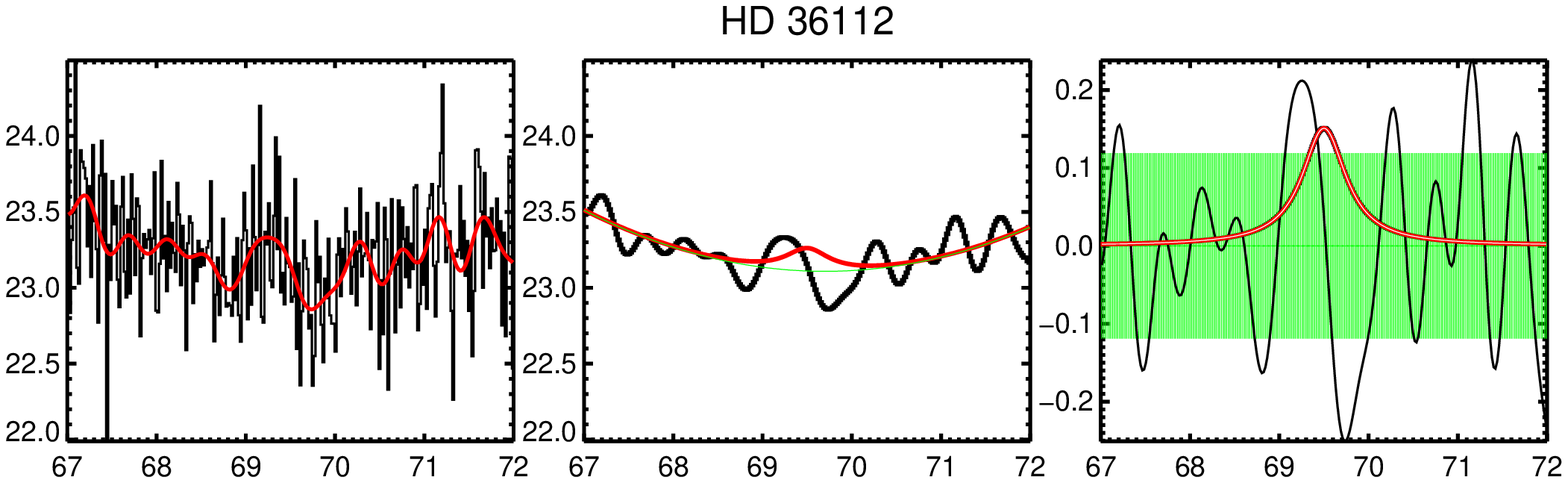}
\label{fig:hd36112-overview01}}
\subfigure{
\includegraphics[scale=.75]{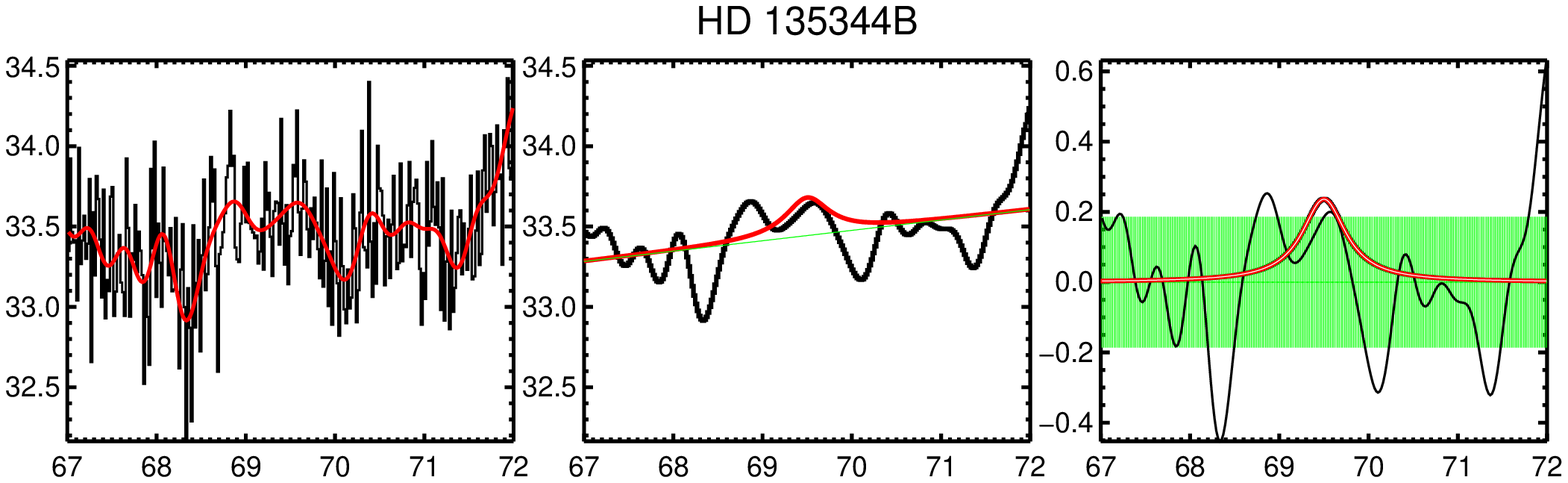}
\label{fig:hd135344-overview01}}
\subfigure{
\includegraphics[scale=.75]{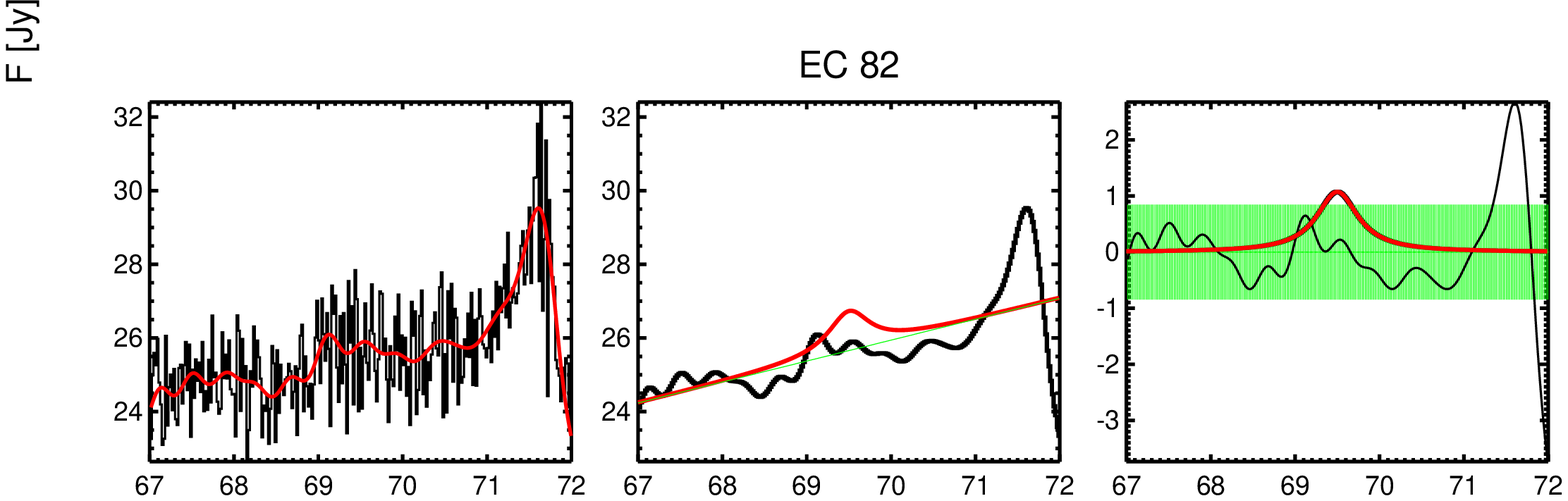}
\label{fig:ec82-overview01}}
\subfigure{
\includegraphics[scale=.75]{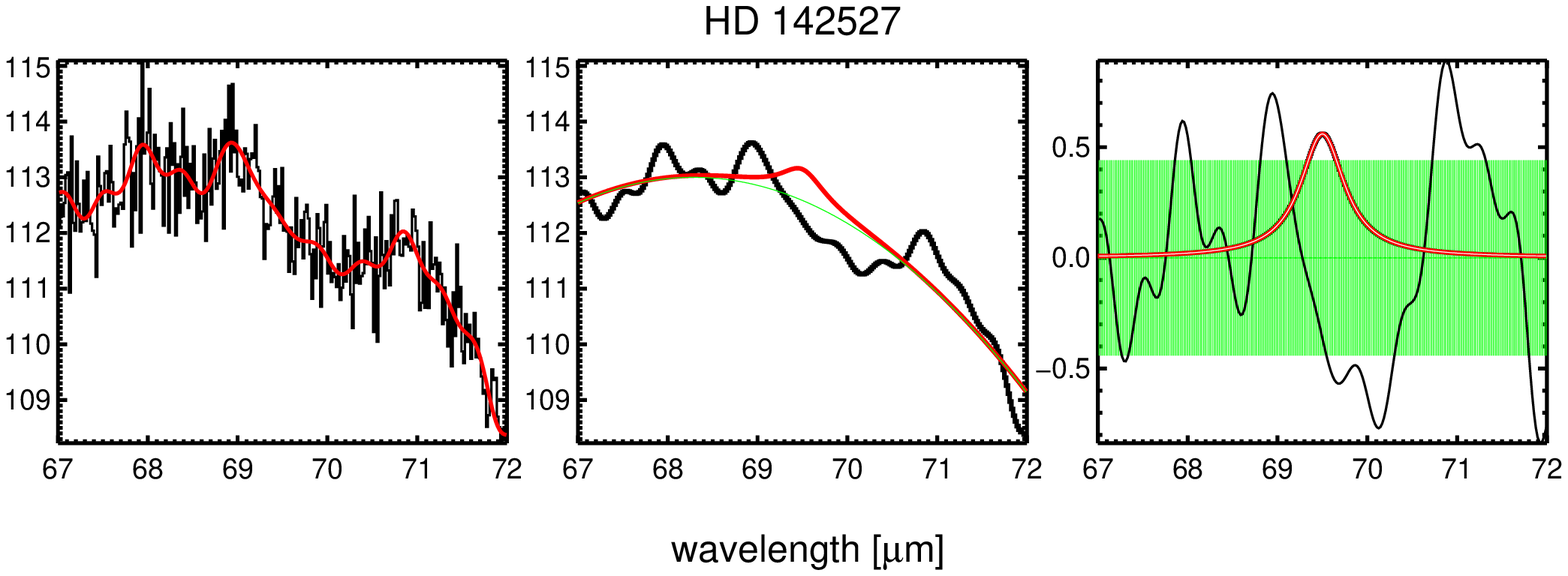}
\label{fig:hd142527-overview01}}
\caption{Same as fig. \ref{fig:overview01}. In the sources in this figure, the 69~$\mu$m forsterite band has not been detected.}
\label{fig:overview01_c}
\end{figure*}

\begin{figure*}
\centering
\subfigure{
\includegraphics[scale=.75]{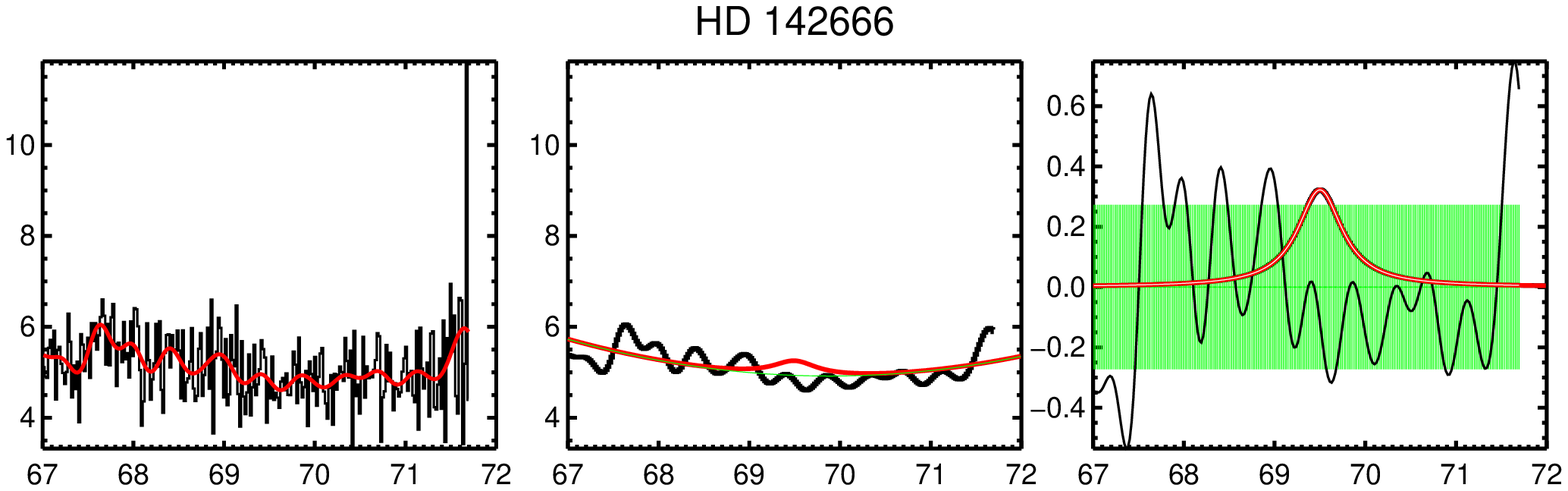}
\label{fig:hd142666-overview01}}
\subfigure{
\includegraphics[scale=.75]{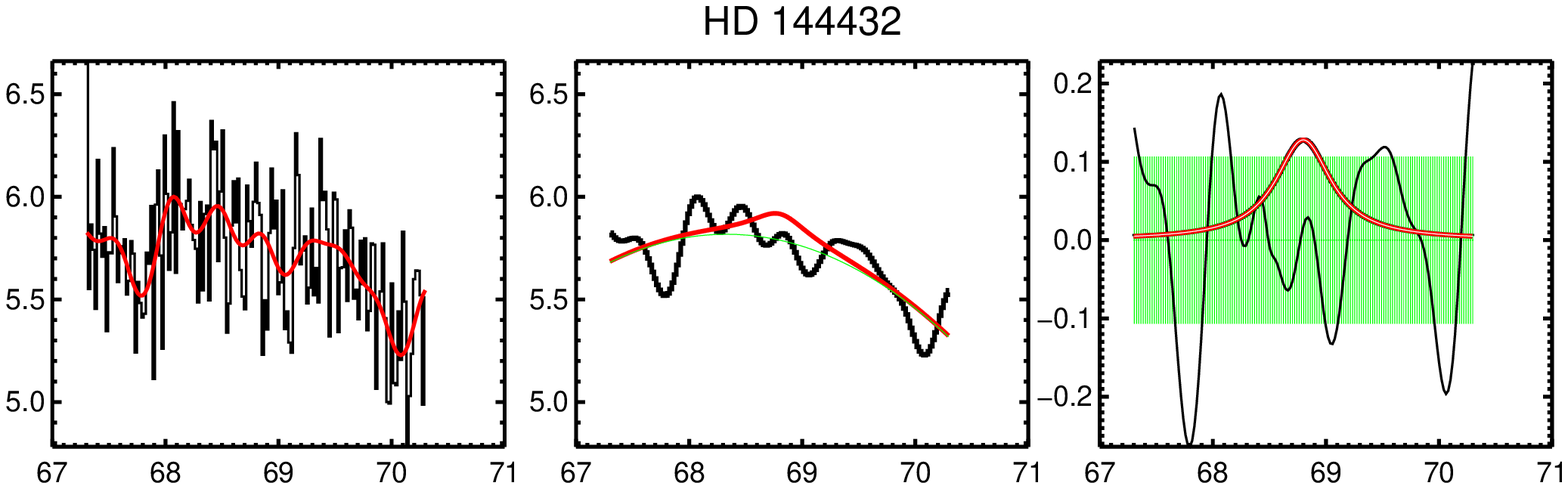}
\label{fig:hd144432-overview01}}
\subfigure{
\includegraphics[scale=.75]{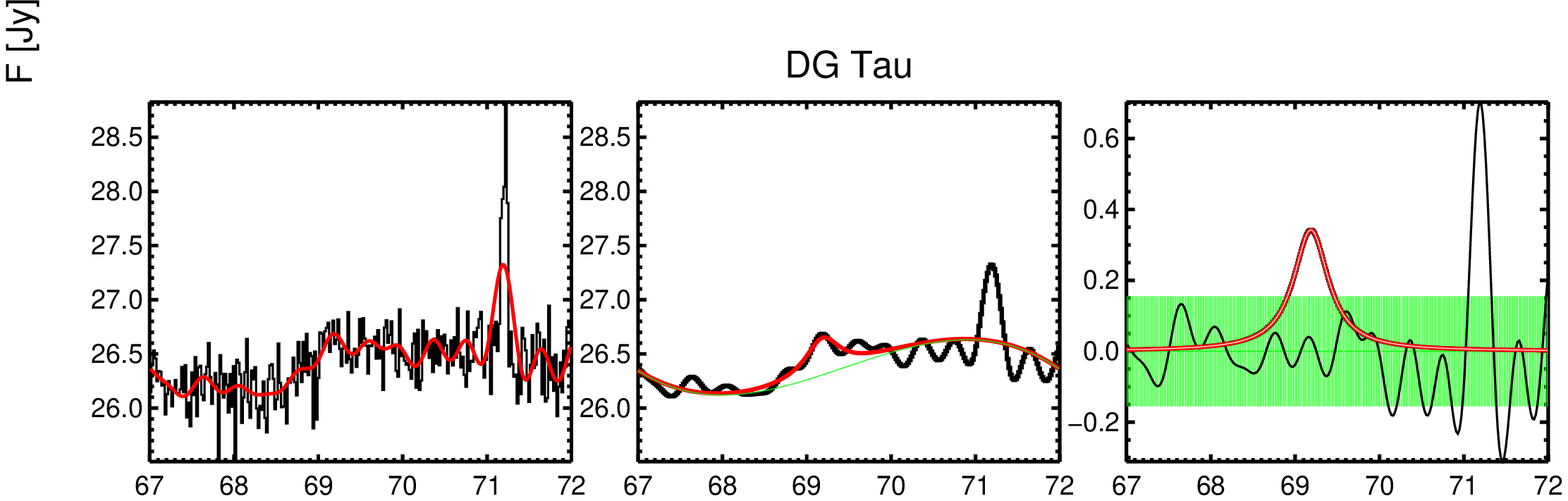}
\label{fig:dgtau-overview01}}
\subfigure{
\includegraphics[scale=.75]{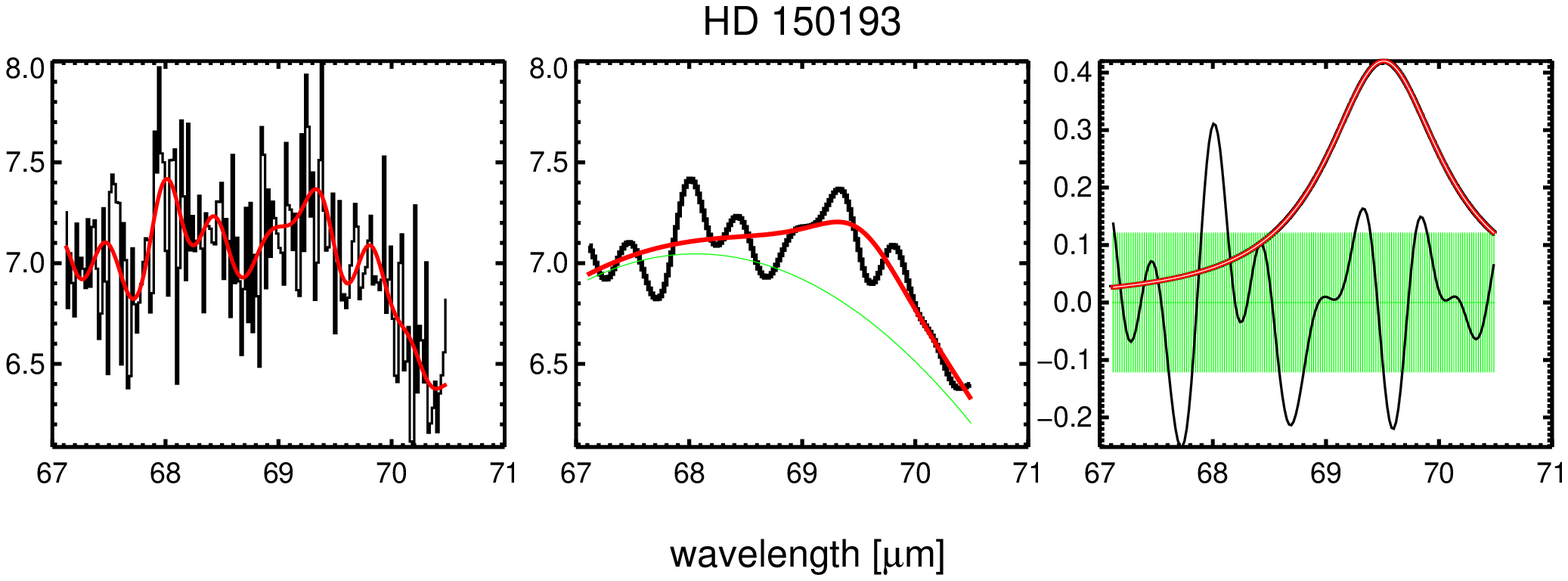}
\label{fig:hd150193-overview01}}
\label{fig:overview01_d}
\caption{Same as fig. \ref{fig:overview01}.  In the sources in this figure, the 69~$\mu$m forsterite band has not been detected.}
\end{figure*}

\begin{figure*}
\centering
\subfigure{
\includegraphics[scale=.75]{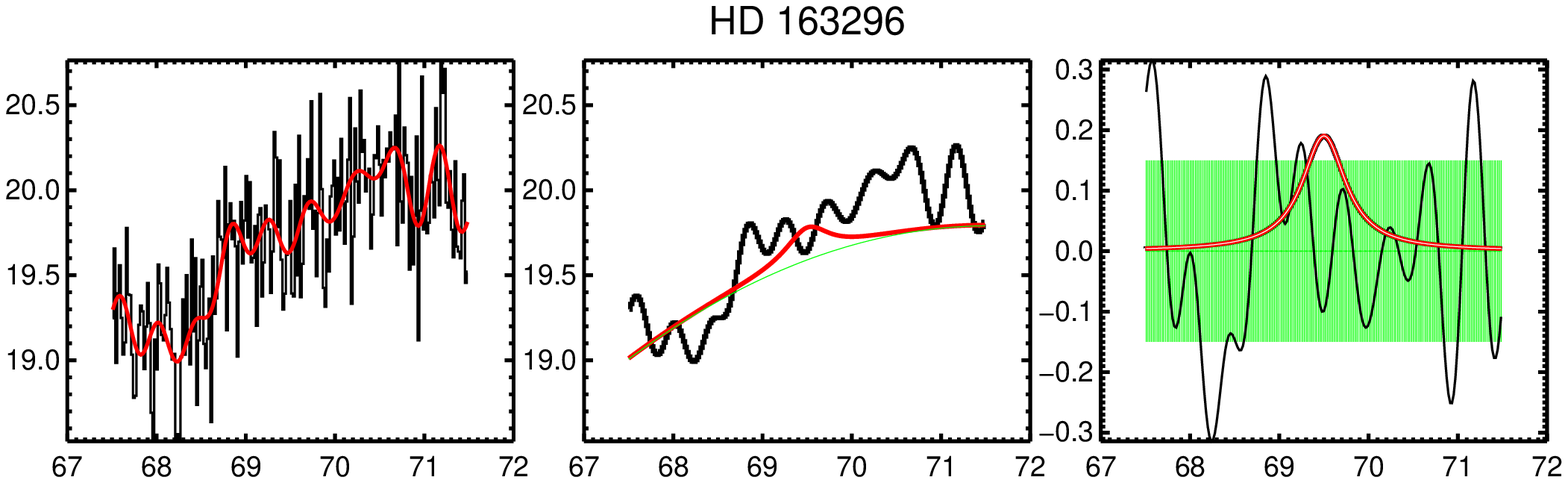}
\label{fig:hd163296-overview01}}
\subfigure{
\includegraphics[scale=.75]{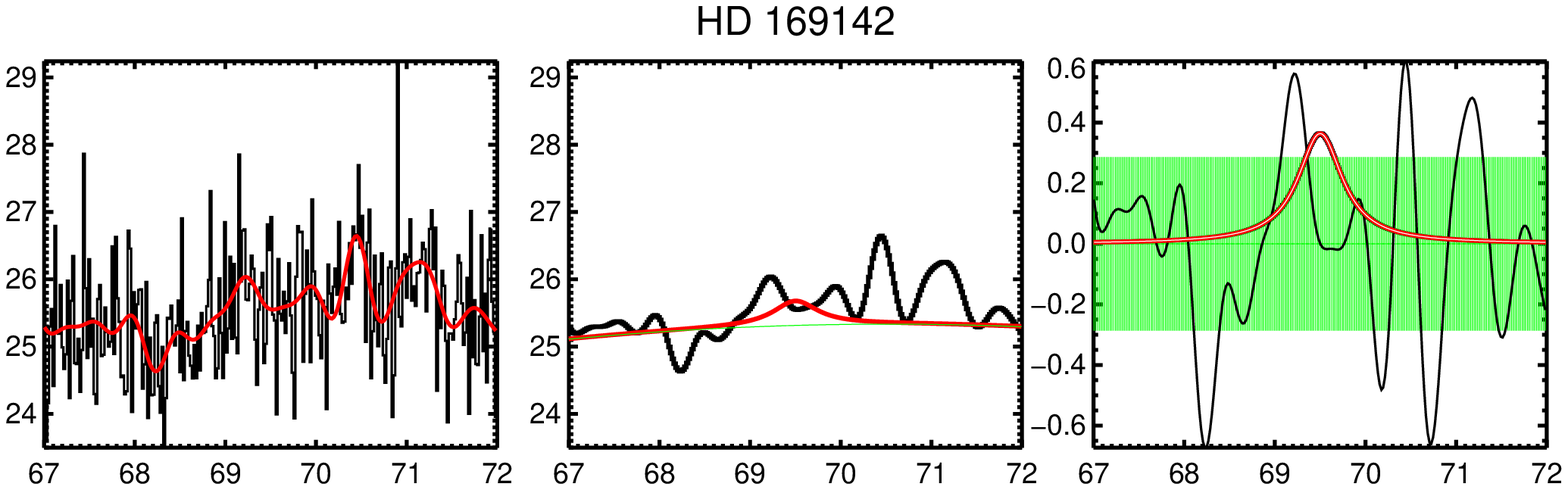}
\label{fig:hd169142-overview01}}
\subfigure{
\includegraphics[scale=.75]{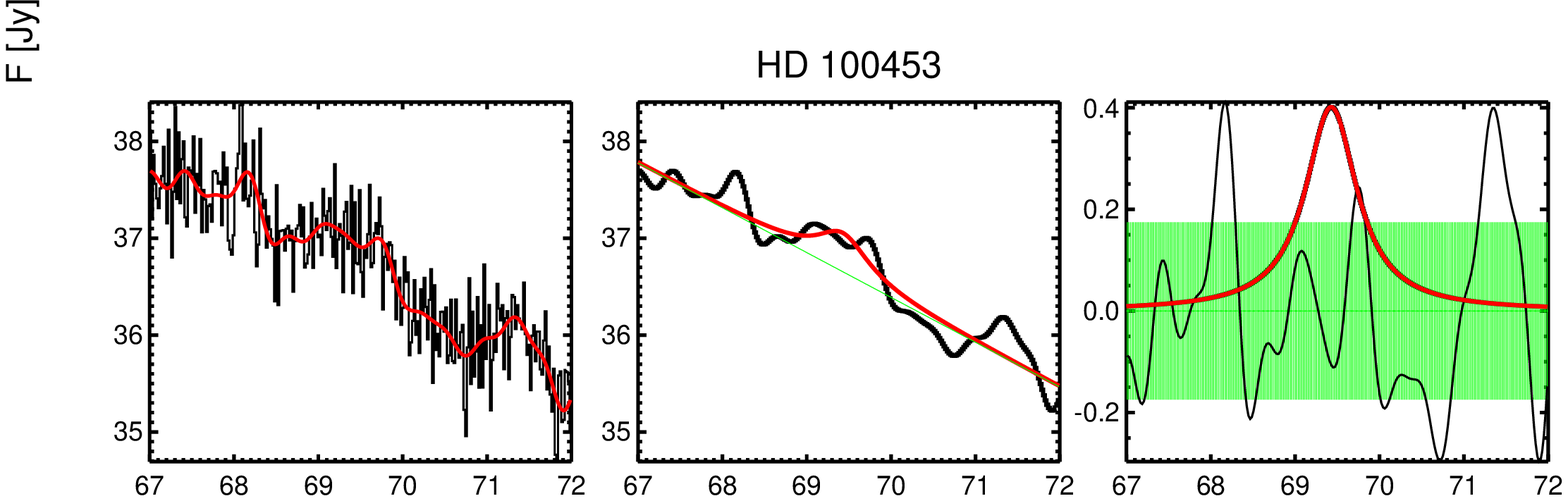}
\label{fig:hd100453-overview01}}
\subfigure{
\includegraphics[scale=.75]{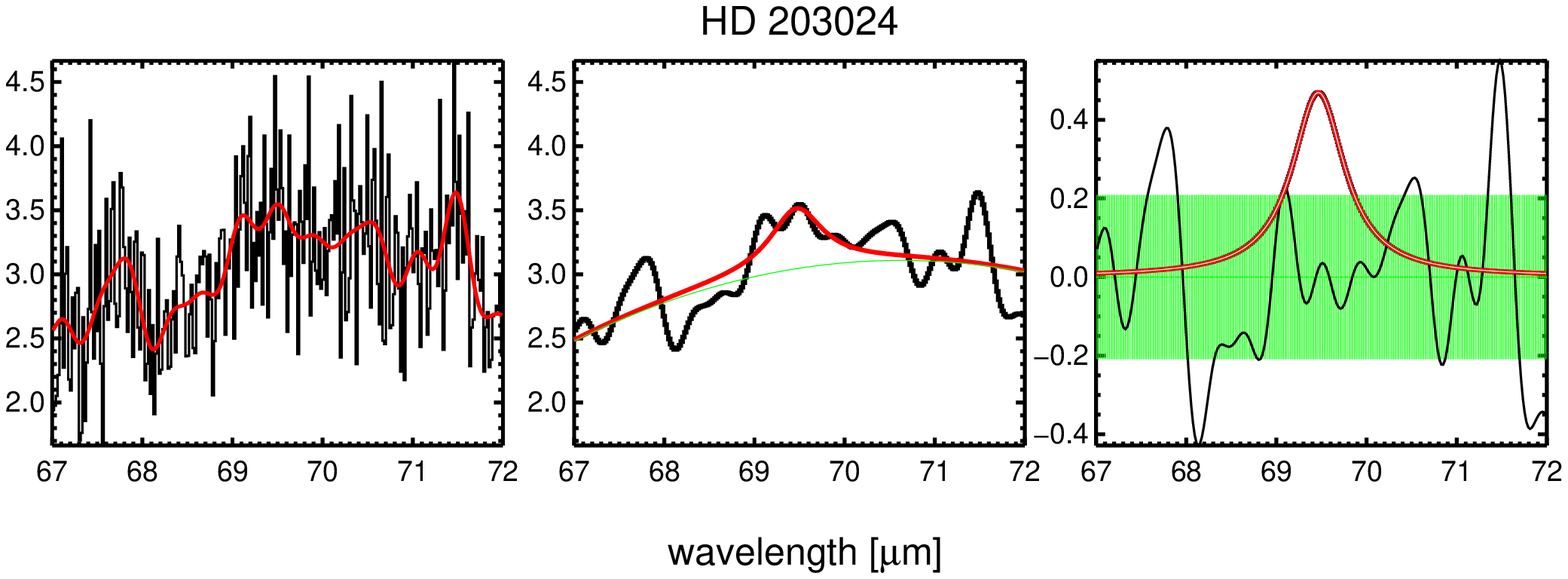}
\label{fig:hd203024-overview01}}
\caption{Same as fig. \ref{fig:overview01}. In the sources in this figure, the 69~$\mu$m forsterite band has not been detected.}
\label{fig:overview01_e}
\end{figure*}

\begin{figure*}
\centering
\subfigure{
\includegraphics[scale=.75]{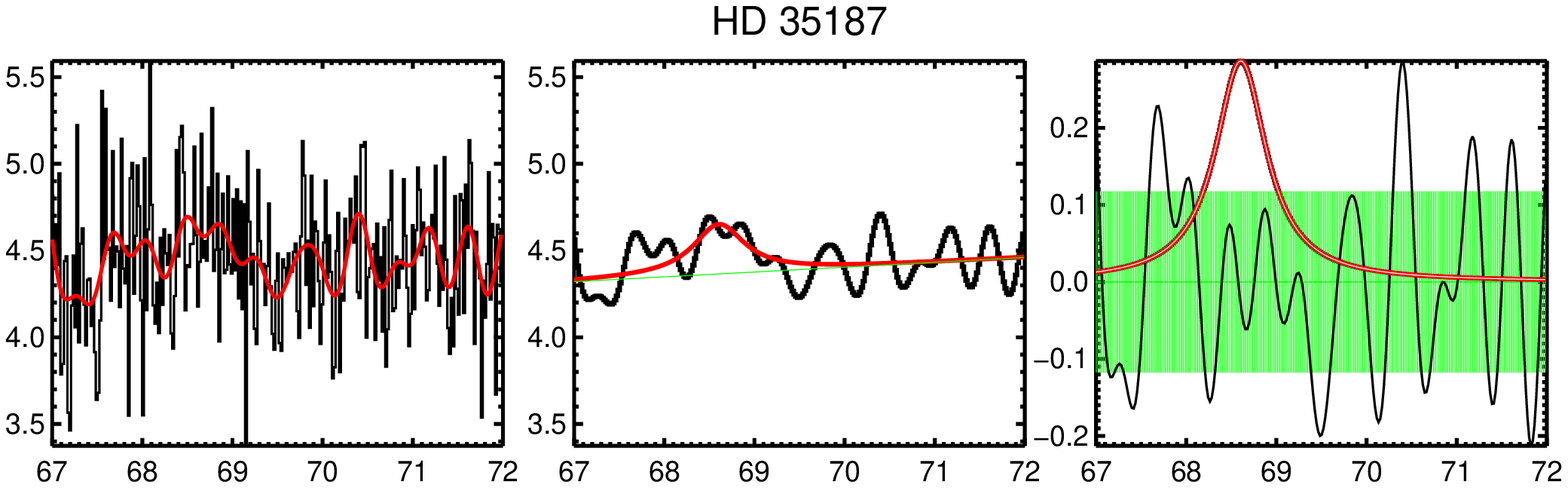}
\label{fig:hd35187-overview01}}
\subfigure{
\includegraphics[scale=.75]{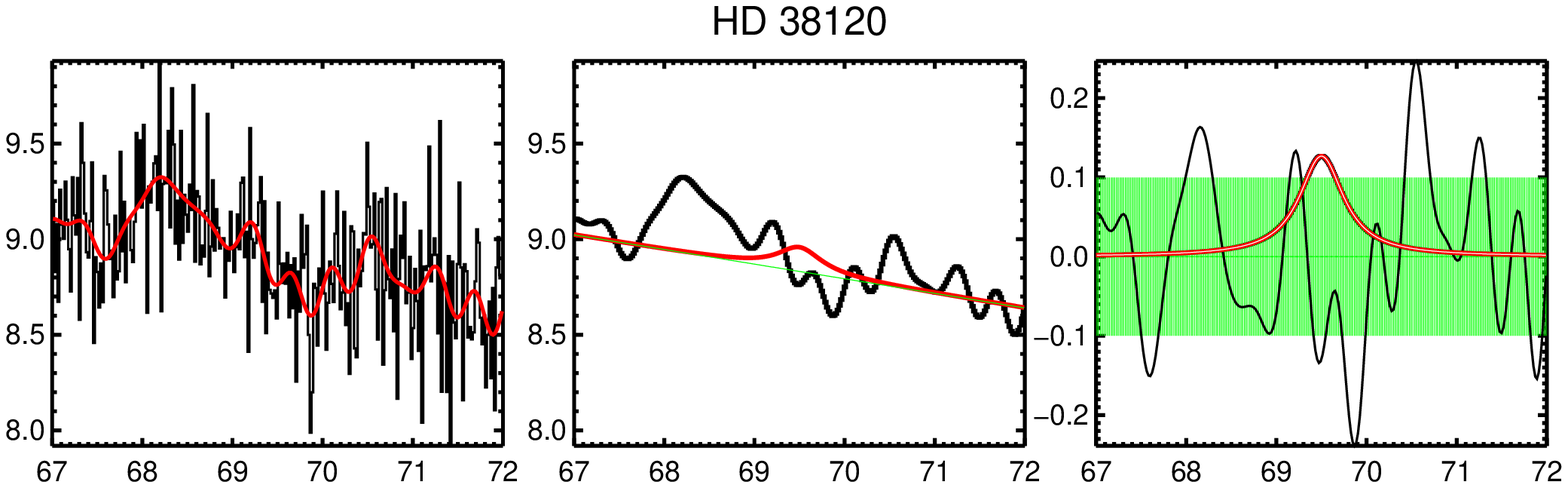}
\label{fig:hd38120-overview01}}
\subfigure{
\includegraphics[scale=.75]{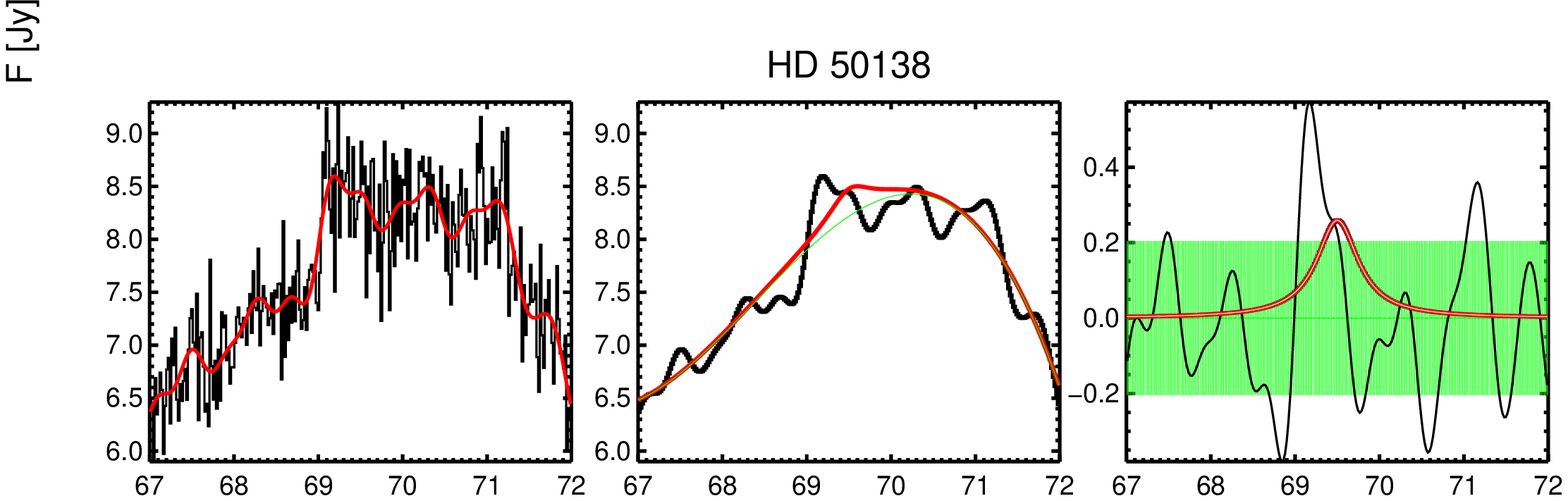}
\label{fig:hd50138-overview01}}
\subfigure{
\includegraphics[scale=.75]{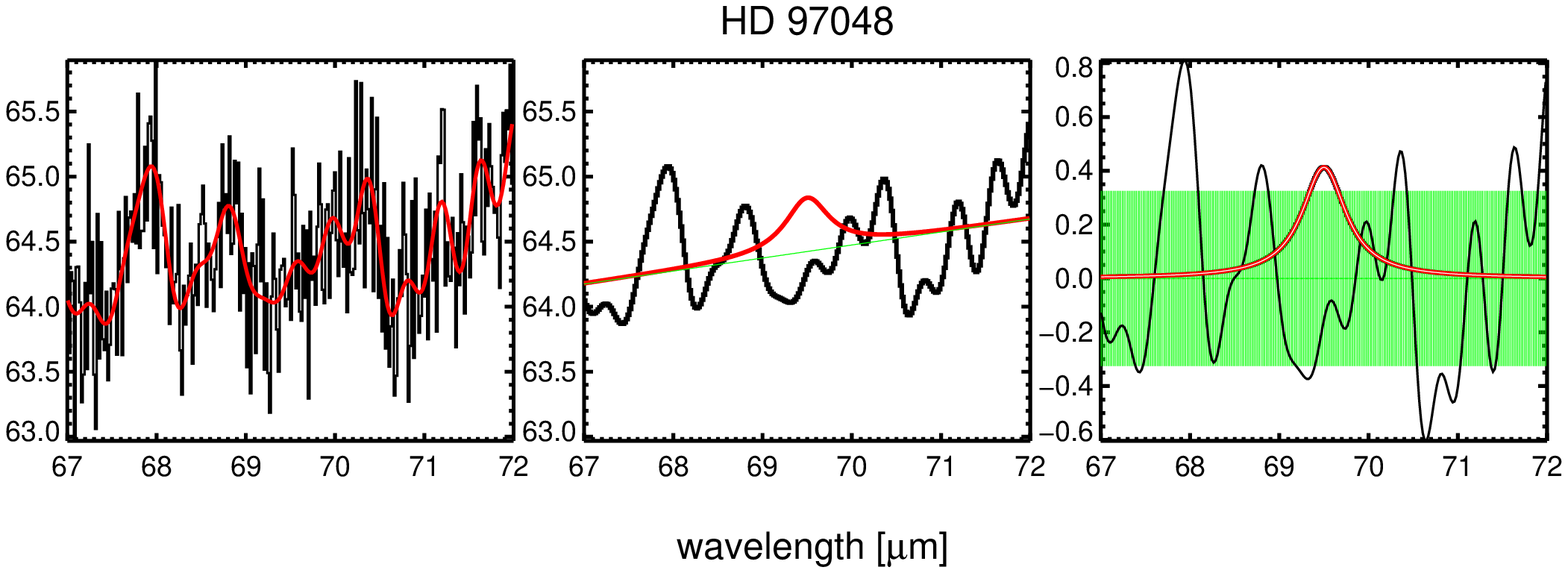}
\label{fig:hd97048-overview01}}
\caption{Same as fig. \ref{fig:overview01}. In the sources in this figure, the 69~$\mu$m forsterite band has not been detected.}
\label{fig:overview01_f}
\end{figure*}

\begin{figure*}
\centering
\subfigure{
\includegraphics[scale=.75]{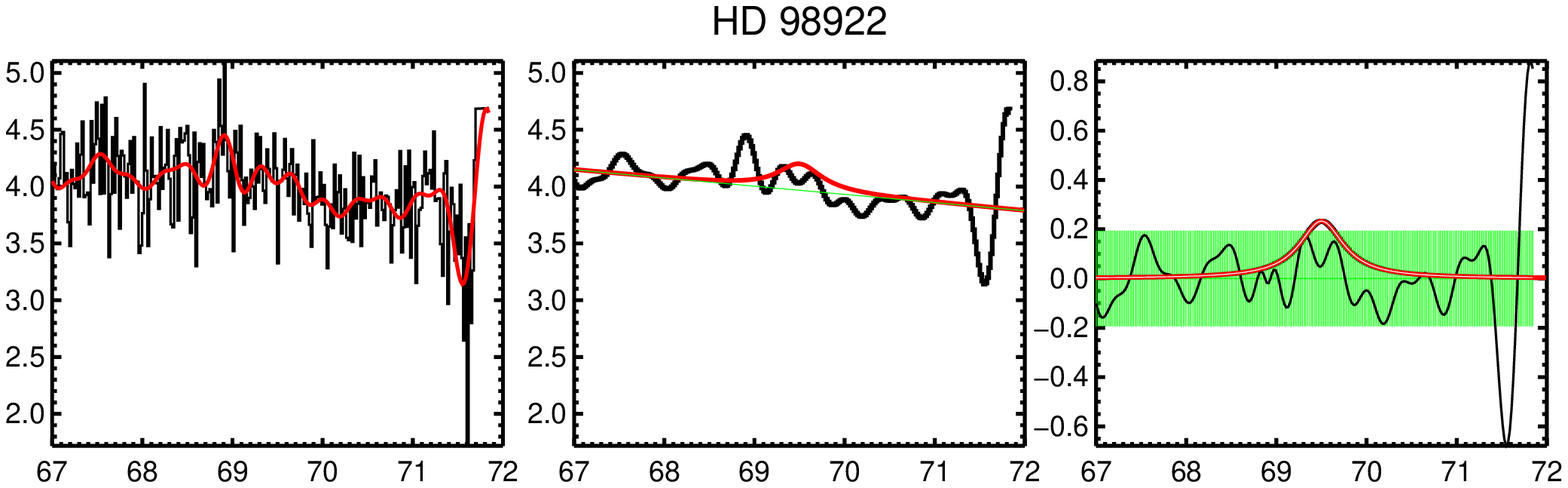}
\label{fig:hd98922-overview01}}
\subfigure{
\includegraphics[scale=.75]{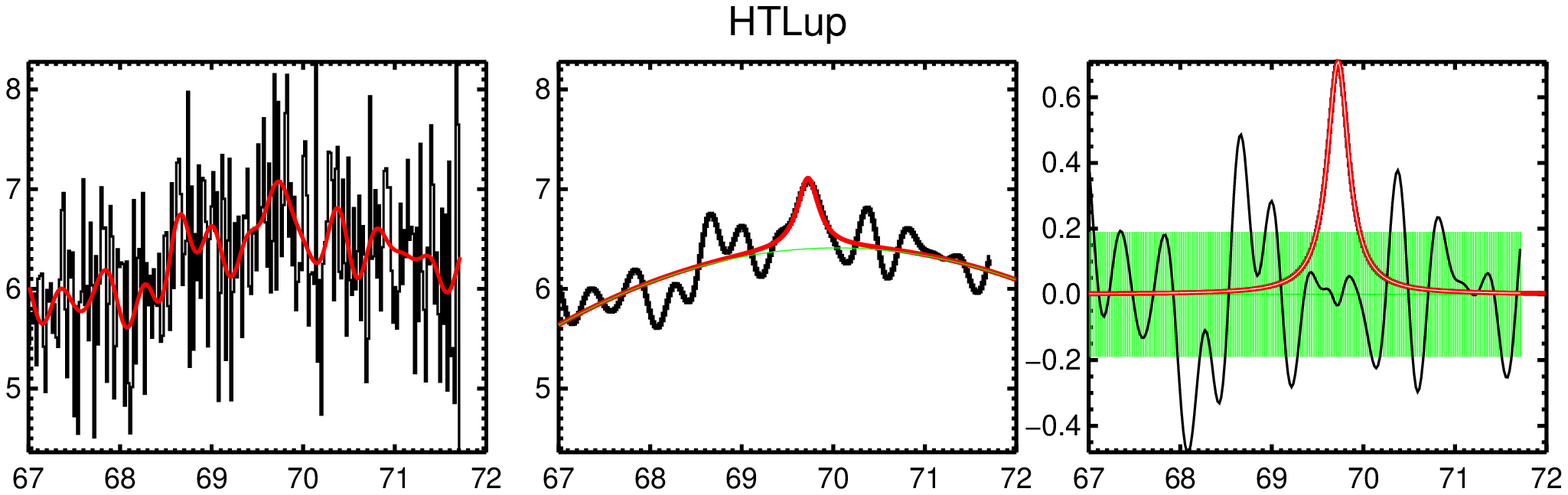}
\label{fig:htlup-overview01}}
\subfigure{
\includegraphics[scale=.75]{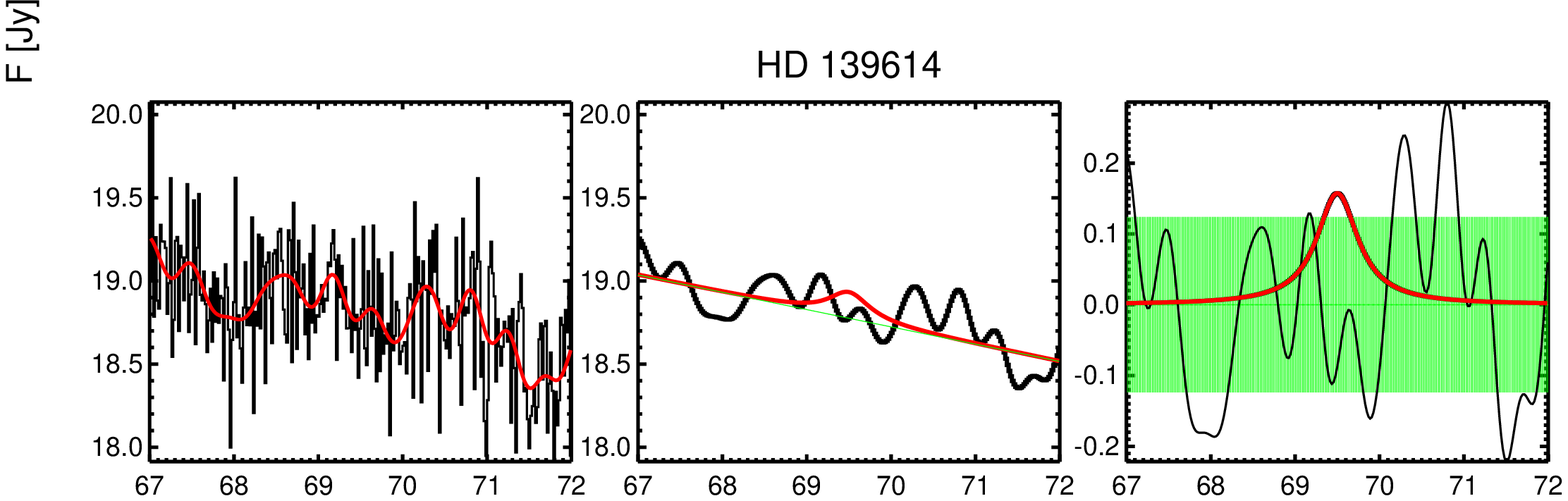}
\label{fig:hd139614-overview01}}
\subfigure{
\includegraphics[scale=.75]{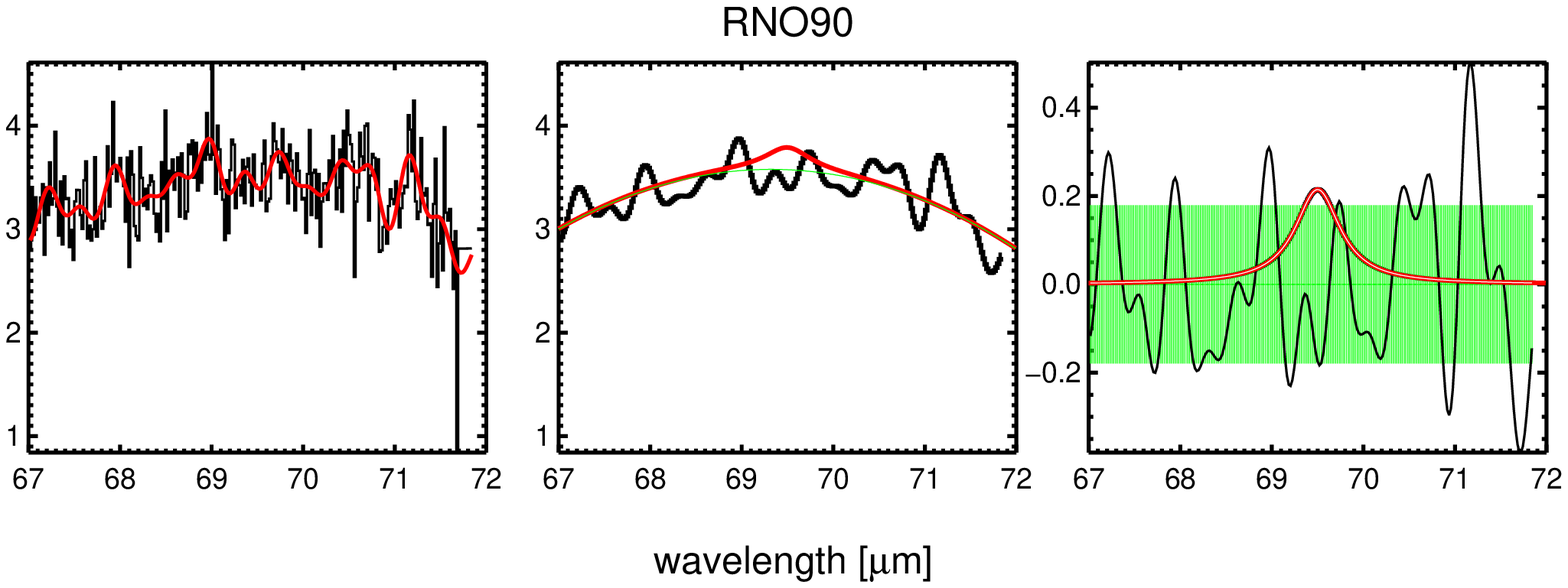}
\label{fig:rno90-overview01}}
\caption{Same as fig. \ref{fig:overview01}. In the sources in this figure, the 69~$\mu$m forsterite band has not been detected.}
\label{fig:overview01_g}
\end{figure*}

\begin{figure*}
\centering
\subfigure{
\includegraphics[scale=.75]{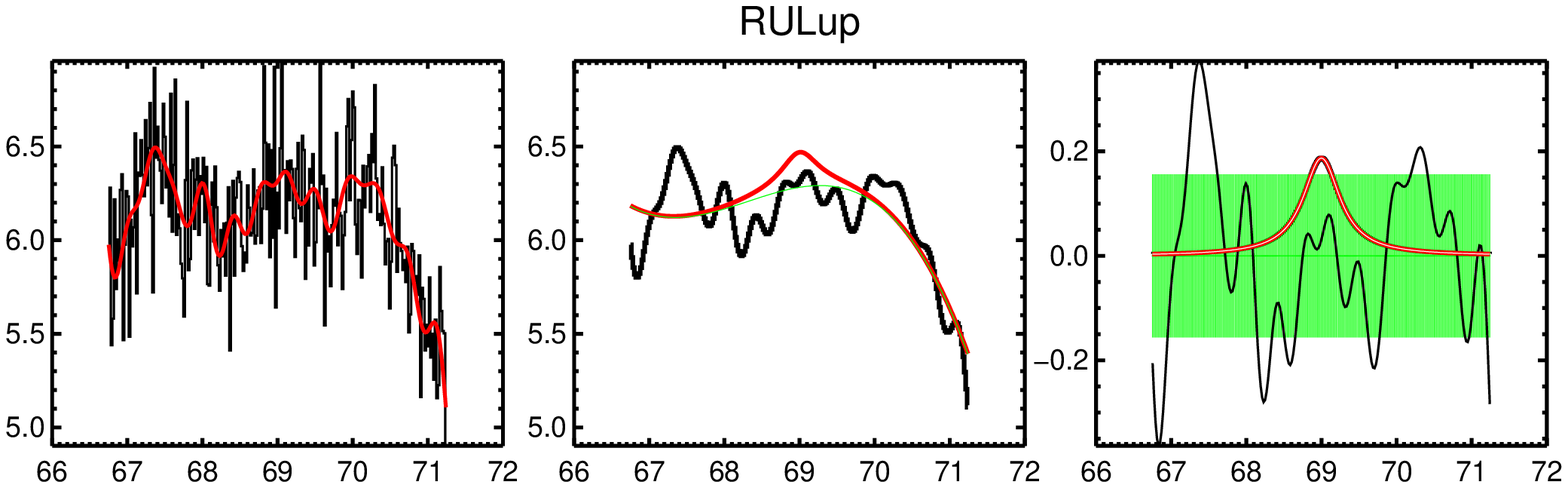}
\label{fig:rulup-overview01}}
\subfigure{
\includegraphics[scale=.75]{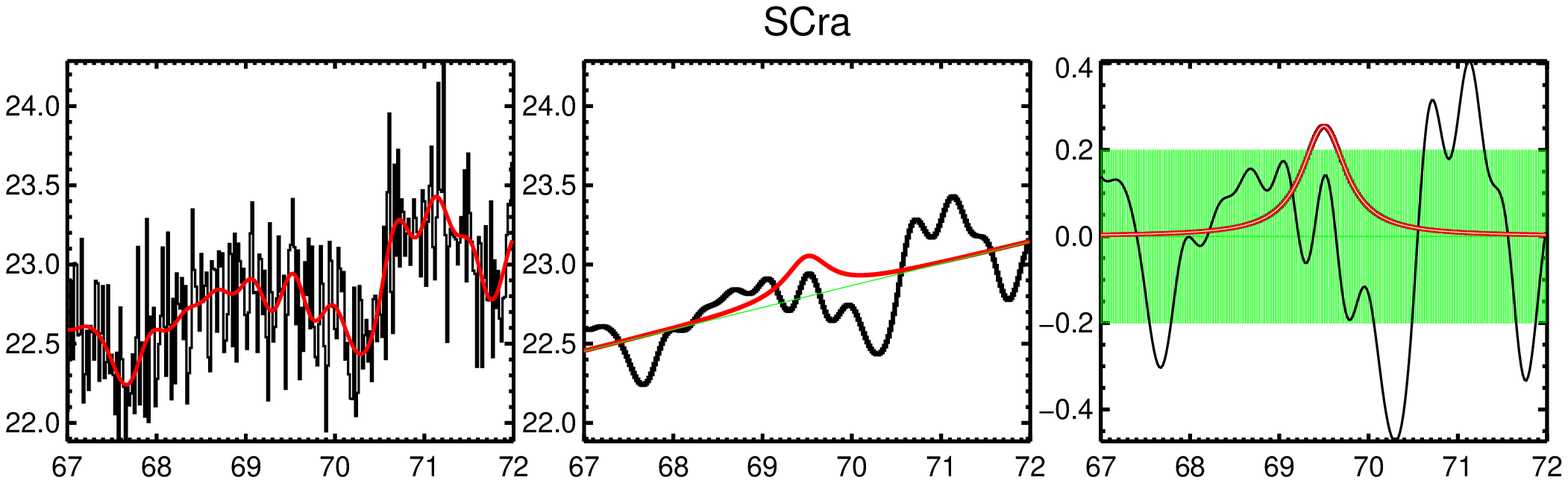}
\label{fig:scra-overview01}}
\subfigure{
\includegraphics[scale=.75]{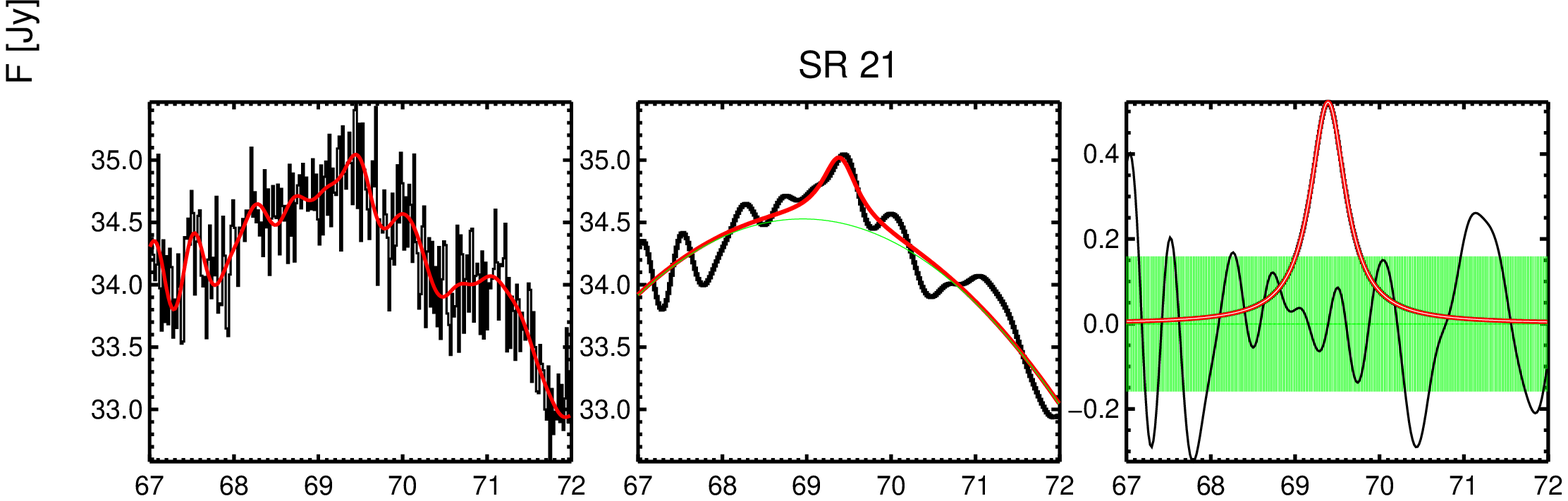}
\label{fig:sr21-overview01}}
\subfigure{
\includegraphics[scale=.75]{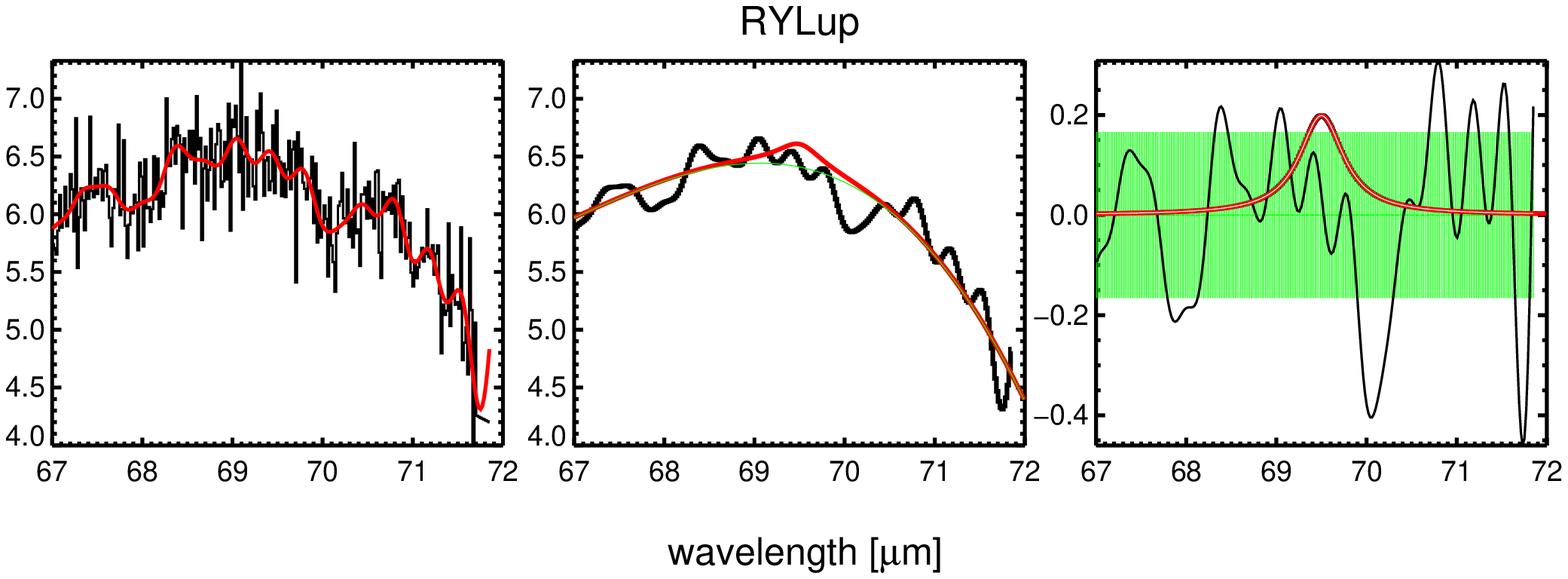}
\label{fig:rylup-overview01}}
\caption{Same as fig. \ref{fig:overview01}. In the sources in this figure, the 69~$\mu$m forsterite band has not been detected.}
\label{fig:overview01_h}
\end{figure*}

\end{appendix}

\end{document}